\documentclass[a4paper]{article}

\usepackage{amsmath,amssymb,color,amsthm,hyperref}
\usepackage{tikz}
\usetikzlibrary{decorations.markings,arrows}
\usepackage{pgf,pgfarrows}

\numberwithin{equation}{section}

\newcommand\gl{\text{gl}}
\newcommand\su{\text{su}}
\newcommand\so{\text{so}}
\renewcommand\sp{\text{sp}}
\newcommand\sgl{\text{sl}}
\newcommand\spl{\text{sp}}
\newcommand\pgl{\text{pgl}}
\newcommand\psl{\text{psl}}
\newcommand\osp{\text{osp}}
\newcommand\tB{{\sf B}}
\newcommand\tG{{\sf G}}

\newcommand\CY{\text{CY}}
\newcommand\AdS{\text{AdS}}
\newcommand\Sphere{\text{S}}
\newcommand\GL{\text{GL}}
\newcommand\OSP{\text{OSP}}
\newcommand\PSU{\text{PSU}}
\newcommand\SO{\text{SO}}
\newcommand\U{\text{U}}
\newcommand\SU{\text{SU}}
\newcommand\SL{\text{SL}}
\newcommand\SP{\text{SP}}

\newcommand\groupG{\text{G}}
\newcommand\groupGp{\text{G}'}
\newcommand\groupH{\text{H}}
\newcommand\groupHp{\text{H}'}
\newcommand\groupU{\text{U}}

\DeclareMathOperator{\Coh}{\text{H}}
\newcommand\CPone{\ensuremath{\text{$\mathbb{CP}$}^{1|2}
}}

\DeclareMathOperator{\Img}{\text{Im}}
\DeclareMathOperator{\Ker}{\text{Ker}}
\DeclareMathOperator{\End}{\text{End}}
\DeclareMathOperator{\Hom}{\text{Hom}}
\DeclareMathOperator{\Ind}{\text{Ind}}
\DeclareMathOperator{\Inv}{\text{Inv}}
\DeclareMathOperator{\tr}{\text{tr}}
\DeclareMathOperator{\str}{\text{str}}
\DeclareMathOperator{\sdim}{\text{sdim}}
\DeclareMathOperator{\Ad}{{\text{Ad}}}
\DeclareMathOperator{\ad}{{\text{ad}}}
\DeclareMathOperator{\id}{{\text{id}}}

\newcommand{\Real}{\mathbb{R}}
\newcommand{\Complex}{\mathbb{C}}

\newcommand{\Integer}{\mathbb{Z}}
\newcommand{\CP}{\mathbb{CP}}

\newcommand{\bartial}{{\bar{\partial}}}
\newcommand{\defeq}{\overset{\text{def}}{=}}

\newcommand{\mat}{\begin{pmatrix}}
\newcommand{\tam}{\end{pmatrix}}
\newcommand{\smat}{\left(\begin{smallmatrix}}
\newcommand{\stam}{\end{smallmatrix}\right)}

\newcommand\SRep{\text{SRep}}

\renewcommand{\b}{\mathfrak{b}}
\newcommand{\g}{\mathfrak{g}}
\newcommand{\h}{\mathfrak{h}}
\renewcommand{\k}{\mathfrak{k}}
\newcommand{\den}{{\underline{0}}}
\newcommand{\deo}{{\underline{1}}}

\newcommand{\cA}{\mathcal{A}}
\newcommand{\cB}{\mathcal{B}}

\newcommand{\cE}{\mathcal{E}}
\newcommand{\cF}{\mathcal{F}}
\newcommand{\cG}{\mathcal{G}}
\newcommand{\cH}{\mathcal{H}}
\newcommand{\cI}{\mathcal{I}}

\newcommand{\cK}{\mathcal{K}}
\newcommand{\cL}{\mathcal{L}}
\newcommand{\cM}{\mathcal{M}}
\newcommand{\cN}{\mathcal{N}}
\newcommand{\cO}{\mathcal{O}}
\newcommand{\cP}{\mathcal{P}}
\newcommand{\cQ}{\mathcal{Q}}

\newcommand{\cS}{\mathcal{S}}
\newcommand{\cT}{\mathcal{T}}
\newcommand{\cU}{\mathcal{U}}
\newcommand{\cV}{\mathcal{V}}

\newcommand{\bF}{\mathbb{F}}

\newcommand{\bK}{\mathbb{K}}
\newcommand{\bS}{\mathbb{S}}

\newcommand\ag{\mathfrak{\hat{g}}}

\newcommand{\bz}{{\bar{z}}}

\newcommand{\bh}{{\bar{h}}}
\newcommand{\bJ}{{\bar{J}}}

\newcommand{\Typ}{\text{TRep}}
\newcommand{\ATyp}{\text{ARep}}

\definecolor{LightRed}{rgb}{1.0,0.5,0.5}
\definecolor{DarkGreen}{rgb}{0.0,0.90,0.0}
\definecolor{LemonChiffon}{rgb}{1.0,0.9,0.8}
\definecolor{LightGreen}{rgb}{.798,1.0,0.4}
\definecolor{Orange}{rgb}{1,0.6,0.0}
\definecolor{grey}{rgb}{.8,0.9,0.9}
\definecolor{pink}{rgb}{1,0.753,0.796}

\title{\Huge Superspace conformal field theory}
\author{Thomas Quella$^1$ and Volker Schomerus$^2$\\[5mm]
  \small $^1$ Universit\"at zu K\"oln, Institut f\"ur Theoretische Physik,
  Z\"ulpicher Stra\ss{}e 77, 50937 Cologne, Germany\\
  \small$^2$ DESY, Theory Group, Notkestrasse 85, Bldg. 2a, 22603
Hamburg, Germany\\[2mm]
\small{\em Emails:} {\sc Thomas.Quella@uni-koeln.de, Volker.Schomerus@desy.de}}

\date{}

\begin{document}
\maketitle

\begin{abstract}
  Conformal sigma models and WZW models on coset superspaces provide
  important examples of logarithmic conformal field theories. They
  possess many applications to problems in string and condensed matter
  theory. We review recent results and developments, including the
  general construction of WZW models on type I supergroups, the
  classification of conformal sigma models and their embedding into
  string theory.
\end{abstract}
\medskip


\date{}

\tableofcontents

\section{Introduction}

  Two-dimensional conformal field theories with central charge $c=0$,
  or other values below the unitarity bound $c=1/2$, play a
  fundamental role for string theory as well as statistical
  mechanics. With a few isolated exceptions, most models within this
  class possess world-sheet correlation functions with logarithmic
  divergencies. With the lack of chiral factorization and other
  familiar features of 2D CFT, such non-unitary models provide a major
  challenge for modern mathematical physics. In this context, internal
  supersymmetry, i.e. the existence of supersymmetry transformations
  that act in field space while leaving world-sheet positions
  invariant, can help to make models more tractable. Our aim here is
  to review recent progress in conformal field theory that is linked
  to internal supersymmetry. Almost all models to be discussed below
  are logarithmic.
 
  Conformal field theories with internal supersymmetries have been a
  topic of considerable interest for the past few decades. Their realm
  of  applications is vast, ranging from string theory to statistical
  physics and condensed matter theory. In the  Green-Schwarz or pure
  spinor type formulation of superstring theory, for example,
  supersymmetries act geometrically as isometries of an underlying
  space-time (target space) supermanifold. Important examples arise in
  the context of AdS/CFT dualities between supersymmetric gauge
  theories and closed strings \cite{Aharony:1999ti}. Apart from string
  theory, supersymmetry has also played a major role in the context of
  quantum disordered systems
  \cite{Parisi:1979ka,Efetov1983:MR708812,Bernard:1995as} and in
  models with non-local degrees of freedom such as polymers
  \cite{Parisi:1982ud,Read:2001pz}. In particular, it seems to be a
  crucial ingredient in the description of the plateaux transitions in
  the spin \cite{Gruzberg:1999dk,Essler:2005ag} and the integer
  quantum Hall effect
  \cite{Weidenmuller:1987gi,Zirnbauer:1999ua,Bhaseen:1999nm,Tsvelik:2007dm}.
  It is well known for instance that observables in the
  Chalker-Coddington network model for the integer quantum Hall effect
  \cite{Chalker:1988} may be expressed as correlation functions in a
  (non-conformal) non-linear $\sigma$-model on
  $\groupU(1,1|2)/\groupU(1|1)\times\groupU(1|1)$
  \cite{Weidenmuller:1987gi,Zirnbauer:1999ua}. However, so far none of
  the attempts \cite{Zirnbauer:1999ua,Bhaseen:1999nm,Tsvelik:2007dm}
  to identify the conformal field theory describing the strong
  coupling fixed point led to a completely satisfactory picture.

  In addition to having such concrete applications, conformal field
  theories with target space (internal) supersymmetry can teach us
  important lessons about logarithmic conformal field theory. As we
  indicated above, most models that possess internal supersymmetry
  exhibit the usual features of non-unitary conformal field theory
  such as the occurrence of reducible but indecomposable\footnote{In
    contrast to some appearances in the physics literature we will use
    the word ``indecomposable'' strictly in the mathematical
    sense. According to that definition also irreducible
    representations are always indecomposable since they cannot be
    written as a direct sum of two other (non-zero) representations.}
  representations and the existence of logarithmic singularities on
  the world-sheet. In this context, many conceptual issues remain to
  be solved, both on the physical and on the mathematical side. These
  include, in particular, the construction of consistent local
  correlation functions \cite{Gaberdiel:1998ps}, the modular
  transformation properties of characters \cite{Semikhatov:2003uc},
  their relation to fusion rules \cite{Fuchs:2003yu}, the treatment of
  conformal boundary conditions \cite{Gaberdiel:2006pp} etc. As we
  shall see below, superspace models provide a wide zoo of theories in
  which such issues can be addressed with an interesting mix of
  algebraic and geometric techniques.

  In addition, the special properties of Lie supergroups allow for
  constructions which are not possible for ordinary groups. For
  instance, there exist several families of coset conformal field
  theories that are obtained by gauging a one-sided action of some
  subgroup rather than the usual adjoint \cite{Metsaev:1998it,
    Berkovits:1999zq,Kagan:2005wt,Babichenko:2006uc}. The same class
  of supergroup $\sigma$-models is also known to admit a new kind of
  marginal deformations that are not of current-current type
  \cite{Berkovits:1999im,Bershadsky:1999hk,Gotz:2006qp}. Finally,
  there seems to be a striking correspondence between the
  integrability of these models and their conformal invariance
  \cite{Bena:2003wd,Young:2005jv, Kagan:2005wt,Babichenko:2006uc}.

  Here we shall review recent developments in this direction along
  with all the required background, in particular from the
  representation theory of Lie superalgebras. The latter will be
  discussed in the next section. Special emphasis is put on atypical
  representations with non-diagonalizable quadratic Casimir because of
  their direct link with logarithms in the conformal field theory
  models to be discussed in Section~\ref{sc:WZW}. There we shall
  provide a comprehensive discussion of Wess-Zumino-Witten (WZW)
  models on target supergroups of type I. These models possess all the
  features mentioned above. The two essential properties which
  facilitate an exact solution are i) the presence of an extended
  chiral symmetry based on an infinite dimensional current
  superalgebra\footnote{Instead of referring to the names ``Kac-Moody
    superalgebra'' or even ``affine Lie superalgebra'' which are
    frequently used in the physics community, we will stick to the
    notion current superalgebra by which we mean a central extension
    of the loop algebra over a finite dimensional Lie superalgebra.}
  and ii) the inherent geometric interpretation. While ii) is common
  to all $\sigma$-models, the symmetries of WZW models are necessary
  to lift geometric insights to the full field theory. Both aspects
  single out supergroup WZW theories among most of the logarithmic
  conformal field theories that have been considered in the past
  \cite{Gurarie:1993xq, Gaberdiel:1998ps,Kausch:2000fu} (see also
  \cite{Gaberdiel:2001tr,Flohr:2001zs} for reviews and further
  references). While investigations of algebraic and mostly chiral
  aspects of supergroup WZW models reach back more than twenty years
  \cite{Rozansky:1992rx,Rozansky:1992td,Maassarani:1996jn,Guruswamy:1999hi,Ludwig:2000em}
  the relevance of geometric methods for our understanding of
  non-chiral issues was realized much later in
  \cite{Schomerus:2005bf,Gotz:2006qp,Saleur:2006tf,Quella:2007hr}.

  WZW models on some special supergroups possess marginal symmetry
  preserving current-current deformations. These are discussed briefly
  at the end of Section~\ref{sc:WZW}. In Section~\ref{sc:String} we
  turn to $\sigma$-models on coset superspaces. We begin by spelling
  out the action and then review what is known about the beta function
  of these models. The results are most complete for $\sigma$-models
  on symmetric superspaces. Some of these $\sigma$-models are used as
  building blocks of exact string backgrounds for strings moving in
  Anti de-Sitter spaces. We shall only outline one example along with
  a short guide to the existing literature. The last subsection
  illustrates how deformed WZW models and $\sigma$-models can appear
  as alternative descriptions of one and the same theory. In order to
  do so, we start with a WZW model for the supergroup $\OSP(2S+2|2S)$
  and deform it through a current-current interaction. Following
  \cite{Mitev:2008yt} we shall argue that, for large coupling
  constant, the model is driven to a weakly curved $\sigma$-model on
  the superspace $\Sphere^{2S+1|2S}$. This does not only provide a
  beautiful link between different parts of this review but could also
  become a paradigm for more general dualities between WZW and
  conformal $\sigma$-models.
  
  As a final step before diving into the main subject we wish to
  provide a brief outline of how this review fits into the Special
  Issue on Logarithmic Conformal Field Theory it belongs to. The
  closest relationship exists to the review by Gainutdinov, Jacobsen,
  Saleur and Vasseur on ``Lattice Regularizations of Logarithmic
  Conformal Field Theories'', which -- among others -- also discusses
  supersymmetric spin systems. Some of our theories can be
  understood as the continuum description of the IR fixed points of
  their lattice models. Besides, the importance of the omni-present
  $c=-2$ ghost system in our treatise provides a natural link to many
  of the other contributions, in particular to Creutzig and Ridout's
  review on ``Logarithmic Conformal Field Theory: Beyond an
  Introduction''. We should also note that, from a physical
  perspective, our review is somewhat complementary to Cardy's in the
  sense that disordered systems can also be described using the
  supersymmetry trick instead of the replica method
  \cite{Efetov1983:MR708812,Bernard:1995as}. However, this is an
  aspect we do not emphasize here. Finally, it is an outstanding
  problem how the results discussed in our review can be embedded into
  the rigorous mathematical setting described by Adamovic and Milas on
  the one hand and Huang and Lepowsky on the other.

\section{Lie superalgebras}

  The theory of Lie superalgebras was developed by Kac
  \cite{Kac:1977em} and, independently, by Nahm, Scheunert and
  Rittenberg \cite{Scheunert:1976uf}. Quite a few notions from
  ordinary Lie theory carry over with only minor changes. On the other
  hand, there are also substantial differences, mostly related to the
  existence of reducible but indecomposable representations and to
  the possibility of choosing inequivalent systems of simple
  roots. With this section we lay the mathematical foundations which
  will enable us to carve out the peculiarities of superspace
  conformal field  theories in subsequent sections.

  We assume that the reader has some familiarity with the structure
  and the representation theory of ordinary Lie algebras. If this is
  not the case, we recommend to consult one of the references
  \cite{FrancescoCFT,Fuchs:1995,FuchsSchweigert}. A comprehensive
  overview, covering most of the more standard aspects of Lie
  superalgebras is the ``Dictionary on Lie superalgebras''
  \cite{Frappat:1996pb}.

\subsection{\label{sc:LieDef}Definition and examples of Lie
  superalgebras}

\begin{table}
\begin{center}
\begin{tabular}{ccccccc}
  Name & Alternative name & $\g_\den$ & $\g_\deo$ & $g^\vee$ & Defect\\\hline\hline\\[-1em]
  $A(m,n)$ & $\sgl(m{+}1|n{+}1)$ & $A_m\oplus A_n\oplus T_1$ & $(\omega_1,\omega_n)_+\oplus(\omega_m,\omega_1)_-$ & $|m{-}n|$ & $\min(m,n)$\\
  $A(n,n)$ & $\psl(n|n)$ & $A_n\oplus A_n$ & $(\omega_1,\omega_n)\oplus(\omega_n,\omega_1)$ & $0$ & $n$\\
  $C(n)$ & $\osp(2|2n{-}2)$ & $C_{n-1}\oplus T_1$ & $(\omega_1)_+\oplus(\omega_1)_-$ & $n{-}1$ & $1$\\\hline
  $F(4)$ & --- & $A_1\oplus B_3$ & $(\omega_1,\omega_3)$ & $3$ & $1$\\
  $G(3)$ & --- & $A_1\oplus G_2$ & $(\omega_1,\omega_2)$ & $2$ & $1$\\
  $B(m,n)$ & $\osp(2m{+}1|2n)$ & $B_m\oplus C_n$ &
  $(\omega_1,\omega_1)$ & $f_{m,n}$ & $\min(m,n)$\\
  $D(m,n)$ & $\osp(2m|2n)$ & $D_m\oplus C_n$ & $(\omega_1,\omega_1)$ &
  $g_{m,n}$ & $\min(m,n)$\\
  $D(2,1;\alpha)$ & --- & $A_1\oplus A_1\oplus A_1$ & $(\omega_1,\omega_1,\omega_1)$ & $0$ & $1$\\\hline
\end{tabular}
  \caption{\label{tab:Classification}Classical Lie superalgebras with
    non-degenerate even invariant form, sorted according to whether
    they are of type I (top) or type II (bottom). The symbol $T_1$
    denotes the abelian Lie algebra $\gl(1)$ of dimension one,
    $g^\vee$ is the dual  Coxeter number, see text after eq.\
    \eqref{eq:Casimir}, and the definition of `defect' is given in the
    text following eq.\  \eqref{eq:AtypicalityConstraint}. The
    $\omega_i$ are fundamental weights in the conventions of
    \cite{FrancescoCFT}, while the subscripts $\pm$ refer to two
    mutually dual non-trivial representations of $T_1$. Furthermore
    one has $f_{m,n}=2(m-n)-1$ and $g_{m,n}=2(m-n-1)$ for $m>n$ and
    $f_{m,n}=n-m+1/2$ and $g_{m,n}=n-m+1$ for $m\leq n$ (for the data
    see \cite{Kac:1994kn}).}
\end{center}
\end{table}

  The notion of a Lie superalgebra is a relatively straightforward
  generalization of that of a Lie algebra, taking into account the
  possibility of having {\em fermionic} (or {\em odd}) generators. As
  a consequence, an arbitrary Lie superalgebra $\g$ splits into an
  even part $\g_\den$ and an odd part $\g_\deo$. A Lie superalgebra is
  thus a $\Integer_2$-graded vector space $\g=\g_\den\oplus\g_\deo$
  with a bilinear Lie bracket $[\,\cdot\,,\,\cdot\,]:\g\otimes\g\to\g$
  which satisfies some natural graded modifications of the axioms for
  a Lie algebra. First of all, the bracket is consistent with the
  grading in the sense that
  $[\g_{\underline{i}},\g_{\underline{j}}]\subset\g_{\underline{i+j}}$.
  This condition ensures that $\g_\deo$ can be regarded as a
  representation of $\g_\den$. We shall later see that the precise
  nature of $\g_\deo$ as a $\g_\den$-module determines many of the
  most important properties of $\g$. Secondly, the bracket is graded
  anti-symmetric, i.e.\ $[X,Y]=-(-1)^{d_Xd_Y}[Y,X]$ for all
  homogeneous elements $X,Y\in\g$ and with
  $d_\bullet:\g_\den\cup\g_\deo\to\Integer_2$ denoting the grade function. And
  last but not least, for homogeneous elements $X,Y,Z\in\g$ one has
  the graded Jacobi identity
\begin{align}
  \label{eq:Jacobi}
  \bigl[X,[Y,Z]\bigr]
  +(-1)^{d_Z(d_X+d_Y)}\bigl[Z,[X,Y]\bigr]
  +(-1)^{d_X(d_Y+d_Z)}\bigl[Y,[Z,X]\bigr]
  \ =\ 0\ \ .
\end{align}
  The Jacobi identity may be viewed as a constraint which ensures the
  existence of an adjoint representation, the representation of $\g$
  on itself (see below). In our paper, we will only be interested in
  Lie superalgebras over the field of complex numbers.

  In most, if not all, physical applications it is essential to have a
  metric to measure lengths and distances. In the context of Lie
  superalgebras, the relevant notion is that of a non-degenerate
  symmetric even invariant bilinear form
  $\langle\cdot,\cdot\rangle:\g\otimes\g\to\Complex$. Invariance
  refers to the condition
\begin{align}
  \bigl\langle[X,Y],Z\bigr\rangle
  \ =\ \bigl\langle X,[Y,Z]\bigr\rangle\ \ ,
\end{align}
  while symmetry requires $\langle X,Y\rangle=(-1)^{d_Xd_Y}\langle
  Y,X\rangle$. The form is even if
  $\langle\g_{\underline{i}},\g_{\underline{j}}\rangle=0$ whenever we
  pair the even and the odd part ($\underline{i}\neq\underline{j}$).

  Let us now discuss a few standard examples of Lie superalgebras. For
  this purpose we consider a complex {\em vector superspace} $V=V_\den\oplus
  V_\deo$ of dimension $m|n$. In other words, $m=\dim(V_\den)$ and
  $n=\dim(V_\deo)$ and the combinations $m+n$ and $m-n$ are known as
  the {\em dimension} and the {\em superdimension} of $V$,
  respectively. The space
  of linear maps $\End(V)$ naturally inherits the structure of a
  vector superspace (actually of an associative super{\em algebra}) by
  using the assignment
\begin{align}
  \End(V)_\den
  &\ \defeq\ \Hom(V_\den,V_\den)\oplus\Hom(V_\deo,V_\deo)
  \quad\text{ and }\quad\\[2mm]
  \End(V)_\deo
  &\ \defeq\ \Hom(V_\den,V_\deo)\oplus\Hom(V_\deo,V_\den)\ \ .
\end{align}
  In fact, $\End(V)$ can easily be seen to give rise to a Lie
  superalgebra by identifying the Lie bracket with the
  (anti-)commutator
\begin{align}
  \label{eq:GLCom}
  [X,Y]\defeq XY-(-1)^{d_Xd_Y}YX\ \ .
\end{align}
  It is a straightforward exercise to verify all relevant axioms of a
  Lie superalgebra. Frequently, the Lie superalgebra above is also
  referred to as $\gl(V)$ or $\gl(m|n)$. It is called the {\em general
    linear superalgebra}.

  The elements of $\gl(m|n)$ can be conveniently expressed as
  matrices. If one uses an adapted basis with the first part of basis
  elements from $V_\den$ and the remaining ones from $V_\deo$, the
  matrices $M\in \gl(m|n)$ will have the block form $M=\smat
  A&B\\C&D\stam$ with $A,D$ being even and $B,C$ being
  odd.\footnote{In order to avoid confusion we should stress that the
    entries of all matrices are complex numbers. There are no
    Grassmann variables involved here.} In this basis, the {\em
    supertrace}
  is defined by $\str\smat A&B\\C&D\stam=\tr(A)-\tr(D)$. As a slight
  variation of $\gl(m|n)$ we can consider the subspace $\sgl(m|n)$ of all
  matrices with vanishing supertrace. It can be checked that this subspace
  is closed under the Lie bracket \eqref{eq:GLCom} and that it hence
  gives rise to a Lie superalgebra as well, also known as the {\em special
  linear superalgebra}.
  As it turns out, the case of $\sgl(n|n)$ is rather special. While
  $\sgl(m|n)$ is simple for $m\neq n$, i.e.\ there are no non-trivial
  ideals, the Lie superalgebra $\sgl(n|n)$ has an abelian ideal $\gl(1)$
  which is generated by multiples of the identity matrix. The quotient
  $\psl(n|n)=\sgl(n|n)/\gl(1)$ is simple. Due to its peculiar properties
  it will accompany us throughout most of this review.

  One more important family of Lie superalgebras, the orthosymplectic
  superalgebras $\osp(m|2n)$, can be introduced as a subspace of
  $\gl(m|2n)$ if one demands that the maps leave invariant a quadratic
  form which is orthogonal on $\Complex^{m|0}$ and symplectic on
  $\Complex^{0|2n}$. Such a form is typically chosen to be represented
  by the block matrix $\smat\id&0&0\\0&0&\id\\0&-\id&0\stam$.

  One of the most important concepts in the theory of Lie
  superalgebras are Casimir operators, in particular quadratic Casimir
  operators. Given any metric $\langle\cdot,\cdot\rangle$ one can
  define the graded symmetric invariant tensor $\kappa^{ab}=\langle
  T^a,T^b\rangle$ with respect to some homogeneous basis. The
  associated {\em quadratic Casimir operator} is then defined by the
  equation
\begin{align}
  \label{eq:Casimir}
  C\ =\ \kappa_{ba}T^aT^b\ \ ,
\end{align}
  where $\kappa_{ab}$ denotes the inverse of the matrix
  $\kappa^{ab}$. For simple Lie superalgebras, the quadratic Casimir
  operator is unique up to normalization. In the standard
  normalization where long roots have length~2, half the value of the
  quadratic Casimir operator in the adjoint representation is known as
  the dual Coxeter number $g^\vee$ of the Lie superalgebra $\g$ (see
  \cite{FrancescoCFT}).

  Setting $\ad_X=[X,\,\cdot\,]\in\End(\g)$, let us finally recall the
  notion of the {\em Killing form}
\begin{align}
  K(X,Y)
  \ =\ \str(\ad_X\circ\ad_Y)\ \ .
\end{align}
  The Killing form plays an essential role in the structure theory of
  {\em ordinary} Lie algebras since semi-simple Lie algebras can be
  characterized by the property that the Killing form is
  non-degenerate. In contrast, even for simple Lie superalgebras the
  Killing form may vanish identically. This is the case for the series
  $\psl(n|n)$ and $\osp(2n{+}2|2n)$. Surprisingly, as we shall see,
  these Lie superalgebras are in a sense the physically most
  interesting ones.

\paragraph{Example: The Lie superalgebra $\gl(1|1)$.}
  The Lie superalgebra $\gl(1|1)$ is generated by two even generators
  $E,N$ and two fermionic generators $\psi^\pm$. While $E$ is central,
  the generator $N$ can be interpreted as a fermion counting
  operator. The non-trivial Lie brackets read
\begin{align}
  [N,\psi^\pm]
  \ =\ \pm\psi^\pm
  \quad\text{ and }\qquad
  [\psi^+,\psi^-]
  \ =\ E\ \ .
\end{align}
  In the fundamental representation, the generators can be represented
  by
\begin{align}
  E\ =\ \mat1&0\\0&1\tam\ ,\quad
  N\ =\ \frac{1}{2}\mat1&0\\0&-1\tam\ ,\quad
  \psi^+
  \ =\ \mat0&1\\0&0\tam\ ,\quad
  \psi^-
  \ =\ \mat0&0\\1&0\tam\ \ .
\end{align}
  One can easily check that the most general metric is specified by
\begin{align}
  \langle E,N\rangle
  \ =\ a\ ,\quad
  \langle\psi^+,\psi^-\rangle
  \ =\ a\ ,\quad
  \langle N,N\rangle
  \ =\ b\ \ ,
\end{align}
  with arbitrary constants $a\neq0$ and $b$. In contrast, the Killing
  form is degenerate and easily seen to give rise to $K(N,N)=-2$, with
  all other components being trivial. The most general quadratic
  Casimir of $\gl(1|1)$ can be expressed as an arbitrary linear
  combination of the two elements
\begin{align}
  \label{eq:Casimirgl}
  C\ =\ 2EN-\psi^+\psi^-+\psi^-\psi^+
  \qquad\text{ and }\qquad
  \tilde{C}\ =\ E^2\ \ .
\end{align}

\subsection{\label{sc:Classification}Classification of Lie superalgebras}

  Simple Lie superalgebras have been classified by Kac in
  \cite{Kac:1977em}. Just as in the case of ordinary Lie algebras,
  there are several infinite families and a few exceptional
  cases.\footnote{We adopt the convention that when using the term Lie
    superalgebra we always mean a Lie superalgebra with non-trivial
    odd part $\g_\deo$. The other Lie superalgebras are referred to as
    ordinary Lie algebras.} In all what follows, we shall limit
  ourselves to {\em classical} Lie superalgebras, i.e.\ simple Lie
  superalgebras whose odd part $\g_\deo$ is completely reducible under
  the action of $\g_\den$. This excludes the {\em Cartan type} Lie
  superalgebras $W(n)$, $S(n)$, $\tilde{S}(n)$ and $H(n)$. The
  classical Lie superalgebras furthermore split into the {\em basic}
  ones, i.e.\ those admitting a metric in the sense of
  Section~\ref{sc:LieDef}, and the two {\em strange} families $P(n)$
  and $Q(n)$ for which such a metric does not exist. For physical
  reasons, we will restrict our attention to the basic Lie
  superalgebras whose complete list can be found in
  Table~\ref{tab:Classification}, together with a few of their
  properties. For a detailed description of all simple Lie
  superalgebras we refer the reader to
  \cite{Kac:1977em,Frappat:1996pb}. Apart from the series introduced
  in Section~\ref{sc:LieDef} we only wish to highlight the family of
  exceptional Lie superalgebras $D(2,1;\alpha)$ which can be regarded
  as a deformation of $\osp(4|2)$.

  It may be shown that the even part $\g_\den$ of a classical Lie
  superalgebra $\g$ is reductive. In other words, it decomposes into a
  direct sum of simple or abelian Lie algebras. One then distinguishes
  furthermore Lie superalgebras of type~I and type~II. In the type~II
  case, $\g_\deo$ is irreducible under the action of $\g_\den$ while
  it splits into two irreducibles if $\g$ is of type~I. We shall see
  below that the properties of Lie superalgebras of type~I and~II,
  respectively, are significantly different.

  In the context of string theory and quantum field theory one mainly
  encounters two classes of Lie superalgebras: Poincar\'e
  superalgebras and conformal superalgebras \cite{Nahm:1977tg}. Both
  of them may be used to describe space-time supersymmetries, i.e. the
  superisometries of space-time manifolds such as Minkowski space and
  Anti-de Sitter space (including Killing spinors). The conformal
  superalgebras for a $d$-dimensional flat space all have a bosonic
  subalgebra of the form $\so(d+1,1)$ or $\so(d,2)$ (for Euclidean and
  Lorentzian signature, respectively). They can thus be regarded as
  specific real forms of the complex simple Lie superalgebras
  $\osp(m|2n)$ and $\sgl(m|n)$ discussed above.\footnote{Recall the
    isomorphisms $\su(2)=\so(3)$, $\so(4)=\su(2)\oplus\su(2)$ and
    $\so(6)=\su(4)$.} In contrast, the Poincar\'e superalgebras are
  not semi-simple and hence they fall outside of the previous
  classification. In particular, we suspect that many of the problems
  in the covariant quantization of superstrings are actually due to
  the lack of a non-degenerate invariant form for the Poincar\'e
  superalgebras. For this reason, certain aspects of strings on AdS
  spaces (with an underlying simple isometry supergroup and hence the
  existence of a metric) might actually be easier to deal with than
  the corresponding ones on flat space, see
  e.g.~\cite{Berkovits:2008ga}.

\subsection{Representation theory}

\subsubsection{\label{sc:RepTheory}Introduction and overview}

\begin{table}
\begin{center}
\begin{tabular}{cl}
  Symbol & Meaning \\\hline\hline
  $\Complex_\lambda$ & One-dimensional module of the Cartan subalgebra
  $\h$\\\hline
  $L_\lambda$ & Simple module of $\g_\den$ \\
  $V_\lambda$ & Verma module (induced from $\Complex_\lambda$)\\\hline
  $\cL_\lambda$ & Simple module of $\g$\\
  $\cK_\lambda$ & Kac module (covering $\cL_\lambda$)\\
  $\cP_\lambda$ & Projective cover of $\cL_\lambda$\\
  $\cB_\lambda$ & Projective module (induced from $L_\lambda$)\\
  $\cV_\lambda$ & Verma module (induced from $\Complex_\lambda$)\\\hline
\end{tabular}
  \caption{\label{tab:Reps}Different types of representations
    featuring in this article.}
\end{center}
\end{table}

  A representation of $\g$ on a vector superspace $\cM$ is, by
  definition, a Lie superalgebra homomorphism from $\g$ to
  $\gl(\cM)$. If the relevant homomorphism is clear from the context we
  will simply refer to $\cM$ when describing the representation. We
  should also mention that we will use the notions representation and
  module
  interchangeably. Elementary examples of representations are the
  trivial representation $\cL_0:\g\to\Complex$ with kernel $\g$ and
  the adjoint representation $\ad:\g\to\g$ in terms of the
  homomorphism $X\mapsto[X,\,\cdot\,]$. We reserve the symbol
  $\cL_\lambda$ to denote irreducible representations of $\g$ while
  $L_\lambda$ refers to irreducible representations of the even
  subalgebra $\g_\den$. In what follows we shall exclusively be
  interested in weight modules. These are modules on which the
  elements of the Cartan subalgebra $\h\subset\g$ act diagonally. In
  the cases of interest, the Cartan subalgebra of $\g$ may be
  identified with the Cartan subalgebra of its bosonic subalgebra
  $\g_\den$.

  There are many invariants which help to characterize
  representations. However, for the purposes of our paper it will
  mostly be sufficient to work with the following two: The quadratic
  Casimir and characters. Since the quadratic Casimir operator
  \eqref{eq:Casimir} is an even element which commutes with all
  generators of $\g$, it will be represented as a multiple of the
  identity matrix in any irreducible representation
  $\cL_\lambda$. This follows from a superalgebra variant of Schur's
  Lemma \cite{Kac:1977em}. The corresponding constant of
  proportionality will be denoted by $C(\lambda)$. Representations
  with different values of $C(\lambda)$ are certainly not isomorphic
  while, a priori, not much can be said if the values agree.

  Just as for ordinary Lie algebras,
  one can define the direct sum and the tensor product of two
  representations. For the tensor product one needs to take into
  account that one is working with graded structures which lead to
  natural sign factors appearing every now and then. One of the main
  differences between simple Lie algebras and Lie superalgebras is
  the occurrence of indecomposable representations which are not
  irreducible. This leads to a veritable zoo of representations, the
  most important of which are listed in
  Table~\ref{tab:Reps}. Eventually, the existence of representations
  which are not fully reducible may be traced back to the nil-potency
  of certain odd generators. More details and examples will be
  provided below. The only series of classical Lie superalgebras for
  which all finite dimensional representations are fully reducible is
  $B(0,n)=\osp(1|2n)$.

  The natural method of constructing non-trivial representations for a
  Lie superalgebra $\g$ is using induction from a subalgebra
  $\k$. Here one starts from a representation $\cM_\k$ of $\k$ and
  uses the known commutation relations between the generators of $\k$
  and $\g$ in order to build up a new representation $\cM_\g$ of
  $\g$. This procedure uses the natural action of $\g$ on the
  universal enveloping superalgebra $\cU(\g)$. More formally, one has
\begin{align}
  \cM_\g
  \ =\ \Ind_\k^\g(\cM_\k)
  \ \defeq\ \cU(\g)\otimes_\k\cM_\k\ \ ,
\end{align}
  where the symbol $\otimes_\k$ refers to the fact that the tensor
  product is transparent with regard to generators from $\k$, i.e.\
  $XY\otimes_\k Z=X\otimes_\k YZ$ for all $Y\in\k$. Due to the
  infinite dimensionality of $\cU(\g)$, induction generally gives rise
  to representations which are very large and not necessarily
  irreducible. The power of induction thus heavily relies on a
  smart choice of subalgebra $\k$.

  For Lie superalgebras, a variety of subalgebras have been suggested,
  each of them clarifying different aspects of the representation
  theory. Two of the examples that will be used in this paper are
  associated with the root space decomposition and the splitting of
  $\g$ into its even and odd part
\begin{align}
  \text{a)}\quad\g=\h\oplus\bigoplus_{\alpha\neq0}\g_\alpha
  \qquad\text{ and }\qquad
  \text{b)}\quad\g=\g_\den\oplus\g_\deo\ \ .
\end{align}
  While the root space decomposition a) paves the road to highest
  weight theory, the second decomposition b) has certain advantages
  with regard to disclosing the categorial properties of
  representations. Moreover, in the case of Lie superalgebras of
  type~I and type~II, respectively, it is custom to work with the
  distinguished $\Integer$-gradings
\begin{align}
  \label{eq:ZGrading}
  \text{c)}\quad\g=\g_1\oplus\g_0\oplus\g_{-1}
  \qquad\text{ and }\qquad
  \text{d)}\quad\g=\g_2\oplus\g_1\oplus\g_0\oplus\g_{-1}\oplus\g_{-2}\ \ .
\end{align}
  where it is understood that $[\g_i,\g_j]\subset\g_{i+j}$ for all
  $i,j\in\Integer$. In both cases, the grading is assumed to be
  consistent with the underlying $\Integer_2$-grading in the sense
  that $\g_{\text{even}}\subset\g_\den$ and
  $\g_{\text{odd}}\subset\g_\deo$. As we shall see below, the
  decomposition c) is actually the most convenient one for
  constructing irreducible representations. In contrast, the
  decomposition d) is only natural if one is interested in specific
  types of infinite dimensional representations.

\paragraph{Highest weight modules.}
  Highest weight theory is the easiest route towards the
  classification of (finite dimensional) irreducible representations
  and for the determination of Casimir invariants. The starting point
  is a one-dimensional representation $\Complex_\lambda$ of the Cartan
  subalgebra $\h$. It is fully specified in terms of a weight
  $\lambda\in\h^\ast$ by the assignment $H\mapsto\lambda(H)$. This
  one-dimensional representation is then extended to a representation
  of the Borel subalgebra $\b_>=\h\oplus\bigoplus_{\alpha>0}\g_\alpha$
  by letting all positive roots act trivially.\footnote{It will not be
    important for our presentation, but we wish to remark that for Lie
    superalgebras there exist various choices of inequivalent simple
    root systems (and associated Dynkin diagrams). This is in stark
    contrast to the case of ordinary Lie algebras.} Finally, the
  negative roots are used to generate new states. This procedure
  results in the {\em Verma module} 
\begin{align}
  \cV_\lambda
  \ \defeq\ \Ind_{\b_>}^\g(\Complex_\lambda)\ \ .
\end{align}
  Verma modules are infinite dimensional as soon as there is at least
  one even root. Irreducible representations $\cL_\lambda$ are
  recovered as the quotient $\cV_\lambda/\cM_\lambda$ of a Verma
  module by its maximal proper submodule $\cM_\lambda$.

  Verma modules have the advantage that they are easy to define and
  that their characters are trivial to write down. Also, the root
  space decomposition leads to a simple expression for the eigenvalue
  of the quadratic Casimir operator ($\rho$ is the Weyl vector),
\begin{align}
  \label{eq:CasimirEigenvalue}
  C(\lambda)
  \ =\ (\lambda,\lambda+2\rho)
\end{align}
  in terms of the naturally induced invariant form $(\cdot,\cdot)$ on
  $\h^\ast$. On the other hand, the identification of the maximal
  submodule $\cM_\lambda$ and the derivation of character formulas for
  the irreducible quotients $\cL_\lambda$ is rather cumbersome.

\paragraph{Kac modules $\cK_\lambda$.}
  Kac modules are finite dimensional representations which have been
  designed to approximate irreducibility as closely as possible
  \cite{Kac1977:MR0444725,Kac:1977MR519631}. The definition is rather
  involved for Lie superalgebras of type~II. For this reason, we will
  restrict our attention to Lie superalgebras of type~I here, where
  Kac modules are extremely well behaved. In this case, we can use the
  distinguished $\Integer$-grading~c) from eq.~\eqref{eq:ZGrading} in
  order to split the fermions into two dual spaces $\g_{\pm1}$, each
  of which forms an irreducible representation of $\g_0=\g_\den$. It
  is then straightforward to extend any $\g_\den$-module $L_\lambda$
  to a representation of $\g_0\oplus\g_1$ by letting the odd
  generators from $\g_1$ act trivially. Since the remaining fermions
  in $\g_{-1}$ are all nilpotent, the induced module
\begin{align}
  \cK_\lambda
  \ =\ \Ind_{\g_0\oplus\g_1}^\g L_\lambda
  \ =\ \cU(\g)\otimes_{\g_0\oplus\g_1}L_\lambda
\end{align}
  is finite dimensional. As a representation of $\g_\den$ one
  immediately finds
\begin{align}
  \label{eq:KacModuleContent}
  \cK_\lambda\bigr|_{\g_\den}
  \ =\ L_\lambda\otimes\bigwedge(\g_{-1})\ \ ,
\end{align}
  and this expression leads to a straightforward formula for the
  character of $\cK_\lambda$
  \cite{Kac1977:MR0444725,Kac:1977MR519631}. Just as for Verma
  modules, one can associate a simple module $\cL_\lambda$ to every
  Kac module $\cK_\lambda$ by taking the quotient
  $\cL_\lambda=\cK_\lambda/\cN_\lambda$ by its maximal submodule
  $\cN_\lambda$. It can be shown that all simple modules can be
  obtained this way. Dual Kac modules $\cK_\mu^\ast$ are obtained by
  replacing $L_\lambda$ with its dual $L_\lambda^\ast$ and by
  exchanging the roles of $\g_{\pm1}$.

\paragraph{Typical versus atypical representations.}
  Kac modules are of particular relevance since they typically turn
  out to be irreducible. If this is the case, the associated simple
  module $\cL_\lambda=\cK_\lambda$ is called {\em typical}, otherwise
  {\em atypical} (this definition only applies to type~I
  superalgebras).\footnote{In physics terminology one would call them
    non-BPS and BPS representations or long and \mbox{(semi-)}\-short
    representations, respectively. Equality
    \eqref{eq:AtypicalityConstraint} defines the equivalent of a BPS
    bound.} There are different ways of characterizing whether a
  module $\cL_\lambda$ is typical or atypical. For instance, a weight
  $\lambda$ is leading to a typical representation if there is an
  isotropic fermionic root $\alpha$ such that \cite{Kac:1977MR519631}
\begin{align}
  \label{eq:AtypicalityConstraint}
  (\lambda+\rho,\alpha)\ =\ 0\ \ .
\end{align}
  The root is called isotropic if $(\alpha,\alpha)=0$. One can even
  define the {\em degree of atypicality} $d(\lambda)$ by counting (in
  a suitable way) the solutions $\alpha$ to this equation (see, e.g.,
  \cite{Cheng:2012MR3012224}). The larger
  $d(\lambda)$, the more complicated the representation
  $\cK_\lambda$ and the resolution of $\cL_\lambda$ in terms of Kac
  modules. The maximal value of $d(\lambda)$ when taken over all
  finite dimensional representations is known as the {\em defect}
  $d(\g)$ of the Lie superalgebra $\g$ \cite{Kac:1994kn}. It is a
  measure for how pathological the representation theory of $\g$
  is. Lie superalgebras with $d(\g)\leq1$ are under good control (see
  also Table~\ref{tab:Classification}). In contrast, a defect
  $d(\g)>1$ indicates that the representation theory is of wild type,
  i.e.\ the classification of {\em all} finite dimensional
  representations (as opposed to merely irreducible representations)
  is believed to be beyond reach (see \cite{Germoni1998:MR1659915} for
  a detailed discussion in connection with $\sgl(m|n)$).

  One of the most fundamental problems in the representation theory of
  Lie superalgebras is to find explicit character formulas for
  atypical irreducible representations. Recent progress on this
  essential issue can be found in the references
  \cite{Serganova1998:MR1648107,Brundan2001:MR1937204,Brundan:MR2881300,Gruson:2010MR2734963,Musson:2011MR2806501}.
  In particular, the concept of ``super duality'' appears to be a
  very powerful new tool
  \cite{Cheng:2011MR2755062,Cheng:2012MR3012224}. For some Lie
  superalgebras of physical interest, most notably $\sgl(n|1)$,
  $\psl(2|2)$ and $D(2,1;\alpha)$, it is also possible to
  consult more specific literature (see, e.g.,
  \cite{Germoni1998:MR1659915,Germoni2000:MR1840448,Gotz:2005jz,Gotz:2005ka,Gotz:2006qp}
  and references therein). A detailed discussion of representations of
  conformal Lie superalgebras is available in
  \cite{Dobrev:2004tk,Bianchi:2006ti,Dolan:2008vc,Dobrev:2012me}.

 In Section~\ref{sc:Cat} we will introduce
  an alternative (but less explicit) characterization of typicality
  which is emphasizing category theoretic aspects.

\paragraph{Projective modules \texorpdfstring{$\cB_\lambda$}{B(lambda)}.}
  Let us next investigate induced representations which result from
  the decomposition~b) of $\g$. Inducing from an irreducible
  representation $L_\lambda$ of the even subalgebra $\g_\den$ we
  obtain the modules
\begin{align} \label{defB} 
  \cB_\lambda
  \ =\ \Ind_{\g_\den}^\g(L_\lambda)\ \ .
\end{align}
  Even though they are too large to give rise to irreducible
  representations of $\g$ directly, they have a number of very
  convenient properties. In particular, they are always finite
  dimensional (if $L_\lambda$ is) and they inherit the property of
  being projective from $L_\lambda$ (see below) since induction is a
  right-exact functor. As we shall discuss in
  Section~\ref{sc:HarmonicAnalysis}, the modules $\cB_\lambda$ play a
  key role in the harmonic analysis on supergroups.

\paragraph{Characterization of indecomposable representations:
  Composition series.}
  Indecomposable representations which are not irreducible can
  (partially) be characterized by means of {\em composition
  series}.\footnote{Frequently, one also encounters the name
    ``Jordan-H\"older series'' \cite{AndersonFuller}.}  By a
  composition series of an indecomposable module $\cM$ we shall mean a
  filtration
\begin{align}
  \{0\}=\cM_0\subset\cM_1\subset \dots \subset \cM_n=\cM
\end{align}
  such that all quotient modules $\cQ_i=\cM_i/\cM_{i-1}$ are
  simple, i.e.\ irreducible. A composition
  series can be obtained iteratively, by letting $\cM_{n-1}$ be a
  maximal submodule of $\cM=\cM_n$, $\cM_{n-2}$ be a maximal
  submodule of $\cM_{n-1}$ and so on. The simple modules $\cQ_{i}$
  are known as composition factors. Their respective multiplicities
  are invariants of $\cM$, i.e.\ they do not depend on the exact
  choice of composition series \cite{AndersonFuller}. In this paper,
  we will use the pictorial description
\begin{align}
  \cM:\quad\cQ_n\to\cQ_{n-1}\to\cdots\to\cQ_{1}\ \ ,
\end{align}
  in order to visualize the composition series of indecomposable
  representations. Moreover, we will frequently work with a slight
  variant of the composition series, the so-called {\em socle series}
  where the quotient modules $\cQ_{i}$ are maximal semi-simple
  (instead of merely being simple). In that case, $\cQ_n$ is known as
  the {\em head} of the module $\cM$, while $\cQ_1$ is known as its
  {\em socle}.

  It should be noted that there is also another variant of the
  previous definition, where the quotients are required to be direct
  sums of Kac modules. In that case one speaks about a Kac composition
  series. While every module admits a composition series, this is not
  necessarily so regarding Kac composition series (a trivial example
  being any simple module that is obtained as a non-trivial quotient
  of a Kac module).

  The existence of composition series with regard to two specific sets
  of modules, so-called standard and co-standard modules, is the
  foundation of tilting theory and it leads to profound consequences.
  For type~I Lie superalgebras, the role of standard and co-standard
  modules is played by Kac modules \cite{Brundan2004:MR2100468}. Since
  we will not need it here, we refrain from being more explicit.

\paragraph{Example: Representations of \texorpdfstring{$\gl(1|1)$}{gl(1|1)}.}
  The bosonic subalgebra of $\gl(1|1)$ is spanned by $E$ and $N$ and
  is isomorphic to $\gl(1)\oplus \gl(1)$. At the same time, it forms
  the Cartan subalgebra of $\gl(1|1)$. Its simple modules are all
  one-dimensional and given by $L_{(e,n)}=\Complex_{(e,n)}$ where $e$
  and $n$ are the eigenvalue of $E$ and $N$, respectively. Starting
  from $L_{(e,n)}$ one can induce the Kac module $\cK_{(e,n)}$ of
  $\gl(1|1)$ by letting $\psi^+$ act trivially on an
  arbitrary non-zero vector $|e,n\rangle$ of $L_{(e,n)}$. The
  resulting module is spanned by the two states $v_1=|e,n\rangle$ and
  $v_2=\psi^-|e,n\rangle$. The second vector has weight $(e,n-1)$. By
  applying $\psi^+$ to $v_2$ one can
  easily verify that $\cK_{(e,n)}$ is irreducible,
  $\cL_{(e,n)}=\cK_{(e,n)}$, if and only if $e\neq0$. For $e=0$, on
  the other hand, $v_2$ spans an invariant submodule since $v_1$
  cannot be reached anymore by the action of $\psi^+$. As a
  consequence, the module $\cK_{(0,n)}$ has a composition series of
  length two which may be depicted as
\begin{align}
  \cK_{(0,n)}:\quad\cL_{n}\to\cL_{n-1}\ \ ,
\end{align}
  where $\cL_n:=\cL_{(0,n)}$ refers to the one-dimensional simple
  module on which $N$ acts as multiplication by $n$ and all the
  remaining generators, $E,\psi^\pm$, act trivially. Finally, the
  modules $\cB_{(e,n)}$ are obtained by acting on
  $L_{(e,n)}=\Complex_{(e,n)}$ with all possible fermionic
  operators. This gives rise to a four-dimensional space which
  decomposes as $\cB_{(e,n)}=\cL_{(e,n)}\oplus\cL_{(e,n+1)}$ for
  $e\neq0$. In contrast, for $e=0$, the module $\cB_{(0,n)}$ turns out
  to be indecomposable with a socle series\footnote{Alternatively, it
    also has a Kac composition series
    $\cB_{(0,n)}:\cK_{(0,n)}\to\cK_{(0,n+1)}$.}
\begin{align}
  \cB_{(0,n)}:\quad
  \cL_{n}\to\cL_{n+1}\oplus\cL_{n-1}\to\cL_{n}\ \ .
\end{align}
  Actually, $\cB_{(0,0)}$ just corresponds to the adjoint
  representation of $\gl(1|1)$.

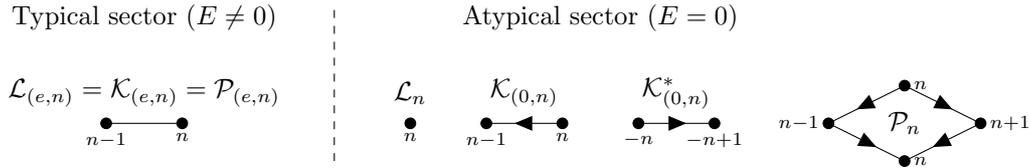
\begin{figure}
\begin{center}
\begin{tikzpicture}[decoration={
  markings,
  mark=at position .63 with {\arrow[line width=.5pt]{triangle 45};}}
]
  \draw (.5,1.4) node {$\cL_{(e,n)}=\cK_{(e,n)}=\cP_{(e,n)}$};
  \draw (.5,2.4) node {Typical sector ($E\neq0$)};
  \draw (0,1) -- (1,1);
  \draw[fill] (0,1) circle (2pt) node[below] {$\scriptstyle n{-}1$};
  \draw[fill] (1,1) circle (2pt) node[below] {$\scriptstyle n$};
  \draw[dashed] (3,.5) -- (3,2.6);
\begin{scope}[xshift=1cm]
  \draw (5.5,2.4) node {Atypical sector ($E=0$)};
  \draw (3,1.4) node {$\cL_n$};
  \draw[fill] (3,1) circle (2pt);
  \draw (3,1) node[below] {$\scriptstyle n$};
  \draw (4.5,1.4) node {$\cK_{(0,n)}$};
  \draw[postaction={decorate}] (5,1) -- (4,1);
  \draw[fill] (4,1) circle (2pt);
  \draw[fill] (5,1) circle (2pt);
  \draw (4,1) node[below] {$\scriptstyle n-1$};
  \draw (5,1) node[below] {$\scriptstyle n$};
  \draw (6.5,1.4) node {$\cK_{(0,n)}^\ast$};
  \draw[postaction={decorate}] (6,1) -- (7,1);
  \draw[fill] (6,1) circle (2pt);
  \draw[fill] (7,1) circle (2pt);
  \draw (6,1) node[below] {$\scriptstyle-n$};
  \draw (7,1) node[below] {$\scriptstyle-n+1$};
  \draw (9.5,1) node {$\cP_n$};
  \draw[postaction={decorate}] (9.5,1.5) -- (8.5,1);
  \draw[postaction={decorate}] (9.5,1.5) -- (10.5,1);
  \draw[postaction={decorate}] (8.5,1) -- (9.5,.5);
  \draw[postaction={decorate}] (10.5,1) -- (9.5,.5);
  \draw[fill] (8.5,1) circle (2pt);
  \draw[fill] (9.5,.5) circle (2pt);
  \draw[fill] (9.5,1.5) circle (2pt);
  \draw[fill] (10.5,1) circle (2pt);
  \draw (9.5,1.5) node[right] {$\scriptstyle n$};
  \draw (8.5,1) node[left] {$\scriptstyle n-1$};
  \draw (10.5,1) node[right] {$\scriptstyle n+1$};
  \draw (9.5,.5) node[right] {$\scriptstyle n$};
\end{scope}
\end{tikzpicture}
  \caption{\label{fig:GLReps}The weight diagrams of the most important
    representations of $\gl(1|1)$. The horizontal axis denotes the
    eigenvalues of $N$. The arrows indicate maps into invariant
    subspaces.}
\end{center}
\end{figure}

  In addition to the representations just mentioned, the Lie
  superalgebra $\gl(1|1)$ admits a whole zoo of finite dimensional
  indecomposable representations. The complete classification can be
  found in \cite{Gotz:2005jz}, including the full set of tensor
  product decompositions. Anticipating the notation
  $\cP_n=\cB_{(0,n)}$, an illustration of the most important
  representations of $\gl(1|1)$ is provided in
  Figure~\ref{fig:GLReps}.\footnote{These pictures should not be
    confused with the composition or socle series even though -- in
    the particular case of $\gl(1|1)$, where the even subalgebra
    coincides with the Cartan subalgebra -- these notions are closely
    related.}

\subsubsection{\label{sc:Cat}Category theoretic perspective}

  The previous section was mainly concerned with defining an
  interesting set of representations. Here we would like to clarify
  the relations betweens the modules $\cL_\lambda$, $\cK_\lambda$ and
  $\cB_\lambda$ and bring some structure into the zoo of
  representations. This will be achieved by means of concepts from
  category theory, see \cite{AndersonFuller} for a helpful
  reference. There are several categories of representations at our
  disposal. With regard to Verma modules, the most appropriate one is
  category $\cO$
  \cite{Humphreys:2008MR2428237,Musson:2012MR2906817,Cheng:2012MR3012224}.
  From the perspective of Kac modules, it is more natural to restrict
  one's attention to the category of representations with
  diagonalizable action of the Cartan subalgebra $\h$. In particular,
  the previous category includes the category of finite
  dimensional weight modules over $\g$ \cite{Zou1996:MR1378540}. All
  statements in this section refer to the latter.\footnote{Since we
    will later also be dealing with non-compact forms of supergroups
    associated to the Lie superalgebra $\g$, some of our statements
    should be taken with a pinch of salt. To be precise, we will feel
    free to extrapolate results from a finite dimensional to an
    infinite dimensional setting where the corresponding mathematical
    results have not (yet) been rigorously established.}

  Much of the subsequent exposition is based on the concept of a
  projective module. A $\g$-module $\cP$ is called projective (in a
  given category of representations) if and only if for every
  surjective $\g$-homomorphism $f:\cM\twoheadrightarrow\cP$ there
  exists a $\g$-homomorphism $h:\cP\to\cM$ such that $f\circ
  h=\id$. In other words, in case $\cM$ is a cover of $\cP$ then it
  contains $\cP$ as a direct summand and the map $f$ can be thought of
  as the projection onto $\cP$. In the present context, i.e.\
  restricting all considerations to the category of finite-dimensional
  $\g$-modules, projective modules also satisfy the dual property of
  being injective \cite{Germoni1998:MR1659915}. One could then replace
  our previous definition by the requirement that any projective
  submodule $\cP$ of an arbitrary module $\cM$ always appears as a
  direct summand.

  Let $\SRep(\g)$ denote the set of (equivalence classes of) all
  finite dimensional simple modules $\cL_\lambda$. As we have seen in
  Section~\ref{sc:RepTheory}, elements from this
  set will be indexed by $\lambda$, with $\lambda$ running through
  some set of weights (a weight of $\g$ and $\g_\den$ at the same
  time). As we have reviewed in Section~\ref{sc:RepTheory}, the simple
  modules fall into two classes, typical and atypical
  representations. We split the set of weights accordingly into
  typical ones and atypical ones, $\Typ(\g)$ and $\ATyp(\g)$. For
  finite-dimensional representations, it can be shown that the typical
  sector is characterized by the property that its simple modules are
  projective (in the sense described in the previous paragraph) while
  for the atypical representations this is not the
  case.\footnote{According to \cite{Kac:1977MR519631}, typical
    representations split in any finite-dimensional
    representation (this property may actually be taken as the
    definition of typicality). Hence they need to be their own
    projective cover.}

  To each weight $\lambda$ one can not only associate the simple
  module $\cL_\lambda$ but rather also further indecomposable modules
  which contain $\cL_\lambda$ is a simple quotient. Besides the Kac
  module $\cK_\lambda$, the most important is the projective cover
  $\cP_\lambda$ (see e.g.\
  \cite[Lemma\,3.2]{Brundan2004:MR2100468}). The projective cover
  $\cP_\lambda$ is the (unique)
  indecomposable projective module which has $\cL_\lambda$ as a simple
  quotient. A representation $\cL_\lambda$ is typical if and only if
  $\cL_\lambda\cong\cP_\lambda$, i.e.\ if $\cL_\lambda$ is already
  projective itself. Otherwise $\cP_\lambda$ is strictly larger than
  $\cL_\lambda$. Every projective module may be decomposed into a
  direct sum of projective covers $\cP_\lambda$ of simple modules.
  It should be noted that there are at least two independent
  intertwiners from $\cP_\lambda$ to itself as long is $\lambda$ is
  atypical. The first one is the identity map. The other maps the head
  of $\cP_\lambda$ to its socle and is thus nilpotent.\footnote{That
    the socle is equal to the head can be inferred from the equality
    $\dim\Hom_\g(\cL_\lambda,\cP_\lambda)=\dim\Hom_\g(\cP_\lambda^\ast,\cL_\lambda^\ast)=1$.
    The second equation just means that $\cL_\lambda^\ast$ constitutes
    the head of $\cP_\lambda^\ast$. To complete the argument one also
    has to use that the socle and the head do not sit on top of each
    other, i.e.\ that $\cP_\lambda$ is not simple.} In all examples we
  are aware of, the latter is given by a suitable power of the
  quadratic Casimir operator.

  After these preparations we are finally able to address the
  structure of the modules
  $\cB_\lambda=\Ind_{\g_\den}^{\g}(L_\lambda)$ which are induced from
  finite dimensional simple $\g_\den$-modules $L_\lambda$. Since
  $L_\lambda$ is projective and induction preserves this property, the
  module $\cB_\lambda$ is projective as well (see, e.g.,
  \cite{Zou1996:MR1378540}). Hence it possesses a decomposition
\begin{align}
  \label{eq:BRep}
  \cB_\lambda
  \ =\ \bigoplus_{\mu\in\SRep(\g)}m_{\lambda\mu}\,\cP_\mu
\end{align}
  into projective covers. Due to the relation
  $\dim\Hom_\g\bigl(\cP_\lambda,\cL_\mu\bigr)=\delta_{\lambda\mu}$ (``the head
  of $\cP_\lambda$ is $\cL_\lambda$''), the
  multiplicities can be obtained as the dimension
\begin{align}
  \label{eq:Multiplicities}
  m_{\lambda\mu}
  \ =\ \dim\Hom_\g\bigl(\cB_\lambda,\cL_\mu\bigr)
\end{align}
  of a suitable space of intertwiners.

  Finally, we introduce an equivalence relation on the set
  $\SRep(\g)$. Two weights $\mu$ and $\nu$ are said to be in the
  same block if there exists a non-split extension
\begin{align}
  0\rightarrow\cL_\mu\rightarrow\cA\rightarrow\cL_\nu\rightarrow0\ \ .
\end{align}
  In other words, if $\cL_\mu$ and $\cL_\nu$ can be obtained as a
  submodule and a quotient of $\cA$, respectively, but nevertheless
  $\cA$ is not isomorphic to $\cL_\mu\oplus\cL_\nu$. The division of
  weights into blocks defines an equivalence relation which can be
  represented as a graph. The vertices are just the simple modules
  and two vertices are connected by a line if they admit a non-split
  extension. The blocks are the connected components of the resulting
  graph. There are a number of theorems on properties of blocks. For
  instance, it can be shown that the degree of atypicality is constant
  on blocks
  \cite{Germoni1998:MR1659915,Serganova:Blocks,Cheng:2012MR3012224}.
  Moreover, different blocks are separated by central characters
  (i.e.\ the joint set of eigenvalues of {\em all} Casimir
  operators).

  We will use the symbol $[\sigma]$ to denote the block a given weight
  $\sigma$ belongs to. A weight is typical if and only if the
  corresponding block contains precisely one element. The symbol
  $\text{ABlocks}(\g)$ will be reserved for the set of blocks obtained
  from {\em atypical} modules $\sigma$.

\paragraph{Example: The blocks of \texorpdfstring{$\mathbf{\gl(1|1)}$}{gl(1|1)}.}
  The simple representations of $\gl(1|1)$ fall into the blocks
  $[(e,n)]=\{(e,n)\}$ with $e\neq0$ and
  $[n]\defeq[(0,n)]=\{(0,n+m)|m\in\Integer\}$, see the illustration
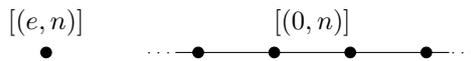
\begin{center}
\begin{tikzpicture}
  \draw[fill] (0,0) circle (2pt);
  \draw (1.5,0) node {\tiny$\cdots$}
        (5.5,0) node {\tiny$\cdots$};
  \draw (1.7,0) -- (5.3,0);
  \draw[fill] (2,0) circle (2pt)
              (3,0) circle (2pt)
              (4,0) circle (2pt)
              (5,0) circle (2pt);
  \draw (0,0) node[above=1mm] {$[(e,n)]$};
  \draw (3.5,0) node[above=1mm] {$[(0,n)]$};
\end{tikzpicture}
\end{center}
  Indeed, the Kac module $\cK_{(0,n)}$
  provides a non-split extension
\begin{align}
  0\to\cL_{n-1}\to\cK_{(0,n)}\to\cL_{n}\to0\ \ .
\end{align}
  The projective covers $\cP_n\defeq\cP_{(0,n)}$ of atypical simple
  modules $\cL_n=\cL_{(0,n)}$ coincide (for $\gl(1|1)$) with the
  modules $\cB_{(0,n)}$. We note in passing that the quadratic Casimir
  operator $C$ acts in a non-diagonalizable fashion on the atypical
  projective covers $\cP_n$ \cite{Rozansky:1992rx}.

\subsection{\label{sc:HarmonicAnalysis}Harmonic analysis on
  supermanifolds}

  In this section, the harmonic analysis on supergroups and
  superspheres is discussed. In contrast to the purely bosonic case,
  the Laplacian turns out to be non-diagonalizable in the former
  case, thereby establishing a natural link to logarithmic conformal 
  field theory. Our presentation will follow closely the logic of the
  article \cite{Mitev:2011zza}.

\subsubsection{Guide to the literature on supergeometry}

  Roughly speaking, a supermanifold is a manifold which can locally be
  described by flat bosonic and fermionic -- Grassmann algebra
  valued -- coordinates. There are several competing definitions of
  how this physicist's intuition can be made mathematically precise.
  Since our main focus rests on physical applications, we are not
  capable of giving a complete account here, only a rather brief
  outline of some of the most important concepts. Readers interested
  in gaining a mathematically rigorous understanding of supermanifolds
  are invited to consult the original literature or some of the recent
  books on this subject. More physically minded readers may also start
  with Witten's recent exposition \cite{Witten:2012bg}.

  A very prominent approach follows the logic of Berezin, Kostant
  and Leites (BKL) and uses the language of algebraic geometry, in
  particular the concept of locally ringed spaces. As introductory
  books we can recommend
  \cite{Leites:1980MR565567,Varadarajan:2004MR2069561,Carmeli:2011MR2840967,Alldridge:Book}.
  Some more advanced aspects of the theory are treated in
  \cite{Manin:1998,Deligne:1999MR1701597}. An alternative route,
  focusing more on differential geometric concepts and building up the
  theory from the local descriptions of supermanifolds, is covered
  in the books
  \cite{Bartocci:1991MR1175751,DeWitt:1992MR1172996,Tuynman:2004MR2102797,Rogers:2007MR2320438}.
  The two approaches give rise to equivalent notions of supermanifolds
  (in the sense of a categorical equivalence). At the level of
  computations in local coordinates, they are virtually
  identical. However,  the BKL approach is more amenable to the
  application of advanced methods from algebraic geometry. Moreover,
  being more abstract, it is more flexible; for instance, only minor
  modifications lead to $\Integer$-graded manifolds as they arise in
  mathematical treatments of the BV formalism.

\subsubsection{Supergroups}

  A mathematically rigorous definition of a Lie supergroup $G$ can be
  obtained from a super Harish-Chandra pair $(\groupG_\den,\g)$, i.e.\
  a pair of an ordinary Lie group $\groupG_\den$ and a Lie
  superalgebra $\g$ such that $\g_\den$ is the Lie algebra associated
  with $\groupG_\den$. Moreover, there should be an action
  of $\groupG_\den$ on $\g$ which restricts to the adjoint action on
  $\g_\den$ and whose differential close to the identity extends to
  the adjoint action of $\g_\den$ on $\g$ (see, e.g.,
  \cite{Carmeli:2011MR2840967}).

  In our review, we shall adopt a more ad-hoc perspective, thereby
  ignoring, to a large extent, potential mathematical subtleties. We
  shall simply assume that the idea of exponentiating a Lie algebra to
  obtain a Lie group carries over to Lie superalgebras if odd
  generators are paired with fermionic coordinates. For our purposes,
  a Lie supergroup $\groupG$ is then a fermionic extension of an
  ordinary Lie group $\groupG_\den$ by fermionic coordinates which
  transform in a suitable representation of $\groupG_\den$. All the
  properties of a supergroup are inherited from this data. For this
  reason, we will review the well-known harmonic analysis on ordinary
  compact Lie groups first. The goal of harmonic analysis is to learn
  about the structure of a group from studying the action of invariant
  differential operators -- Lie derivatives and Laplace operators --
  on its algebra of functions.

  Let us consider a compact, simple, simply-connected Lie group
  $\groupG_\den$. According to the Peter-Weyl theorem, the algebra
  $\cF(\groupG_\den)$ of square integrable functions on $\groupG_\den$
  (with respect to the Haar measure) decomposes as
\begin{align}
  \label{eq:PeterWeyl}
  \cF(\groupG_\den)
  \ \cong\ \bigoplus_{\mu\in\SRep(\groupG_\den)}L_\mu\otimes L_\mu^\ast
\end{align}
  under the left-right regular action $(l,r)\cdot f:g\mapsto
  f(l^{-1}gr)$ of $\groupG_\den\times \groupG_\den$, where the sum
  is over all finite dimensional irreducible representations of
  $\groupG_\den$ and the asterisk $\ast$ refers to the dual
  representation. Each individual term in the decomposition can be
  thought of as being associated with representation matrices
  $\rho^{(\mu)}(g)\in\End(L_\mu)$ with $g\in\groupG_\den$. The statement
  of the Peter-Weyl theorem is that these matrix elements can be used
  to approximate any function on $\groupG_\den$ with arbitrary
  precision. In eq.~\eqref{eq:PeterWeyl} it has been assumed
  implicitly that we consider the closure of the right hand side.

  The extension to supergroups is straightforward. Compared to an
  ordinary group, a supergroup comes with additional Grassmann algebra
  valued coordinates which generate the exterior algebra
  $\bigwedge(\g_\deo^\ast)$. This space admits an obvious action of
  $\g_\den$ by the Lie bracket. The algebra of functions on the
  supergroup $\groupG$ is the induced module (with respect to the
  right action of $\groupG_\den$)
\begin{align}
  \cF(\groupG)
  \ =\ \Ind_{\g_\den}^{\g}\cF(\groupG_\den)
  \ =\ \cF(\groupG_\den)\otimes\bigwedge(\g_\deo^\ast)\ \ .
\end{align}
  This definition has a natural interpretation arising from formally
  expanding functions on~$\groupG$ in a Taylor series in the odd
  coordinates. Apart from the right action of $\groupG$,
  $\cF(\groupG)$ also admits a left action, just as in the bosonic
  case. Our goal is to understand the decomposition of this algebra as
  a $\g\oplus\g$-module (with respect to the left and right regular
  action). The result will provide a super-analogue of the Peter-Weyl
  theorem.

  Since all finite dimensional representations of a reductive Lie
  algebra $\g_\den$ are projective the same will be true for the
  induced module $\cF(\groupG)$. Hence, as a right $\g$-module,
  $\cF(\groupG)$ has the decomposition
\begin{align}
  \label{eq:FGright}
  \cF(\groupG)
  \ =\ \bigoplus_{\mu\in\SRep(G)} M_\mu\otimes\cP^\ast_\mu\ \ ,
\end{align}
  where the sum is over all projective covers of $\g$ and the $M_\mu$
  are some multiplicity spaces. As a left $\g$-module, $\cF(\groupG)$ has
  precisely the same decomposition. Indeed, the algebra of functions
  has to be isomorphic with respect to the left and the right regular
  action due to the existence of the isomorphism
  $\Omega:\cF(\groupG)\to\cF(\groupG)$ which acts as  $\Omega(f):g\mapsto
  f(g^{-1})$ and which intertwines the left and right
  regular actions.

  In the typical sector, $\cP_\mu$ agrees with $\cL_\mu$. Given the
  symmetry between the left and the right action it is then obvious
  that $M_\mu\cong\cL_\mu$ as vector spaces. We will now show that
  this is indeed always the case, not only in the typical sector but
  also in the atypical sector. First of all we notice that, by
  definition \eqref{defB} and eq.~\eqref{eq:BRep}, the algebra of 
  functions on $\groupG$ has the form
\begin{align}
  \label{eq:SuperPeterWeylRight}
  \cF(\groupG)
  \ =\ \bigoplus_{\mu\in\SRep(\g_\den)}L_\mu\otimes\cB_\mu^\ast
  \ =\ \bigoplus_{\mu\in\SRep(\g_\den)}m_{\mu\nu}\,L_\mu\otimes\cP_\mu^\ast
\end{align}
  as a $\g_\den\oplus\g$-module. We then employ Frobenius reciprocity to
  rewrite eq.~\eqref{eq:Multiplicities} as
\begin{align}
  m_{\mu\nu}
  \ =\ \dim\Hom_{\g_\den}\bigl(L_\mu,\cL_\nu\bigr)\ \ ,
\end{align}
  which proves our assertion, given that all $\g_\den$-modules are fully
  reducible.

  The result just obtained suggests that we have a factorization
  $\cL_\mu\otimes\cL^\ast_\mu$ of the individual contributions in the
  typical sector, just as in the case of $\cF(\groupG_\den)$. In the
  atypical sector, however, such a factorization is not possible since
  the projective covers $\cP_\mu$ are strictly larger than the simple
  modules $\cL_\mu$. For this reason, the left and right modules in
  the atypical sector are entangled in a complicated way, arranging
  themselves in one infinite dimensional non-factorizable (non-chiral)
  indecomposable $\g\oplus\g$-module $\cI_{[\sigma]}$ for each
  individual {\em block} $[\sigma]$. We finally find\footnote{Versions
    of this result appear to be known among mathematicians specialized
    on Lie superalgebras, even though no specific reference seems to
    exist. In the physics literature, the result was first noted in
    \cite{Quella:2007hr} and, from the more general perspective
    adopted here, in \cite{Mitev:2011zza}.}
\begin{align}
  \label{eq:SuperPeterWeyl}
  \cF(\groupG)
  \ =\ \bigoplus_{\mu\in\Typ(\g)}\cL_\mu\otimes\cL^\ast_\mu
       \oplus\bigoplus_{[\sigma]\in\text{ABlocks}(\g)}\cI_{[\sigma]}
\end{align}
  for the decomposition of $\cF(\groupG)$ as a
  $\g\oplus\g$-module. For Lie supergroups of type~I an alternative
  derivation based on BGG duality \cite{Zou1996:MR1378540} has been
  worked out in great detail in \cite{Quella:2007hr}. In that case,
  the distinguished $\Integer$-grading \eqref{eq:ZGrading} allows to
  split the fermions into two separate sets which transform
  non-trivially only under the left and right action of $\groupG_\den$
  on $\groupG$, respectively. Needless to say, this decomposition is
  perfectly adapted to the natural bimodule structure of $\cF(G)$.

  It should be emphasized that the Laplace operator $\Delta$ -- which
  can be thought of as half the quadratic Casimir operator acting on
  $\cF(\groupG)$ -- is not diagonalizable on the atypical projective
  covers $\cP_\mu$ and on the non-chiral modules $\cI_{[\sigma]}$,
  certainly for type~I supergroups \cite{Quella:2007hr} but probably
  beyond. For $\sigma$-models on supergroups this implies that they
  are generally logarithmic conformal field theories (see
  \cite{Schomerus:2005bf} for an explicit derivation of logarithmic
  correlation functions). Exceptions may occur for small volumes where
  the spectrum of the CFT can be truncated in such a way that the
  modules $\cI_{[\sigma]}$ no longer contribute
  \cite{Saleur:2006tf,Mitev:2008yt}.

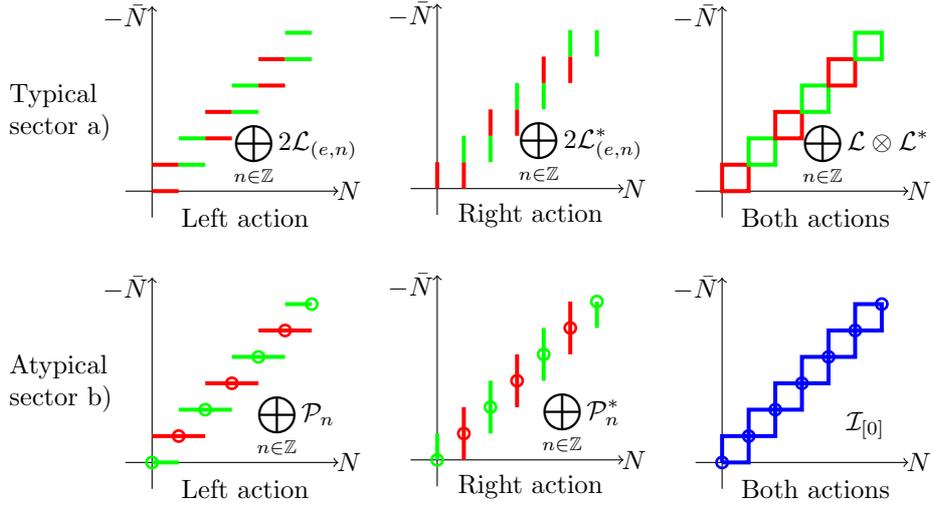
\begin{figure}
\begin{center}
\begin{tikzpicture}[scale=.7]
  \draw (0,2) node[left,text width=1.4cm] {Typical\\sector a)};
  \draw[->] (.5,0) -- (.5,4);
  \draw[->] (0,.5) -- (4,.5);
  \draw[red,line width=1.5pt] (.5,.5) -- (1,.5);
  \draw[red,line width=1.5pt] (.5,1) -- (1,1);
  \draw[green,line width=1.5pt] (1,1) -- (1.5,1);
  \draw[green,line width=1.5pt] (1,1.5) -- (1.5,1.5);
  \draw[red,line width=1.5pt] (1.5,1.5) -- (2,1.5);
  \draw[red,line width=1.5pt] (1.5,2) -- (2,2);
  \draw[green,line width=1.5pt] (2,2) -- (2.5,2);
  \draw[green,line width=1.5pt] (2,2.5) -- (2.5,2.5);
  \draw[red,line width=1.5pt] (2.5,2.5) -- (3,2.5);
  \draw[red,line width=1.5pt] (2.5,3) -- (3,3);
  \draw[green,line width=1.5pt] (3,3) -- (3.5,3);
  \draw[green,line width=1.5pt] (3,3.5) -- (3.5,3.5);
  \draw (2.25,0) node(y) {Left action};
  \draw (4.2,.5) node(y) {$N$};
  \draw (0,3.8) node(y) {$-\bar{N}$};
  \draw (3.2,1.2) node {$\displaystyle\bigoplus_{n\in\Integer}2\cL_{(e,n)}$};
\end{tikzpicture}
\begin{tikzpicture}[scale=.7]
  \draw[->] (.5,0) -- (.5,4);
  \draw[->] (0,.5) -- (4,.5);
  \draw[red,line width=1.5pt] (.5,.5) -- (.5,1);
  \draw[red,line width=1.5pt] (1,.5) -- (1,1);
  \draw[green,line width=1.5pt] (1,1) -- (1,1.5);
  \draw[green,line width=1.5pt] (1.5,1) -- (1.5,1.5);
  \draw[red,line width=1.5pt] (1.5,1.5) -- (1.5,2);
  \draw[red,line width=1.5pt] (2,1.5) -- (2,2);
  \draw[green,line width=1.5pt] (2,2) -- (2,2.5);
  \draw[green,line width=1.5pt] (2.5,2) -- (2.5,2.5);
  \draw[red,line width=1.5pt] (2.5,2.5) -- (2.5,3);
  \draw[red,line width=1.5pt] (3,2.5) -- (3,3);
  \draw[green,line width=1.5pt] (3,3) -- (3,3.5);
  \draw[green,line width=1.5pt] (3.5,3) -- (3.5,3.5);
  \draw (2.25,0) node(y) {Right action};
  \draw (4.2,.5) node(y) {$N$};
  \draw (0,3.8) node(y) {$-\bar{N}$};
  \draw (3.2,1.2) node {$\displaystyle\bigoplus_{n\in\Integer}2\cL_{(e,n)}^\ast$};
\end{tikzpicture}
\begin{tikzpicture}[scale=.7]
  \draw[->] (.5,0) -- (.5,4);
  \draw[->] (0,.5) -- (4,.5);
  \draw[red,line width=1.5pt] (.5,.5) rectangle (1,1);
  \draw[green,line width=1.5pt] (1,1) rectangle (1.5,1.5);
  \draw[red,line width=1.5pt] (1.5,1.5) rectangle (2,2);
  \draw[green,line width=1.5pt] (2,2) rectangle (2.5,2.5);
  \draw[red,line width=1.5pt] (2.5,2.5) rectangle (3,3);
  \draw[green,line width=1.5pt] (3,3) rectangle (3.5,3.5);
  \draw (2.25,0) node(y) {Both actions};
  \draw (4.2,.5) node(y) {$N$};
  \draw (0,3.8) node(y) {$-\bar{N}$};
  \draw (3.2,1.2) node {$\displaystyle\bigoplus_{n\in\Integer}\cL\otimes\cL^\ast$};
\end{tikzpicture}
\end{center}
\begin{center}
\begin{tikzpicture}[scale=.7]
  \draw (0,2) node[left,text width=1.4cm] {Atypical\\sector b)};
  \draw[->] (.5,0) -- (.5,4);
  \draw[->] (0,.5) -- (4,.5);
  \draw[green,line width=1.5pt] (.5,.5) -- (1,.5);
  \draw[green,line width=1pt] (.5,.5) circle (3pt);
  \draw[red,line width=1.5pt] (.5,1) -- (1.5,1);
  \draw[red,line width=1pt] (1,1) circle (3pt);
  \draw[green,line width=1.5pt] (1,1.5) -- (2,1.5);
  \draw[green,line width=1pt] (1.5,1.5) circle (3pt);
  \draw[red,line width=1.5pt] (1.5,2) -- (2.5,2);
  \draw[red,line width=1pt] (2,2) circle (3pt);
  \draw[green,line width=1.5pt] (2,2.5) -- (3,2.5);
  \draw[green,line width=1pt] (2.5,2.5) circle (3pt);
  \draw[red,line width=1.5pt] (2.5,3) -- (3.5,3);
  \draw[red,line width=1pt] (3,3) circle (3pt);
  \draw[green,line width=1.5pt] (3,3.5) -- (3.5,3.5);
  \draw[green,line width=1pt] (3.5,3.5) circle (3pt);
  \draw (2.25,0) node(y) {Left action};
  \draw (4.2,.5) node(y) {$N$};
  \draw (0,3.8) node(y) {$-\bar{N}$};
  \draw (3.2,1.2) node {$\displaystyle\bigoplus_{n\in\Integer}\cP_n$};
\end{tikzpicture}
\begin{tikzpicture}[scale=.7]
  \draw[->] (.5,0) -- (.5,4);
  \draw[->] (0,.5) -- (4,.5);
  \draw[green,line width=1.5pt] (.5,.5) -- (.5,1);
  \draw[green,line width=1pt] (.5,.5) circle (3pt);
  \draw[red,line width=1.5pt] (1,.5) -- (1,1.5);
  \draw[red,line width=1pt] (1,1) circle (3pt);
  \draw[green,line width=1.5pt] (1.5,1) -- (1.5,2);
  \draw[green,line width=1pt] (1.5,1.5) circle (3pt);
  \draw[red,line width=1.5pt] (2,1.5) -- (2,2.5);
  \draw[red,line width=1pt] (2,2) circle (3pt);
  \draw[green,line width=1.5pt] (2.5,2) -- (2.5,3);
  \draw[green,line width=1pt] (2.5,2.5) circle (3pt);
  \draw[red,line width=1.5pt] (3,2.5) -- (3,3.5);
  \draw[red,line width=1pt] (3,3) circle (3pt);
  \draw[green,line width=1.5pt] (3.5,3) -- (3.5,3.5);
  \draw[green,line width=1pt] (3.5,3.5) circle (3pt);
  \draw (2.25,0) node(y) {Right action};
  \draw (4.2,.5) node(y) {$N$};
  \draw (0,3.8) node(y) {$-\bar{N}$};
  \draw (3.2,1.2) node {$\displaystyle\bigoplus_{n\in\Integer}\cP_n^\ast$};
\end{tikzpicture}
\begin{tikzpicture}[scale=.7]
  \draw[->] (.5,0) -- (.5,4);
  \draw[->] (0,.5) -- (4,.5);
  \draw[blue,line width=1.5pt] (.5,.5) -- (1,.5) -- (1,1.5) -- (2,1.5)
  -- (2,2.5) -- (3,2.5) -- (3,3.5) -- (3.5,3.5) -- (3.5,3) -- (2.5,3)
  -- (2.5,2) -- (1.5,2) -- (1.5,1) -- (.5,1) -- (.5,.5);
  \draw[blue,line width=1pt] (.5,.5) circle (3pt);
  \draw[blue,line width=1pt] (1,1) circle (3pt);
  \draw[blue,line width=1pt] (1.5,1.5) circle (3pt);
  \draw[blue,line width=1pt] (2,2) circle (3pt);
  \draw[blue,line width=1pt] (2.5,2.5) circle (3pt);
  \draw[blue,line width=1pt] (3,3) circle (3pt);
  \draw[blue,line width=1pt] (3.5,3.5) circle (3pt);
  \draw (2.25,0) node(y) {Both actions};
  \draw (4.2,.5) node(y) {$N$};
  \draw (0,3.8) node(y) {$-\bar{N}$};
  \draw (3.2,1.2) node {$\cI_{[0]}$};
\end{tikzpicture}
  \caption{\label{fig:GL}(Color online) Sketch of the harmonic
    analysis on supergroups using the example of $\GL(1|1)$. It
    is shown how the space of functions organizes itself with respect
    to the left and the right action of $\g$ (the axes correspond to
    eigenvalues of the respective Cartan subalgebra) and with respect
    to the simultaneous action of both. In the typical sector a) we
    observe a factorization of representation (green and red) while in
    the atypical sector b) the functions organize themselves in
    infinite dimensional non-factorizing representations (blue) due to
    the extension of simple modules into projective covers.}
\end{center}
\end{figure}

\paragraph{Example: Harmonic analysis on $\GL(1|1)$.}
  Since our previous result has been very abstract, let us explain it
  in more detail using the example of $\GL(1|1)$ (first
  considered in \cite{Schomerus:2005bf}). As coordinates we choose
  $x,y,\eta,\bar{\eta}$ which are used to represent a general group
  element $g\in\GL(1|1)$ in the form
\begin{align}
  g\ =\ e^{i\eta\psi^+}\,e^{ixE+iyN}\,e^{i\bar{\eta}\psi^-}\ \ .
\end{align}
  The space of functions on $\GL(1|1)$ is spanned by the plane
  waves $e^{-i(ex+ny)}$, multiplied by a polynomial in the fermionic
  coordinates $\eta,\bar{\eta}$. On this space, the regular action of
  $\GL(1|1)$ on itself leads to the following two mutually (graded) 
  commuting copies of the Lie superalgebra $\gl(1|1)$,
\begin{align}
  \label{eq:Liegl11}
  E
  &=i\partial_x\ ,&
  N
  &=i\partial_y-\eta\partial_\eta\ ,&
  \psi^+
  &=-i\partial_\eta\ ,&
  \psi^-
  &=ie^{iy}\partial_{\bar{\eta}}-\eta\partial_x\\[2mm]
  \bar{E}
  &=-i\partial_x\ ,&
  \bar{N}
  &=-i\partial_y+\bar{\eta}\partial_{\bar{\eta}}\ ,&
  \bar{\psi}^-
  &=-i\partial_{\bar{\eta}}\ ,&
  \bar{\psi}^+
  &=ie^{iy}\partial_{\eta}-\bar{\eta}\partial_x\ \ .
\end{align}
  We note that the generators $E,\bar{E}$ and $N,\bar{N}$ reduce to
  momentum operators when restricted to purely bosonic functions. The
  space of functions on $\GL(1|1)$ is spanned by the following vectors
  with quantum numbers
\begin{center}
\begin{tabular}{c|cccc}
  & $e^{-i(ex+ny)}$ & $e^{-i(ex+ny)}\eta$ & $e^{-i(ex+ny)}\bar{\eta}$ & $e^{-i(ex+(n+1)y)}\eta\bar{\eta}$ \\\hline\hline\\[-1em]
  $(E,\bar{E})$ & $(e,-e)$ & $(e,-e)$ & $(e,-e)$ & $(e,-e)$ \\[2mm]
  $(N,\bar{N})$ & $(n,-n)$ & $(n-1,-n)$ & $(n,-n+1)$ & $(n,-n)$
\end{tabular}
\end{center}
  Restricting our attention to integer values of $n$ and fixed value
  of $e$ for simplicity of illustration, the resulting weight diagram
  reduces to the patterns sketched in Figure~\ref{fig:GL}.
  In the typical sector $e\neq0$, the states organize themselves into
  two-dimensional simple modules $\cL$, both with respect to the left and
  with respect to the right action (red and green lines). Under the combined
  action they combine into four-dimensional representations (red and green
  boxes) which correspond to the tensor product
  $\cL\otimes\cL^\ast$. In the atypical sector $e=0$, however, the
  picture is very different. Here the states organize themselves in
  four-dimensional projective covers $\cP$ if only one of the two
  actions is considered. One should imagine the diamond from
  Figure~\ref{fig:GLReps} but now perpendicular to the plane, with two
  of the vertices being located in the plane. In this case the states
  cannot be organized in a tensor product under the combined action
  for obvious reasons. Instead, they have to combine into infinite
  dimensional indecomposable multiplets (blue), one for each value of
  $n\text{ mod }1$, i.e.\ one for each block.

  Finally, we comment on the action of the Laplace operator
  $\Delta$. It is obtained by expressing the quadratic Casimir $C$
  from \eqref{eq:Casimirgl} in terms of the Lie derivatives
  \eqref{eq:Liegl11} and it reads
\begin{align}
  \Delta
  \ =\ C/2
  \ =\ -\partial_x\partial_y-i/2\partial_x-e^{iy}\partial_\eta\partial_{\bar{\eta}}\
  \ .
\end{align}
  An explicit inspection of the two-dimensional $\Delta$-invariant
  subspace spanned by the functions $e^{-i(n+1)}\eta\bar{\eta}$ and
  $e^{-in}$ shows that the Laplacian cannot be diagonalized on the
  atypical sector with $e=0$ (which is annihilated by $\partial_x$)
  \cite{Schomerus:2005bf}.

\subsubsection{\label{sc:HASupercoset}Harmonic analysis on supercosets}

  The harmonic analysis on a supercoset $\groupG/\groupH$ where the
  supergroup elements are identified according to the rule $g\sim gh$
  with $h\in\groupH$ can be deduced from that of the supergroup
  case. Indeed, the algebra of functions on $\groupG/\groupH$ can be
  thought of as the space of $\groupH$-invariant functions on
  $\groupG$,
\begin{align}
  \label{eq:InvSpace}
  \cF(\groupG/\groupH)
  \ =\ \Inv_{\groupH}\cF(\groupG)\ \ ,
\end{align}
  where an element $h$ of the supergroup $\groupH$ acts on
  $f\in\cF(\groupG)$ according to
\begin{align}
  h\cdot f(g)
  \ =\ f(gh)\ \ .
\end{align}
  It is obvious that the space $\groupG/\groupH$ and hence also the
  algebra of functions $\cF(\groupG/\groupH)$ still admits an action of
  $\groupG$. The isometry supergroup of $\groupG/\groupH$ might be
  bigger than $\groupG$ but for simplicity we will only consider the
  symmetry $\groupG$.

  Writing the invariant subspace \eqref{eq:InvSpace} explicitly as a
  direct sum over indecomposable $\groupG$-modules turns out to be
  rather involved in the general case. The main reason is that the
  modules over Lie superalgebras are not fully reducible. On the
  one hand such modules already appear in $\cF(\groupG)$, as
  $\groupG$-modules with respect to the right regular action, see
  eq.~\eqref{eq:FGright}. On the other hand, they may also arise when
  decomposing simple $\groupG$-modules after restricting the action to
  the supergroup $\groupH$. Finally, the invariants that need to be
  extracted when restricting from $\cF(\groupG)$ to
  $\cF(\groupG/\groupH)$ can be either true $\groupH$-invariants of
  $\cF(\groupG)$ (i.e.\ simple $\groupH$-modules) or they can sit in a
  larger indecomposable $\groupH$-module. A general solution to this
  intricate problem is currently beyond reach. Nevertheless,
  eq.~\eqref{eq:SuperPeterWeylRight} teaches us one general lesson: In
  case $\groupH$ is purely bosonic, the algebra of functions
  $\cF(\groupG/\groupH)$ necessarily involves projective covers of
  simple $\groupG$-modules, with a non-diagonalizable action of the
  Laplace operator. In particular, this observation is relevant for
  the supercosets describing $\AdS$ backgrounds in string theory.

  The problem sketched in the previous paragraph can be circumvented
  when working on the level of characters since these are not
  sensitive to the indecomposable structure of modules. For many
  purposes this will be sufficient, as the following example shows.

\paragraph{Example: The supersphere \texorpdfstring{$\Sphere^{3|2}$}{S(3|2)}.}

  Instead of treating the general case, we discuss one example in some
  detail, namely the supersphere $\Sphere^{3|2}$. This supersphere
  can be thought of as being embedded into the flat superspace
  $\Real^{4|2}$, i.e.\ we have four bosonic coordinates $x^i$ and two
  fermionic coordinates $\eta_1,\eta_2$ subject to the constraint
  $\vec{x}^2+2\eta_1\eta_2=R^2$. For our purposes, we can identify the
  algebra of functions $\cF(\Sphere^{3|2})$ with the polynomial
  algebra $S(\Real^{4|2})$ in the six coordinates $(X^a)=(x^i,\eta_\nu)$ 
  modulo the ideal generated by
  $\vec{X}^2=\vec{x}^2+2\eta_1\eta_2=R^2$.

  These coordinates transform
  in the vector representation of $\SO(4)$ and the defining
  representation of $\SP(2)$, respectively. Moreover, such bosonic
  transformations leave the constraint invariant. In addition, one can
  consider transformations which mix bosons and fermions. The
  resulting supergroup of isometries is $\OSP(4|2)$. Since the
  stabilizer of an arbitrary point on $\Sphere^{3|2}$ is
  $\OSP(3|2)$, this confirms that the supersphere possesses a
  representation as a supercoset $\OSP(4|2)/\OSP(3|2)$.

  In order to determine the character for the $\OSP(4|2)$-module
  $\cF(\Sphere^{3|2})$ we proceed as follows. We first identify the
  Cartan subalgebra of $\osp(4|2)$ with the Cartan subalgebra of
  $\su(2)\oplus\su(2)\oplus\su(2)$, where we use the
  identifications $\so(4)\cong\su(2)\oplus\su(2)$
  and $\sp(2)\cong\su(2)$. We then choose linear
  combinations of the six coordinates such that we can assign the
  weights $(\epsilon,\eta,0)$ and $(0,0,\epsilon)$ to them, with
  $\epsilon,\eta=\pm1$. In addition we introduce a quantum number for
  the polynomial grading. Each of the six coordinates $X^a$ then
  contributes a term $z_1^{m_1}z_2^{m_2}z_3^{m_3}t$ to the character,
  where $m_i\in\{0,\pm1\}$ have to be chosen according to the
  respective weights and $t$ keeps track of the polynomial grade. For
  a product of coordinates, these individual contributions have to be
  multiplied with each other since the quantum numbers and the
  polynomial degree add up.

  After these preparations it is very simple to write down the
  character of all polynomial functions in the coordinates
  $X^a$. Dividing out the ideal $\vec{X}^2=R^2$ is taken into account
  by multiplying the previous character with $1-t^2$ since the
  constraint relates every polynomial of degree $n\geq2$ to a
  polynomial of degree $n-2$ with the same $\osp(4|2)$-weight. At the
  end we take the limit $t\to1$ since the grade is not a good quantum
  number once we impose the constraint. The total character is thus
  given by
\begin{align}
  \label{eq:ZSS1}
  Z_{\cF(\Sphere^{3|2})}(z_1,z_2,z_3)
  \ =\ \lim_{t\to1}\frac{(1-t^2)(1+tz_3)(1+t/z_3)}{(1-tz_1z_2)(1-tz_1/z_2)(1-tz_2/z_1)(1-t/z_1z_2)}\ \ ,
\end{align}
  where all terms in the numerator should be expanded in a geometric
  series before the limit is taken \cite{Mitev:2008yt}. The resulting
  expression can be represented as a linear combination of characters
  of $\osp(4|2)$. Without going into the details we just write down
  the result
\begin{align}
  Z_{\cF(\Sphere^{3|2})}(z_1,z_2,z_3)
  \ =\ \chi_{[0,0,0]}(z_1,z_2,z_3)
       +\sum_{k=0}^\infty\chi_{[1/2,k/2,k/2]}(z_1,z_2,z_3)\
  \ ,
\end{align}
  where the labels $[j_1,j_2,j_3]$ refer to simple modules (see
  \cite{Mitev:2008yt}). Since all labels correspond to atypical
  irreducible representations in different blocks, this character
  decomposition at the same time yields the result for the harmonic
  analysis on $\Sphere^{3|2}$ (see also
  \cite{Coulembier:2012arXiv1202.0668C}). The considerations of this
  section will be used in Section~\ref{sc:SupersphereDuality} when we
  discuss the spectrum of freely moving open strings on
  $\Sphere^{3|2}$.

\subsection{Cohomological reduction}

  We shall now review a tool which is capable of extracting part of
  the information of interest from a simplified setup. In the context
  of (logarithmic) conformal field theory it can be used to reduce the
  calculation of certain observables to those in theories of either
  free bosons or symplectic fermions~\cite{Candu:2010yg}. The method
  has been applied successfully to a non-perturbative proof of
  conformal invariance for symmetric superspace $\sigma$-models
  \cite{Candu:2010yg}. Combining it with the idea of quasi-abelian
  perturbation theory, see Section~\ref{sc:Deformations}, it has also
  been used to conjecture exact D-brane spectra on projective
  superspaces~\cite{Candu:2009ep}.

\subsubsection{General formalism}

  The procedure of cohomological reduction starts with the choice of a
  BRST charge~$Q$ in the Lie superalgebra~$\g$. By definition, the
  BRST charge is a fermionic element $Q\in\g$ which satisfies
  $[Q,Q]=0$. The existence of such a~$Q$ is guaranteed for all Lie
  superalgebras for which atypical representations exist, e.g.~for all
  simple Lie superalgebras except for $\osp(1|2n)$. It implies the
  existence of a nilpotent operator $Q_\cM=\rho_\cM(Q)$ in any
  representation $\rho_\cM:\g\to\End(\cM)$. The equation $Q_\cM^2=0$
  then allows us to define the cohomology classes
\begin{align}
  \Coh_Q(\cM)
  \ =\ \Ker Q_\cM  / \Img Q_\cM \ \ .
\end{align}
  Before we proceed with analyzing the general structure of the
  cohomology classes $\Coh_Q(\cM)$ it is useful and instructive to
  focus on the adjoint representation $\cM=\g$ for a second. Using
  just the Jacobi identity \eqref{eq:Jacobi}, it is then possible to
  establish that \cite{Candu:2010yg}
\begin{enumerate}
\item[1)] the subspaces $\Ker\ad_Q$ and $\Img\ad_Q$
  are subalgebras of $\g$,
\item[2)] the subalgebra $\Img\ad_Q$ is an ideal of
  $\Ker\ad_Q$,
\item[3)] the quotient space $\Coh_Q(\g)$ is a Lie
  superalgebra.
\end{enumerate}
  The element $Q$ thus defines a decomposition of $\g$ into three
  vector superspaces $\mathfrak{h},\mathfrak{e},\mathfrak{f}$,
\begin{equation}
\begin{split}
  \label{eq:CohDeco}
  \g
  &\ =\ \h \oplus \mathfrak{e} \oplus \mathfrak{f} \ \ \quad \quad \text{ such that} \\[2mm]
  \mathfrak{e}
   &\ =\ \Img\ad_Q \ ,\quad
   \h \oplus \mathfrak{e} = \Ker\ad_Q
  \quad\text{ and }\quad
  \h\ =\ \Coh_Q(\g)\ \ .
\end{split}
\end{equation}
  Moreover, any metric $\langle\cdot,\cdot\rangle$ on $\g$ restricts
  to a non-degenerate form on $\h\subset\g$. The Lie sub-superalgebras
  $\mathfrak{e}$ and $\mathfrak{f}$, on the other hand, are isotropic,
  i.e.\ $\langle\mathfrak{e} ,\mathfrak{e}\rangle = 0 =
  \langle\mathfrak{f},\mathfrak{f}\rangle$.

  We also note that $\mathfrak{e}$ and $\mathfrak{f}$ both carry an
  action of the Lie superalgebra $\h$. In other words, the direct sum
  decomposition of $\mathfrak{g}$ in eq.~\eqref{eq:CohDeco}
  can be regarded as an isomorphism of $\h$-modules. The Lie
  superalgebra $\h$ has been computed for various choices of $\g$ and
  any $Q \in \g$. The results may be summarized as follows
\begin{align}
  \h\bigl(\gl(M|N)\bigr)
  &\ \simeq\ \gl(M-r_Q|N-r_Q)\ ,\nonumber\\[2mm]
  \h\bigl(\sgl(M|N)\bigr)
  &\ \simeq\ \sgl(M-r_Q|N-r_Q)\ ,\\[2mm]
  \h\bigl(\osp(R|2N)\bigr)
  &\ \simeq\ \osp(R-2 r_Q|2N- 2 r_Q)\ .\nonumber
\end{align}
  The answer depends on $Q$ only through an integer $r_Q
  \geq 1$ that can be looked up in \cite{Candu:2010yg}. In all three
  cases listed above, there exist elements $Q$ with minimal rank
  $r_Q=1$.

  The relevance of our previous discussion stems from the fact that
  all the linear spaces $\Coh_Q(\cM)$ come equipped with an action of
  the Lie sub-superalgebra $\h \subset \g$. First, notice that there
  is an $\h$-stable filtration
\begin{equation}
  \label{eq:d_stab_filtr}
  \cM\ \supset\  \Ker Q_\cM \ \supset \ \Img Q_\cM \ .
\end{equation}
  Indeed, $\cM$ is an $\h$-module by restriction, while $\Ker Q_\cM$
  and $\Img Q_\cM$ are $\h$-submodules because $\h\subset\Ker\ad_Q$.
  Finally, $\Ker Q_\cM\supset \Img Q_\cM$ follows from $Q_\cM^2=0$.

  It is not difficult to see that $\cM \rightarrow \Coh_Q(\cM)$ is
  functorial, i.e.\ it is consistent with forming tensor products,
  direct sums and conjugation in the category of $\h$-modules
  \cite{Candu:2010yg}. Even though $\Coh_Q(\cM)$ vanishes for a large
  class of representations, it can certainly contain non-trivial
  elements. A trivial example is the cohomology of the adjoint
  $\g$-module $\cM=\g$ which reduces to $\Coh_Q(\g)=\h$. For a finite
  dimensional $\g$-module $\cM$ one may actually show that
\begin{equation}
  \sdim \Coh_Q(\cM)
  \ =\ \sdim\cM \ \ .
\end{equation}
  We conclude that all modules $\cM$ with non-vanishing superdimension
  give rise to a non-trivial cohomology group $\Coh_Q(\cM)\neq 0$. The
  condition $\sdim\cM\neq0$ is often satisfied for atypical
  representations (short multiplets) and it is these representations
  that will survive the procedure of cohomological reduction. For
  typical representations $\cM$, on the other hand, it can be proven
  that the cohomology $\Coh_Q(\cM)$ is always trivial. More generally,
  $\Coh_Q(\cM)=0$ for all (finite dimensional) projective modules.
  To summarize, each instance of cohomological reduction extracts
  information on certain atypical (short or semi-short) constituents
  of a given representation $\cM$. In this regard it closely resembles
  supersymmetric indices.

\subsubsection{Application: Function spaces on supercosets}

  Let us now discuss the application of cohomological reduction to
  supercosets $\groupG/\groupGp$. For this purpose, we consider a Lie
  superalgebra $\g$ along with a subalgebra $\g' \subset \g$. The
  corresponding Lie supergroups will be denoted by $\groupG$ and
  $\groupGp$, respectively. As before, we want to pick some fermionic
  element $Q \in \g$ with $[Q,Q]=0$. Let us now assume that $Q$ is
  contained in the subalgebra $\g' \subset \g$ so that its cohomology
  defines two Lie sub-superalgebras $\h \subset \g$ and $\h' \subset
  \g'$ with $\h' \subset \h$. We denote the associated Lie supergroups
  by $\groupH$ and $\groupHp$, respectively. Note that the space of
  functions on the coset superspace $\groupG/\groupGp$ carries an
  action of $\g$. In particular, the element $Q$ acts and gives rise
  to some cohomology. The central claim of this section is that the
  cohomology of some geometric object (smooth function, tensor form,
  square integrable function) defined on the coset superspace
  $\groupG/\groupGp$ is equivalent to a similar object defined on
  $\groupH/\groupHp$. This gives rise to isomorphisms of the type
\begin{equation} \label{eq:mainresult}
  \Coh_Q\bigl(\cF(\groupG/\groupGp)\bigr)
  \ \cong \ \cF(\groupH/\groupHp)\ \ ,
\end{equation}
which means that the cohomology of $Q$ in the space of square
integrable functions on $\groupG/\groupGp$ may be interpreted as a space of
square integrable functions on the coset superspace $\groupH/\groupHp$. We
note that $\cF(\groupH/\groupHp)$ carries an action of the Lie superalgebra
$\h = \Coh_Q(\g) \subset \g$. The isomorphism
\eqref{eq:mainresult} is an isomorphism of $\h$-modules.
The derivation of eq.\ \eqref{eq:mainresult} is relatively involved and can
be found in full detail in \cite{Candu:2010yg}.

\paragraph{Example: Cohomological reduction from $\mathbf{\CPone}$ to
  $\mathbf{\Real^{0|2}}$.}

As an example of the above, let us discuss the Lie superalgebra
$\g = \gl(2|2)$. For $Q$ we pick the matrix that contains a
single entry in the upper right corner. It is then easy to check
that
$$
\Ker\ad_Q \ = \ \h \oplus \mathfrak{e} \ \ni \ \left( \begin{array}{cccc}
a_{11} &
a_{12} & b_{11} & b_{12} \\ 0 & a_{22} & b_{21} & b_{22} \\
0 & c_{12} & d_{11} & d_{12} \\
0 & 0 & 0 & a_{11} \end{array} \right) \ , \ \  \Img_Q \g \ = \ \mathfrak{e}
\ \ni \ \left( \begin{array}{cccc} a_{11} &
a_{12} & b_{11} & b_{12} \\ 0 & 0 & 0 & b_{22} \\
0 & 0 & 0 & d_{12} \\
0 & 0 & 0 & a_{11} \end{array} \right)\ .
$$
Consequently,  $\Coh_Q(\g) = \h = \gl(1|1)$ consists of all
supermatrices in which $a_{22}, b_{21}, c_{12}$ and $d_{11}$ are
the only non-vanishing entries. Let us also specify the Lie
sub-superalgebra $\g'$ to consist of all elements in $\g$ with
vanishing entries $b_{11} = b_{21} = d_{12} = d_{21} = c_{11} =
c_{12} = 0$. Hence, $\g' \cong \gl(2|1) \oplus \gl(1)$. The
cohomology $\Coh_Q(\g') = \h' = \gl(1) \oplus \gl(1)$ of $\g'$ can
be read off easily.

In our example, the quotient  $\groupG/\groupGp$ is the complex projective
superspace $\CPone  \cong S^2 \times \mathbb{R}^{0|4}$. Functions
thereon may be decomposed into finite dimensional representations
of $\psl(2|2)$ as follows \cite{Candu:2009ep}
\begin{align}
  \cF(\CPone)
  \ \cong\ \bigoplus_{j=0}^{\infty}\cK_{(j,0)}\ \ .
\end{align}
The representations $\cK_{(j,0)}$ of $\psl(2|2)$ that appear in this
decomposition possess dimension $d_j = 16(2j+1)$. They are Kac modules
generated from the spherical harmonics on the bosonic 2-sphere by
application of four fermionic generators. For $j \neq 0$, the
$\psl(2|2)$-modules $\cK_{(j,0)}$ turn out to be projective (typical long
multiplets) and hence $\Coh_Q(\cK_{(j,0)}) = 0$ for all $j \neq 0$. The
only non-vanishing cohomology comes from the 16-dimensional Kac
module $\cK_{(0,0)}$. The latter is built from three atypical
irreducibles, namely two copies of the trivial representation $\cL_0$
and one copy of the 14-dimensional adjoint representation of
$\psl(2|2)$ \cite{Gotz:2005ka}. Each of these pieces contributes to
cohomology. While the two trivial representations give rise to two
even states, the adjoint representation has an excess of two odd
states which descend to cohomology. In total, we obtain a 4-dimensional
cohomology
\begin{align}
  \Coh_Q\bigl(\cF(\groupG/\groupGp)\bigr)
  \ =\ \Coh_Q\bigl(\cF(\CPone)\bigr)
  \ =\ \Coh_Q(\cK_{(0,0)})
  \ =\ \Real^{2|2} \ \ .
\end{align}
To be more precise, we note that
the linear space $\Real^{2|2}$ carries the 4-dimensional
projective cover of $\gl(1|1)$. According to our general statement,
the cohomology should agree with the space of functions on the
quotient $\groupH/\groupHp = \GL(1|1)/\GL(1) \times \GL(1)$. The quotient
possesses two fermionic coordinates and hence gives rise to a
4-dimensional algebra of functions over it,
\begin{align}
  \cF(\groupH/\groupHp)
  \ =\ \mathbb{R}^{2|2} \ \ .
\end{align}
It indeed agrees with the cohomology in the space of functions
over $\CPone$, as it was claimed in eq.\ \eqref{eq:mainresult}.

\section{\label{sc:WZW}Supergroup WZW models}

  In this section we discuss supergroup WZW models as primary examples
  of logarithmic conformal field theories, emphasizing the connections
  to harmonic analysis. In the first part we focus on supergroups of
  type~I. In this case, bosonic and fermionic degrees can be
  essentially decoupled, leading to the notion of a free fermion
  realization. As an example for the type~II families we introduce the
  free field realization of the orthosymplectic series at level
  $k=1$.

\subsection{The action functional}

  Let us fix a supergroup $\groupG$ together with a metric
  $\langle\cdot,\cdot\rangle$ of its Lie superalgebra $\g$. We assume
  the supergroup to be simply-connected and the invariant form to be
  suitably normalized (see below). The supergroup WZW model is a
  two-dimensional $\sigma$-model describing the propagation of strings
  on $\groupG$. The action functional for maps $g:\Sigma\to\groupG$ is
  given by~\cite{Witten:1983ar}
\begin{equation}
  \label{eq:WZWLagrangian}
  \cS^{\text{WZW}}[g]
  \ =\ -\frac{i}{4\pi}\int_{\Sigma}\langle g^{-1}\partial g,
  g^{-1}\bartial g\rangle\,dz\wedge d\bar{z}
       -\frac{i}{24\pi}\int_{B_3}\langle g^{-1}dg,[g^{-1}dg,g^{-1}dg]\rangle\ \ .
\end{equation}
  The second term is integrated over an auxiliary three-manifold
  $B_3$ which satisfies $\partial B_3=\Sigma$.
  Note that the measure $idz\wedge d\bar{z}$ is real. The
  topological ambiguity of the second term possibly imposes a
  quantization condition on the metric $\langle\cdot,\cdot\rangle$
  or, more precisely, on its bosonic restriction, in order to
  render the path integral well-defined.\footnote{Note that
  for WZW models based on bosonic groups one usually explicitly
  introduces an integer valued constant, the level, which appears
  as a prefactor of the Killing form. For supergroups the Killing
  form might vanish. Hence there is no canonical normalization
  of the metric. Moreover, we would like to include models whose
  metric renormalizes non-multiplicatively (see below). Under
  these circumstances  it is not particularly convenient to
  display the level explicitly and we assume instead that all
  possible parameters are contained in the metric $\langle\cdot,
  \cdot\rangle$.}

  By construction, every WZW model has a global symmetry $\groupG\times\groupG$
  corresponding to multiplying the field $g(z,\bz)$ by arbitrary group
  elements from the left and from the right. In fact, this symmetry is
  elevated to a current superalgebra symmetry $\hat{\g}$
\begin{align}
  \label{eq:WZWOPE}
  J^\mu(z)\,J^\nu(w)
  &\ =\ \frac{\langle J^\mu,J^\nu\rangle}{(z-w)^2}
        +\frac{i{f^{\mu\nu}}_\lambda\,J^\lambda(w)}{z-w}
        +\text{non-singular}
\end{align}
  and a corresponding anti-holomorphic symmetry if one allows these
  elements to depend holomorphically and antiholomorphically on $z$,
  respectively. In the last formula, the symbols $f$ denote the
  structure constants of $\g$ and the currents are defined by
  $J\propto-\partial gg^{-1}$ and $\bJ\propto g^{-1}\bartial g$. The
  equations of motion guarantee that they are holomorphic and
  antiholomorphic, respectively.

  Just as for bosonic WZW models, the current algebra
  \eqref{eq:WZWOPE} implies the existence of a conformal
  energy-momentum tensor which is obtained by the standard Sugawara
  construction. If $\g$ is simple then the metric is uniquely
  determined up to a scale factor, the level $k$. In that case, the
  central charge of the WZW model is given by
\begin{align}
  c\ =\ \frac{k\sdim(\groupG)}{k+g^\vee}\ \ ,
\end{align}
  where $g^\vee$ is the dual Coxeter number (see
  Table~\ref{tab:Classification}) and
  $\sdim\groupG=\dim(\g_\den)-\dim(\g_\deo)$ is the superdimension of
  $\g$. Since our treatment is supposed to cover simple as well as
  non-simple Lie supergroups, we will refrain from working with the
  level from now on.

  One peculiarity of supergroup WZW models is that the indefinite
  signature of the metric implies that one part of the bosonic
  subgroup should be regarded as non-compact while the other one
  should be regarded as compact. One of the most familiar examples is
  the WZW model based on $\PSU(2|2)$ which features two $\SU(2)$ WZW
  models at positive and negative level, respectively. The most
  natural interpretation here corresponds to a supersymmetrization of
  the $\SL(2,\Real)\times\SU(2)$ WZW model (corresponding to
  $\PSU(1,1|2)$). An exception only occurs for small values of the
  level where quantum renormalization might help to bring a small
  negative level into the positive regime. In this case, supergroup
  WZW models may be used to describe a compact target space. An
  example of this type will be discussed in
  Section~\ref{sc:SupersphereDuality}.

\subsection{\label{sc:FFR}Free fermion realization for type~I
  supergroups}

  In order to unravel the peculiarities of supergroup WZW models we
  first of all concentrate on the special case of Lie supergroups of
  type~I. These are singled out by the fact that they admit a
  distinguished $\Integer$-grading \eqref{eq:ZGrading}. This grading
  permits to split the fermionic generators into two sets of
  generators $S_1^a$ and $S_{2b}$ ($a,b=1,\ldots,r$) which are dual in
  the sense that $\langle S_1^a,S_{2b}\rangle=\delta_b^a$. Moreover we
  assume that the bosonic generators $K^i$ give rise to a metric
  $\langle K^i,K^j\rangle=\kappa^{ij}$ of $\g_\den$. Under these
  circumstances, the complete set of non-trivial relations can be
  chosen to read~\cite{Quella:2007hr}
\begin{align}
  \label{eq:TypIComm}
  [K^i,K^j]&\ =\ i{f^{ij}}_l\,K^l\ ,&
  [K^i,S_1^a]&\ =\ -{(R^i)^a}_b\,S_1^b\ ,\\[2mm]
  [S_1^a,S_{2b}]&\ =\ -{(R^i)^a}_b\,\kappa_{ij}\,K^j\ ,&
  [K^i,S_{2a}]&\ =\ {S_{2b}\,(R^i)^b}_a\ \ .
\end{align}
  We note that the fermions $S_1^a$ and $S_{2a}$ transform in an
  $r$-dimensional representation $R$ of $\g_\den$ and its dual
  $R^\ast$, respectively. The symbol $R^i$ is an abbreviation for the
  corresponding representation matrix $R(K^i)$. The structure
  constants which appear in the commutator $[S_1^a,S_{2b}]$ are
  uniquely determined by the requirement that the metric
  $\langle\cdot,\cdot\rangle$ is invariant.

  Given the Lie superalgebra $\g$, we can combine its generators
  with elements of a (suitably large) Grassmann algebra in order to
  obtain a Lie algebra which can be exponentiated. Following our
  previous strategy for $\GL(1|1)$, we shall define the supergroup
  $\groupG$ to be given by elements
\begin{equation}
  \label{eq:Para}
  g\ =\ e^{\theta}\,g_B\,e^{\bar{\theta}}
\end{equation}
  with $\theta=\theta^aS_{2a}$ and
  $\bar{\theta}=\bar{\theta}_bS_1^b$. This particular way of
  distributing the fermions facilitates the decoupling of bosons and
  fermions and, at the same time, holomorphic factorization. The
  parametrization has been termed ``chiral superspace''
  in \cite{Gates:1983nr}. The coefficients $\theta^a$ and
  $\bar{\theta}_b$ are independent Grassmann
  variables while $g_B$ denotes an element of the bosonic
  subgroup $\groupG_\den\subset\groupG$ obtained by exponentiating the
  Lie algebra generators $K^i$. The attentive reader may have
  noticed that the product of two such supergroup elements
  \eqref{eq:Para} will not again give a supergroup
  element of the same form. We shall close an eye on such
  issues. For us, passing through the supergroup is merely an
  auxiliary step that serves the purpose of constructing a
  WZW-like conformal field theory with Lie superalgebra
  symmetry. Since Lie superalgebras do not suffer from
  problems with Grassmann variables, the resulting conformal
  field theory will be well-defined.

  Given the parametrization \eqref{eq:Para}, the Lagrangian
  \eqref{eq:WZWLagrangian} can be simplified considerably by making
  iterative use of the Polyakov-Wiegmann identity
\begin{equation}
  \label{eq:PW}
  \cS^{\text{WZW}}[gh]
  \ =\ \cS^{\text{WZW}}[g]+\cS^{\text{WZW}}[h]-\frac{i}{2\pi}\int
       \langle g^{-1}\bartial g,\partial hh^{-1}\rangle\,dz\wedge d\bar{z}\ \ .
\end{equation}
  The WZW action evaluated on the individual fermionic bits vanishes
  because the invariant form
  is only supported
  on grade $0$ of the $\Integer$-grading. The final result is then
\begin{equation}
  \label{eq:WZWLagrangianNew}
  \cS^{\text{WZW}}[g] \ = \ \cS^{\text{WZW}}[g_B,\theta]
  \ =\ \cS^{\text{WZW}}[g_B]
       -\frac{i}{2\pi}\int\langle\bartial\theta,g_B\,\partial\bar{\theta}
  \,g_B^{-1}\rangle\,dz\wedge d\bar{z}\ \ .
\end{equation}
  For the correct determination of the mixed bosonic and fermionic
  term it was again necessary to refer to the grading of
  $\g$.

  It is now crucial to realize (see also \cite{Schomerus:2005bf,Gotz:2006qp})
  that we may pass to an equivalent first order description of the WZW
  model above by introducing an additional set of auxiliary fields
  $p_a$ and $\bar p^a$,
\begin{equation}
  \label{eq:FreeFieldInt}
  \begin{split}
    \cS[g_B,p,\theta]
    &\ =\ \cS_{\text{ren}}^{\text{WZW}}[g_B]+\cS_{\text{free}}
           [\theta,\bar{\theta},p,\bar{p}]
          +\cS_{\text{int}}[g_B,p,\bar{p}]\\[2mm]
    &\ =\ \cS_{\text{ren}}^{\text{WZW}}[g_B]
          +\frac{i}{2\pi}\int\Bigl\{\langle p,\bartial\theta\rangle-
   \langle\bar{p},\partial\bar{\theta}\rangle-\langle p,g_B\,\bar{p}\,
   g_B^{-1}\rangle\Bigr\}\,dz\wedge d\bar{z}\ \ .
  \end{split}
\end{equation}
  Here, $\theta, \bar \theta$ and our new fermionic fields
  $p=p_aS_1^a$ and $\bar{p}=\bar{p}^aS_{2a}$ all take values in
  the Lie superalgebra $\g$. Our conventions may look slightly
  asymmetric but as we will see later this just resembles
  the asymmetry in the parametrization \eqref{eq:Para}.
  Up to certain subtleties that are encoded in the subscript
  ``ren'' of the
  first term, it is straightforward to see that we recover
  the original Lagrangian \eqref{eq:WZWLagrangianNew} upon
  integrating out the auxiliary fields $p$ and $\bar{p}$.

  Let us comment a bit more on each term in the action
  \eqref{eq:FreeFieldInt}. Most importantly, we need to specify
  the renormalization of the bosonic WZW model which results
  from the change in the path integral measure
  (cf.~\cite{Polyakov:1984et}). The computation of the relevant
  Jacobian has two important effects. First of all, it turns out
  that the construction of the purely bosonic WZW model entering
  the action \eqref{eq:FreeFieldInt} employs the following
  renormalized metric\footnote{We assume this metric to
  be non-degenerate. Otherwise we would deal with what is known
  as the critical level or, in string terminology, the
  tensionless limit.}
\begin{equation}
  \label{eq:RenMetric}
  \langle K^i,K^j\rangle_{\text{ren}}\ =\ \kappa^{ij}-\gamma^{ij}
  \qquad\text{ with }\qquad
  \gamma^{ij}\ =\ \tr(R^iR^j)\ \ .
\end{equation}
  Note that this renormalization is not necessarily
  multiplicative. Only for simple Lie algebras the renormalized metric
  would be identical to the original one up to a factor, allowing for
  the familiar interpretation as a level shift. For the bosonic
  subalgebra $\g_\den$ of a basic Lie superalgebra $\g$, however, this
  is not the case. First of all, the simple constituents of $\g_\den$
  all come with their own level and each of them may be shifted
  differently. But then there are also examples such as $\gl(m|n)$
  where $\g_\den$ is not reductive and hence the notion of levels is
  not well-defined.

  As a second consequence of the renormalization, the action
  \eqref{eq:FreeFieldInt} may contain a Fradkin-Tseytlin term,
  coupling a non-trivial dilaton to the world-sheet curvature
  $R^{(2)}$,
\begin{equation}
  \label{eq:Dilaton}
  \cS^{\text{WZW}}_{\text{FT}}[g_b] \ = \
  \int_{\Sigma}d^2\sigma\sqrt{h}R^{(2)}\phi(g_B)\ \ \ \ \
  \text{where} \ \ \ \
  \phi(g_B)\ =\ -\frac{1}{2}\,\ln\det R(g_B)\ \ .
\end{equation}
  The same kind of expression has already been encountered in
  the investigation of the $\GL(1|1)$ WZW model, cf.\ %
  \cite{Rozansky:1992rx,Rozansky:1992td,Schomerus:2005bf}.
  It is obvious that $\phi$ vanishes whenever $\g_\den$ is a
  semisimple Lie algebra. A non-trivial dilaton is, however, a
  feature of the series $\osp(2|2n)$, $\sgl(m|n)$ and $\gl(m|n)$ or, in
  other words, of most basic Lie superalgebras of type~I. The precise
  reason for the claimed form of renormalization, i.e.\ the
  modification of the metric and the appearance of the dilaton, will
  become clear in the following section when we discuss the full
  quantum symmetry of the supergroup WZW model. At the moment let us
  just restrict ourselves to the comment that the dilaton is required
  in order to ensure the supergroup invariance of the path integral
  measure for the free fermion realization, i.e.\ the description of
  the WZW model in terms of the Lagrangian \eqref{eq:FreeFieldInt}.

  Before we conclude this subsection, let us quickly return to the
  fermionic terms of the Lagrangian \eqref{eq:FreeFieldInt} which
  may be rewritten in an even more explicit form using
\begin{equation}
  g_B\,\bar{p}\,g_B^{-1}
  \ =\ g_B\,S_{2b}\,\bar{p}^b\,g_B^{-1}
  \ =\ S_{2a}\,{R^a}_b(g_B)\,\bar{p}^b\ \ .
\end{equation}
  The result for the interaction term is
\begin{equation}
  \label{eq:FreeFieldIntTwo}
  \cS_{\text{int}}[g_B,p,\bar p]\ =
   - \frac{i}{2\pi}\int p_a\,{R^a}_b(g_B)\,\bar{p}^b
   \,dz\wedge d\bar{z}\ \ .
\end{equation}
  In an operator formulation, the object ${R^a}_b(g_B)$ should be
  interpreted as a vertex operator of the bosonic WZW model,
  transforming in the representation $R\otimes R^\ast$. We may
  consider the interaction term $p_a\,{R^a}_b(g_B)\,\bar{p}^b$
  as a screening current. The term \eqref{eq:FreeFieldIntTwo} enters
  the perturbative expansion of correlation functions when starting
  from the decoupled theory of bosons and fermions. We would like to
  stress that the series truncates at finite order for any given
  correlator since $p$ and $\bar{p}$ are fermionic fields.

  Note that the screening current entering \eqref{eq:FreeFieldIntTwo}
  is non-chiral by definition, a feature that is not really specific
  to supergroups but applies equally well to free field resolutions of
  bosonic models. Nevertheless, the existing literature on free field
  constructions did not pay much attention to this point. Actually,
  the distinction is not really relevant for purely bosonic WZW models
  because of their simple factorization into left and right movers. In
  the present context, however, a complete non-chiral treatment must
  be enforced in order to capture and understand the special
  properties of supergroup WZW models.

\subsection{Algebraic treatment}

  While our previous considerations took place entirely on a
  Lagrangian level, they also have a purely algebraic counterpart. In
  this section we will show, how the current algebra $\hat{\g}$ can be
  embedded into a theory described by the current algebra $\g_\den$
  and free fermions. This embedding will lead to a natural class of
  representations and, eventually, to a natural modular invariant
  partition function.

\subsubsection{Current algebra}

  In what follows we shall focus on the holomorphic sector of the
  theory. Our first goal is to rephrase the current
  algebra~\eqref{eq:WZWOPE} in terms of the special basis chosen in
  Section~\ref{sc:FFR}. In the bosonic subsector we find
\begin{equation}
  \label{eq:OPEBB}
  K^i(z)\,K^j(w)
  \ =\ \frac{\kappa^{ij}}{(z-w)^2}+\frac{i{f^{ij}}_l\,K^l(w)}{z-w}\ \ .
\end{equation}
  The transformation properties of the fermionic currents are
\begin{align}
  \label{eq:OPEBF}
  K^i(z)\,S_1^a(w)&\ =\ -\frac{{(R^i)^a}_b\,S_1^b(w)}{z-w}&
  &\text{ and }&
  K^i(z)\,S_{2a}(w)&\ =\ \frac{{S_{2b}(w)\,(R^i)^b}_a}{z-w}\ \ .
\end{align}
  Finally we need to specify the operator product of the 
  fermionic currents,
\begin{equation}
  \label{eq:OPEFF}
  S_1^a(z)\,S_{2b}(w)\ =\ \frac{\delta_b^a}{(z-w)^2}-\frac{{(R^i)^a}_b\,
  \kappa_{ij}\,K^j(w)}{z-w}\ \ .
\end{equation}
  The previous operator product expansions (OPEs) are straightforward
  adaptions of the commutation relations \eqref{eq:TypIComm}. The
  central extension is determined by the invariant metric
  $\langle\cdot,\cdot\rangle$.

  The current superalgebra above defines a chiral vertex algebra
  via the Sugawara construction. As usual, the corresponding energy
  momentum tensor is obtained by contracting the currents with the
  inverse of a distinguished invariant and non-degenerate metric. The
  appropriate fully renormalized (hence the subscript ``full-ren'')
  metric is defined by
\begin{equation}
  \label{eq:FulMetric}
  \begin{split}
    \langle K^i,K^j\rangle_{\text{full-ren}}
    &\ =\ (\Omega^{-1})^{ij}\ =\
    \kappa^{ij}-\gamma^{ij}-\frac{1}{2}\,{f^{im}}_n\,{f^{jn}}_m\ \ ,\\[2mm]
    \langle S_1^a,S_{2b}\rangle_{\text{full-ren}}
    &\ =\ {(\Omega^{-1})^a}_b\ =\ \delta_b^a+{(R^i\kappa_{ij}R^j)^a}_b
  \end{split}
\end{equation}
  and it is the result of adding half the Killing form of the Lie
  superalgebra $\g$ to the original classical metric
  $\langle\cdot,\cdot\rangle$.\footnote{Again, this renormalization does not
    need to be multiplicative.} Note that
  some of the terms in the fully renormalized metric
  \eqref{eq:FulMetric} can be identified with the (partially)
  renormalized metric \eqref{eq:RenMetric} which we introduced when
  deriving the free fermion Lagrangian. The energy momentum tensor of
  our theory involves the inverse of the fully renormalized metric,
\begin{equation}
  \label{eq:AffineT}
  T\ =\ \frac{1}{2}\,\bigl[K^i\,\Omega_{ij}\,K^j-S_1^b{\Omega^a}_bS_{2a}+
  S_{2a}{\Omega^a}_bS_1^b\bigr]\ \ .
\end{equation}
  Both, currents and energy momentum tensor, may similarly be defined
  for the antiholomorphic sector. The appearance of a renormalized
  metric in the Sugawara construction is a common feature. Supergroup
  WZW models are certainly not exceptional in this respect.

\subsubsection{\label{sc:AlgFF}Free fermion realization}

  Our next aim is to describe the current superalgebra defined above
  and the associated primary fields in terms of the decoupled system
  of bosons and fermions that appear in the Lagrangian
  \eqref{eq:FreeFieldInt}. As one of our ingredients we shall
  employ the bosonic current algebra
\begin{equation}
  \label{eq:FFBosCurrent}
  K_B^i(z)\,K_B^j(w)
  \ =\ \frac{(\kappa-\gamma)^{ij}}{(z-w)^2}+
  \frac{i{f^{ij}}_l\,K_B^l(w)}{z-w}\ \ ,
\end{equation}
  which is defined using the (partially) renormalized metric
  $\langle\cdot,\cdot\rangle_{\text{ren}}$ which has been introduced
  in eq.\ \eqref{eq:RenMetric}. In addition, we need $r$ free
  fermionic ghost systems, each of central charge $c=-2$. They are
  described by fields $p_a(z)$ and $\theta^a(z)$ with OPE
\begin{equation}
  p_a(z)\,\theta^b(w)\ =\ \frac{\delta_a^b}{z-w}
\end{equation}
  and spins $h=1$ and $h=0$, respectively.
  The fermionic fields are assumed to have trivial operator product
  expansions with the bosonic generators. By construction, the
  currents $K_B^i$ and the fields $p_a$, $\theta^b$ generate the
  chiral symmetry of the field theory whose action is
\begin{equation}
  \label{eq:WZWDecoupled}
  \cS_0[g_B,p,\theta]
  \ =\ \cS_{\text{ren}}^{\text{WZW}}[g_B]+\cS_{\text{free}}
  [\theta,\bar{\theta},p,\bar{p}]\ \ .
\end{equation}
  Our full WZW theory may be considered as a deformation of this
  theory, once we take into account the interaction term between bosons
  and fermions, see eq.\ \eqref{eq:FreeFieldIntTwo}.

  It is easy to see that the decoupled action
  \eqref{eq:WZWDecoupled} defines a conformal field theory with energy
  momentum tensor
\begin{equation}
  \label{eq:FreeT}
  T_0\ =\ \frac{1}{2}\,\Bigl[K_B^i\,\Omega_{ij}\,K_B^j+
  \tr(\Omega R^i)\,\kappa_{ij}\,\partial K_B^j\Bigr]
        -p_a\partial\theta^a\ \ .
\end{equation}
  Note the existence of the dilaton contributions, i.e.\ terms
  linear in derivatives of the currents. In addition to the conformal
  symmetries, the action \eqref{eq:WZWDecoupled} is also invariant
  under a $\hat\g\oplus\hat\g$ current superalgebra. The corresponding
  holomorphic currents are defined by the relations (normal ordering
  is implied)
\begin{equation}
  \label{eq:FFRes}
  \begin{split}
    K^i(z)&\ =\ K_B^i(z)+p_a\,{(R^i)^a}_b\,\theta^b(z)\\[2mm]
    S_1^a(z)&\ =\ \partial\theta^a(z)+{(R^i)^a}_b\,\kappa_{ij}\,
    \theta^bK_B^j(z)-\frac{1}{2}{(R^i)^a}_c\,\kappa_{ij}\,
    {(R^j)^b}_d\,p_b\theta^c\theta^d(z)\\[2mm]
    S_{2a}(z)&\ =\ -p_a(z)\ \ .
  \end{split}
\end{equation}
  It is a straightforward exercise, even though slightly
  cumbersome and lengthy, to check that this set of
  generators reproduces the operator product expansions
  \eqref{eq:OPEBB}, \eqref{eq:OPEBF} and \eqref{eq:OPEFF}.
  Obviously, a similar set
  of currents may be obtained for the antiholomorphic sector.
  Given the representation \eqref{eq:FFRes} for the current
  superalgebra one may also check the equivalence of the
  expressions \eqref{eq:AffineT} and \eqref{eq:FreeT} for the
  energy momentum tensors.

\subsubsection{\label{sc:WZWReps}Chiral representation theory}

  The representation theory of the bosonic current algebra
  $\hat{\g}_\den$ and of the fermions $p\theta$ is well
  understood (see, e.g., \cite{FrancescoCFT,Kausch:2000fu} and
  references therein). It will thus not come as a surprise that the
  free fermion realization outlined in the previous section provides a
  natural route to the construction of representations of
  $\hat{\g}$. In a sense, the conformal embedding
\begin{align}
  \bigl(\hat{\g},\langle\cdot,\cdot\rangle\bigr)
  \ \hookrightarrow\
  \cU\bigl(\bigl(\hat{\g}_\den,\langle\cdot,\cdot\rangle_{\text{ren}}\bigr)
  \,\oplus\,\{p_a,\theta^a\}\ \bigr)
\end{align}
  constructed in eq.~\eqref{eq:FFRes} is to the current superalgebra
  $\hat{\g}$ what the distinguished $\Integer$-grading
  \eqref{eq:ZGrading} was to the underlying horizontal subalgebra $\g$
  -- with the slight complication that the metric needs to be
  renormalized.\footnote{In more abstract terms, the renormalization
    is likely to correspond to the shift with the Weyl vector.}

  In order to substantiate our claim, let us now define the current
  superalgebra analog of Kac modules. The starting point is a simple
  module $\hat{L}_\lambda$ of
  $\bigl(\hat{\g}_\den,\langle\cdot,\cdot\rangle_{\text{ren}}\bigr)$
  and the vacuum module $\hat{\Lambda}$ of the free ghosts
  $\{p_a,\theta^a\}$. The module $\hat{\Lambda}$ is defined by the
  highest weight conditions
\begin{align}
  p_n^a|0\rangle
  \ =\ \theta_{n+1}^a|0\rangle
  \ =\ 0\qquad\text{ for }n\geq0
\end{align}
  and it is the unique irreducible module, assuming integer moding of
  the fermions.\footnote{Other modings are obtained by spectral
    flow, see Section~\ref{sc:WZWSF}.} The current superalgebra
  $\bigl(\hat{\g},\langle\cdot,\cdot\rangle\bigr)$ is then acting in
  the tensor product
\begin{align}
  \label{eq:AffineKac}
  \hat{\cK}_\lambda
  \ =\ \hat{L}_\lambda\otimes\hat{\Lambda}\ \ .
\end{align}
  It can be shown \cite{Quella:2007hr} that this action is typically
  irreducible (i.e.\ $\hat{\cK}_\lambda=\hat{\cL}_\lambda$) and that
  the conformal weight of the associated field is described by the
  (renormalized) Casimir eigenvalues
\begin{align}
  h_\lambda
  \ =\ \frac{1}{2}\,C_\lambda^{\text{full-ren}}\ \ .
\end{align}
  The corresponding Casimir operator is given by
  $C^{\text{full-ren}}=K^i\Omega_{ij}K^j+\tr(\Omega
  R^i)\kappa_{ij}K^j$ and agrees with the zero-mode contribution to
  eq.~\eqref{eq:FreeT}. In cases where $\hat{\cK}_\lambda$ contains
  singular vectors (necessarily fermionic), it will be called
  atypical. In such a case, one can define the irreducible
  representation $\hat{\cL}_\lambda$ as the quotient of
  $\hat{\cK}_\lambda$ by its maximal ideal $\hat{\cN}_\lambda$. It is
  the power of our free fermion realization that it automatically
  takes care of all bosonic null vectors while ignoring the fermionic
  ones (if present).

  Let us dwell a bit on the structure of $\hat{\cK}_\lambda$. First of
  all, it is obvious that the space of ground states of
  $\hat{\cK}_\lambda$ coincides with $\cK_\lambda$ since
  $\hat{L}_\lambda$ is built on top of the $\g_\den$-module
  $L_\lambda$ and $\hat{\Lambda}$ on top of $\bigwedge(\g_{-1})$,
  compare eq.~\eqref{eq:KacModuleContent}. In particular,
  $\hat{\cK}_\lambda$ will certainly be atypical (i.e.\ contain
  fermionic singular vectors) if $\cK_\lambda$ is atypical. On the
  other hand, $\hat{\cK}_\lambda$ can also be atypical if
  $\cK_\lambda=\cL_\lambda$ is typical. In that case, the module
  contains fermionic singular vectors which are located at higher
  energy levels. In all examples we are aware of, the corresponding
  modules $\hat{\cK}_\lambda$ arise as the spectral flow image of a
  different atypical module $\hat{\cK}_\mu$ that has its singular
  vectors on the ground state level. We believe that this is true in
  general.

  We finally note that the character of the Kac modules
  $\hat{\cK}_\lambda$ follows immediately from the known characters of
  the constituents in eq.~\eqref{eq:AffineKac} and modular
  transformations are easily worked out. The story becomes more subtle
  if one is interested in characters of atypical irreducible modules
  $\hat{\cL}_\lambda$. As far as we know, a general theory has not
  been established in that case despite of the existence of certain
  results and conjectures in the mathematical literature
  \cite{Kac:1994kn,Kac:2001,Serganova:2011MR2743764,Gorelik:2011MR2796063,Gorelik:2012MR2968633,Serganova:2012MR2866847}.
  It is nevertheless worth pointing out that there appear to be
  interesting connections to Mock modular forms
  \cite{Zwegers:2008,Folsom:MR2719689,Bringmann:MR2534107}.
  For the low-rank examples of $\gl(1|1)$, $\sgl(2|1)$ and $\psl(2|2)$
  rather explicit expressions for characters and (partially) modular
  transformations have been derived in the physics literature
  \cite{Rozansky:1992td,Semikhatov:2003uc,Gotz:2006qp,Saleur:2006tf,Creutzig:2007jy}.
  Fortunately, knowledge about the characters of atypical
  representations is {\em not} required to prove modular invariance
  for the partition function of (type~I) supergroup WZW models. This
  will be shown in Section~\ref{sc:PartitionFunction}.

\subsubsection{\label{sc:WZWSF}Spectral flow automorphisms}

  In the previous subsection we have silently skipped over one rather
  important element in the representation theory of current
  (super)algebras: The spectral flow automorphisms. As we shall recall
  momentarily, spectral flow automorphisms describe symmetry
  transformations in the representation theory of current
  algebras. Furthermore, they seem to be realized as exact symmetries
  of the WZW models on supergroups, a property that makes them highly
  relevant for the discussion of partition functions.

  Throughout the following discussion, we shall denote (spectral flow)
  automorphisms of the current superalgebra $\hat\g$ by $\omega$. We
  shall mostly assume that the action of $\omega$ is consistent with
  the boundary conditions for currents, i.e.\ that it preserves the
  integer moding of the currents. In the context of representation
  theory, any such spectral flow automorphism $\omega$ defines a map
  on the set of (isomorphism classes of) representations
  $\rho:\hat\g\to\End(\hat{\cM})$ via concatenation,
  $\omega(\rho)=\rho\circ\omega:\hat\g\to\End(\hat{\cM})$. Since the
  automorphism shuffles the modes of the currents, it usually changes
  the energy level of states. It is also well known that
  spectral flow may yield representations whose energy spectrum is not
  bounded from below. Such representations are relevant for the study
  of WZW models on non-compact (super)groups such as $\SL(2,\Real)$ or
  $\PSU(1,1|2)$ \cite{Maldacena:2000hw,Gotz:2006qp}.

  In line with our general strategy, we would like to establish that
  spectral flow automorphisms $\omega$ of the current superalgebra are
  uniquely determined by their action on the bosonic generators.
  A spectral flow automorphism $\omega:\hat\g_\den\to\hat\g_\den$ of the bosonic
  subalgebra $\hat\g_\den$ is, by definition, a linear map
\begin{equation}
  \label{cbos}
  \omega\bigl(K^i(z)\bigr)
  \ = \ {(W_\den)^i}_j(z)\, K^j(z) + w_\den^i\, z^{-1}
\end{equation}
  which is consistent with the OPE \eqref{eq:OPEBB}. The map
  $W_\den(z)=z^{\zeta_\den}$ is defined in terms of an
  endomorphism $\zeta_\den:\g_\den\to\g_\den$ of the horizontal subalgebra.
  While the eigenvalues of $\zeta_\den$ determine how the spectral
  flow shifts the modes of the currents, the vector $w_\den^i$ affects
  only the zero-modes. In order to
  preserve the trivial monodromy under rotations around the origin
  we will assume that $W_\den(z)$ is a meromorphic function, i.e.\ that
  all the eigenvalues of $\zeta_\den$ are integer. Inserting the
  transformation \eqref{cbos} into the operator product expansions
  \eqref{eq:OPEBB} leaves one with the relation
\begin{equation}
  \label{detzeta0}
  {(\zeta_\den)^i}_j
  \ =\ {f^{ik}}_l\,\kappa_{kj}\,w_\den^l
\end{equation}
  while all other constraints are satisfied automatically. Hence the
  only free parameter is the shift vector $w_\den^i$. In the case of a
  semisimple Lie algebra $\g_\den$ (which leads to a non-degenerate
  Killing form) this argument can also be reversed and hence it allows
  to express $w_\den^i$ in terms of $\zeta_\den$.
  Any vector $w_\den^i$ which leads to a matrix $\zeta_\den$
  with integer eigenvalues under the identification \eqref{detzeta0}
  will accordingly be referred to as a
  spectral flow automorphism of $\hat{\g}_\den$ from now on.

  Given the insights of the previous paragraphs it is now fairly
  straightforward to extend the spectral flow automorphism
  $\omega:\hat\g_\den\to\hat\g_\den$ to the full current superalgebra. To this
  end, we introduce the element
\begin{equation}
  \label{defzeta1}
  \zeta_\deo\ =\ - R^i\, \kappa_{ij} \, w^j_\den \ \ .
\end{equation}
  Following the discussion in the bosonic sector, we now introduce a
  function $W_\deo(z)=z^{\zeta_\deo}$. This allows us to define the action
  of the spectral flow automorphism $\omega$ on the fermionic currents
  by
\begin{equation}
\omega\bigl(S_1^a(z)\bigr) \ = \ {(W_\deo)^a}_b (z)\  S^b_1(z) \ \ \ , \ \
\ \ \omega\bigl(S_{2a}(z)\bigr) \ = \ S_{1b}(z) \,  {({\overline W_\deo})^b}_a(z)\ \ ,
\end{equation}
  where $\overline W_\deo$ denotes the inverse of $W_\deo$.

  We would also like to argue that the spectral flow symmetry is
  consistent with the free fermion representation \eqref{eq:FFRes}.
  To be more specific, we shall construct an automorphism on the
  chiral algebra of the decoupled system generated by the currents
  $K_B^i(z)$ and the free ghosts $p_a(z)$ and $\theta^a(z)$
  that reduces to the expressions above if we plug the transformed
  fields into the defining equations \eqref{eq:FFRes}. In this
  context the most important issue is to understand how the
  renormalization of the metric $\kappa \rightarrow \kappa - \gamma$
  affects the action of the spectral flow. As it turns out, the data
  $\zeta_\den$ which gave rise to a spectral flow automorphism
  of $\hat\g_\den$ above, can also be used to define a spectral flow
  automorphism of the renormalized current algebra, i.e.\ of the
  algebra that is generated by $K^j_B$ with operator products given
  in eq.\ \eqref{eq:FFBosCurrent}. Only the shift vector $w_\den^i$ of
  the zero modes needs a small adjustment such that the new spectral
  flow action reads
\begin{equation}
  \label{cBBos}
  \omega\bigl(K^i_B(z)\bigr)
  \ = \ {(W_\den)^i}_j(z)\, K^j_B(z) + w_B^i\, z^{-1} \ \ \
  \text{where}  \ \ \ w_B^i \ = \ w^i_\den + \tr\bigl(\zeta_\deo R^i\bigr)\ \ .
\end{equation}
  Note that $\zeta_\den$ is not changed and hence it has the same
  (integer) eigenvalues as before.
  In order to obtain an automorphism which is compatible
  with the free field construction we finally need to
  introduce the transformations
\begin{align}
  \omega\bigl(p_a(z)\bigr) \ =\ p_b(z) \, {(\overline W_\deo)^b}_a(z)
  \ \ \ \ , \ \ \ \
  \omega\bigl(\theta^a(z)\bigr) \ = \ {(W_\deo)^a}_b(z)  \, \theta^b(z)\ \ .
\end{align}
  It is then straightforward but lengthy to check that the
  previous transformations define an automorphism of the algebra
  generated by $p_a$, $\theta^a$ and $K^j_B$ that descends to the
  original spectral flow automorphism $\omega$ of our current
  superalgebra $\hat\g$. During the calculation one has to be aware
  of normal ordering issues.
\smallskip

  In conclusion we have shown that any spectral flow automorphism
  of the bosonic subalgebra of a current superalgebra (related to
  a Lie superalgebra of type~I) can be extended to the full current
  superalgebra. Furthermore, this extension was seen to be consistent
  with our free fermion realization. Let us remark that even if we
  start with a spectral flow automorphism $\omega$ preserving periodic
  boundary conditions for bosonic currents, the lifted spectral flow
  $\omega$ does not necessarily have the same property on fermionic
  generators. Only those spectral flow automorphisms $\omega:\ag\to
  \ag$ for which $W_\deo$ is meromorphic as well seem to arise as
  symmetries of WZW models on supergroups. Nevertheless, also
  non-meromorphic spectral flows turn out to be of physical
  relevance. They can be used to describe the twisted sectors of
  orbifold theories.

\subsubsection{\label{sc:PartitionFunction}The partition function:
  Decoupled versus interacting theory}

  We now return to the important question of characterizing the bulk
  state space of the WZW  model on the supergroup $\groupG$. With
  regard to the decoupled action $\cS_0$ from
  eq.~\eqref{eq:WZWDecoupled} it is clear that the partition function
  factorizes into the partition function of the (renormalized) bosonic
  WZW model on $\groupG_\den$ (including spectral flow sectors if
  necessary) and the free ghosts
\begin{align}
  \label{eq:PartitionFunction}
  Z\bigl(G,\langle\cdot,\cdot\rangle\bigr)
  \ =\ Z\bigl(G_\den,\langle\cdot,\cdot\rangle_{\text{ren}}\bigr)
       \cdot Z(p_a,\theta^a)\ \ ,
\end{align}
  see \cite{Quella:2007hr} for a more detailed discussion. In this
  form, modular invariance is manifest. What is much less obvious: The
  partition function remains the same if the interaction  term
  $\cS_{\text{int}}$ in eq.\ \eqref{eq:FreeFieldIntTwo} is switched on
  which converts the decoupled theory into the supergroup WZW model.

  In order to understand the logic of the previous statement we have
  to resort to our results on the harmonic analysis on supergroups in
  Section~\ref{sc:HarmonicAnalysis}. Following the standard reasoning
  \cite{Gepner:1986wi}, the space of ground states of a WZW model can
  be described by the algebra of functions $\cF(\groupG)$ in the limit
  of large volume. The action of the current algebra on these
  states reduces to the biregular action of $\groupG$ on
  $\cF(\groupG)$ and the energy is approximated by the eigenvalue of
  the Laplacian which, in turn, corresponds to half of the
  (unrenormalized) quadratic Casimir operator. It may be shown that
  that the Lie derivatives for the left regular action are given by
  \cite{Quella:2007hr}
\begin{equation}
  \label{eq:DiffL}
  \begin{split}
    K^i&\ =\ K_B^i-{(R^i)^a}_b\,\theta^b\,\partial_a
    \hspace{3cm}S_{2a}\ =\ -\partial_a\\[2mm]
    S_1^a&\ =\ {R^a}_b(g_B)\,\bartial^b+{(R^i)^a}_b\,\theta^b\,
   \kappa_{ij}\,K_B^j-\frac{1}{2}\,{(R^i)^a}_c\,\kappa_{ij}\,
   {(R^j)^b}_d\,\theta^c\theta^d\partial_b\ \ ,
  \end{split}
\end{equation}
  As we see, this does not quite correspond to the zero mode action of
  the currents \eqref{eq:FFRes} due to the extra contribution in
  $S_1^a$ which involves
  $\bartial^b=\partial/\partial\bar{\theta}_a$. In view of this issue
  it is tempting to simply drop the troublesome term. Even though that
  might seem a rather arbitrary modification at first, it turns out
  that the corresponding truncated differential operators $\bK^i =
  K^i$, $\bS_{2a}=S_{2a}$ and
\begin{equation}
  \bS_1^a \ = \ {(R^i)^a}_b\,\theta^b\,\kappa_{ij}\,K_B^j-
  \frac{1}{2}\, {(R^i)^a}_c\,\kappa_{ij}\,{(R^j)^b}_d\,\theta^c
  \theta^d\partial_b
\end{equation}
  also satisfy the commutation relations of $\g$. These
  considerations extend to the full biregular action of $\g\oplus\g$.

  We thus conclude that $\cF(\groupG)$ admits two different actions of
  $\g\oplus\g$. The action of the generators $(K^i,S_1^a,S_{2a})$ and
  their barred analogues leads to the
  decomposition~\eqref{eq:SuperPeterWeyl} involving the non-chiral
  representations $\cI_{[\sigma]}$. On the other hand, the action of
  the generators $(\bK^i,\bS_1^a,\bS_{2a})$ leads to a different
  decomposition of $\cF(\groupG)$ in terms of sums over products
  $\cK_\lambda\otimes\cK_\lambda^\ast$. We shall denote the
  corresponding module by $\bF(\groupG)$. The modules $\cF(\groupG)$
  and $\bF(\groupG)$ both have the same character since they are
  isomorphic as $\g_\den\oplus\g_\den$-modules
  \cite{Quella:2007hr}. On the other hand, the
  Laplace operator can be diagonalized on $\bF(\groupG)$ while it is not
  diagonalizable on $\cF(\groupG)$. From a physical perspective,
  $\bF(\groupG)$ corresponds to the space of ground states of the
  decoupled theory while $\cF(\groupG)$ is associated to the full WZW
  model. Our previous comments imply the equality of the respective
  partition functions and the occurrence of logarithmic correlation
  functions in the supergroup WZW model. A more detailed discussion of
  these issues can be found in~\cite{Quella:2007hr}. For completeness
  we should mention that WZW models for non-simply connected
  supergroups can be described by means of an orbifolding procedure
  which is implemented in terms of simple currents (e.g.\ spectral
  flow automorphisms).

  It is instructive to contrast the previous statements with the logic
  of ordinary free field
  resolutions~\cite{Feigin:1990qn,Rasmussen:1998cc}. In that case, the
  role of $\bigl(\hat{\g}_\den,\langle\cdot,\cdot\rangle\bigr)$ is
  played by free scalar fields (associated with the Cartan subalgebra
  $\h$) and the root generators give rise to either free bosonic or
  fermionic ghost systems, depending on the nature of the roots. In
  that case, the total partition function is {\em not} the product of
  the individual partition functions. Rather, one needs to pass over
  to a cohomology which is defined in terms of screening
  currents. This removes the singular vectors from the state
  space. In contrast, the fermionic singular vectors arising from the
  free fermion realization of supergroup WZW models need {\em not} be
  removed and the partition functions of the interacting and the
  non-interacting theory agree. Nevertheless there are still
  differences on the level of the state space. While the
  non-interacting theory is free of logarithms and exhibits a trivial
  holomorphic factorization, the interacting theory is logarithmic and
  has a non-chiral state space.

\subsection{\label{sc:OSP}Free field realization the
  \texorpdfstring{$\OSP(m|2n)$}{OSP(m|2n)} WZW model at level \texorpdfstring{$k=1$}{k=1}}

  The distinguished $\Integer$-grading \eqref{eq:ZGrading} of type~II
  Lie superalgebras permits to mimic the Gauss decomposition of
  Section~\ref{sc:FFR}. However, in practice this turns out
  not to be of particular use since it artificially divides the simple
  $\g_\den$-module $\g_\deo$ into two parts $\g_{\pm1}$ and since the
  corresponding fermions cease to be nilpotent. Instead, following
  \cite{Mitev:2008yt}, we wish to concentrate here on a true free
  field construction for the $\OSP(m|2n)$ WZW model which, however,
  has the drawback that it only works for level $k=1$. On the other
  hand, this model features in some of the most interesting
  applications with orthosymplectic supersymmetry
  \cite{Candu:2008vw,Candu:2008yw,Mitev:2008yt,Candu:2008PhD}.

\subsubsection{\label{sc:OSPCurrents}Current algebra}

  The basic fields which enter the construction are a set of $m$ free
  fermions $\psi$ and $n$ bosonic $\beta\gamma$ systems with OPE
\begin{align}
  \label{eq:FFT}
  \psi^i(z)\,\psi^j(w)
  \ =\ \frac{\delta^{ij}}{z-w}\ ,\quad
  \beta_a(z)\,\gamma^b(w)
  \ =\ \frac{\delta_a^b}{z-w}\ \ .
\end{align}
  These fields all have conformal dimension $h=1/2$ which is
  guaranteed by choosing the energy momentum tensor
\begin{align}
  \label{eq:OSPFreeT}
  T(z)
  \ =\ -\frac{1}{2}\,\psi_i\partial\psi^i(z)
       +\frac{1}{2}\Bigl[\beta_a\partial\gamma^a(z)
       -\gamma^a\partial\beta_a(z)\Bigr]\ \ .
\end{align}
  The central charge of the system is easily seen to take the value
  $c=\tfrac{m}{2}-n$. In the next step, we will verify that the
  bilinears in the free fields realize the current superalgebra
  $\widehat{\osp}(m|2n)$ at level $k=1$ (with positive level for the
  orthogonal part and negative level for the symplectic part). In
  Section~\ref{sc:OSPPartitionFunction} we will furthermore establish
  that the corresponding WZW model is equivalent to a suitable
  orbifold of the free field theory.

  In order to construct the currents of $\widehat{\osp}(m|2n)$ we
  choose a matrix $X\in\osp(m|2n)$ and decompose it into blocks
  according to
\begin{align}
  \label{eq:OspDef}
  X\ =\ \left(\begin{array}{c|cc}\cE & \bar{\cT} &
\cT \\\hline -\cT^t & \cF & \cG\\
\bar{\cT}^t & \bar{\cG} &
-\cF^t\end{array}\right)
\end{align}
  where $\cE$ is antisymmetric and $\cG,\bar{\cG}$ are symmetric. A
  basis for the individual constituents of $X$ is provided by the
  vectors
\begin{align}
  \label{eq:OSPCurrents}
  \cE_{ij}
  &\ =\ e_{ij}-e_{ji}\quad(i<j)\ ,&
  \cG_{ab}
  &\ =\ \bar{\cG}_{ab}
   \ =\ e_{ab}+e_{ba}\\[2mm]
  \cF_{ab}
  &\ =\ e_{ab}&
  \cT_{ia}
  &\ =\ \bar{\cT}_{ia}
   \ =\ e_{ia}\ \ ,
\end{align}
  where the matrices $e_{\alpha\beta}$ have a single unit entry in row
  $\alpha$ and column $\beta$. We agree to denote by $E_{ij}$ the
  supermatrix of the form \eqref{eq:OspDef} where $\cE$ is given by
  $\cE_{ij}$ and all other blocks vanish. The basis elements
  $F_{ab},G_{ab},\bar{G}_{ab},T_{ia},\bar{T}_{ia}$ are defined
  similarly.

  In the next step we define a metric on $\osp(m|2n)$ using the
  assignment $\langle X,Y\rangle=\frac{1}{2}\str(XY)$. Evaluating it
  on the basis elements, the metric is fully specified by
\begin{align}
  \label{eq:OSPMetric}
  \langle E_{ij},E_{kl}\rangle
  &\ =\ -\delta_{ik}\delta_{jl}
  \quad(\text{for }i<j\text{ and }k<l)\ ,&
  \langle F_{ab},F_{cd}\rangle
  &\ =\ -\delta_{ad}\delta_{bc}\nonumber\\[2mm]
  \langle G_{ab},\bar{G}_{cd}\rangle
  &\ =\ -\delta_{ac}\delta_{bd}
  \quad(\text{for }a\neq b\text{ and }c\neq d)\ ,&
  \langle G_{aa}, \bar{G}_{bb}\rangle
  &\ =\ -2\delta_{ab}\\[2mm]
  \langle T_{ia},\bar{T}_{jb}\rangle
  &\ =\ \delta_{ij}\delta_{ab}\ \ .\nonumber
\end{align}

  The free field representation of $\widehat{\osp}(m|2n)$ is defined
  in terms of the assignments
\begin{equation}
  \label{eq:OSPFF}
  \begin{split}
  E_{ij}(z)&\ =\ (\psi_i\psi_j)(z)\ ,\qquad
  F_{ab}(z)\ =\ -(\beta_a\gamma_b)(z)\\[2mm]
  G_{ab}(z)&\ =\ (\beta_a\beta_b)(z)\ ,\qquad \
  \bar{G}_{ab}(z)\ =\ -(\gamma_a\gamma_b)(z)\\[2mm]
  T_{ia}(z)&\ =\ i(\psi_i\beta_a)(z)\ ,\qquad
  \bar{T}_{ia}(z)\ =\ -i(\psi_i\gamma_a)(z)\ \ .
\end{split}
\end{equation}
  The OPEs between these currents present the proper affinization of
  the commutation relations of the corresponding matrices defined in
  \eqref{eq:OSPCurrents}, with a central extension specified by the
  metric~\eqref{eq:OSPMetric} (this is equivalent to having level
  $k=1$). Assuming $m\neq 2n+1$, the energy momentum tensor can then
  be defined in the standard Sugawara form
\begin{align}
  T(z)
  &\ =\ \frac{(J^{\mu}J_{\mu})(z)}{2(k+g^{\vee})}
   \ =\ \frac{1}{2(k+g^{\vee})}
\Big[-\sum_{i<j=1}^m(E_{ij}^2)-\sum_{a,b=1}^n
(F_{ab}F_{ba})-\sum_{a<b=1}^n\big(\left\{G_{ab},\bar{G}_{ab}\right\}\big)
\nonumber\\[2mm]
  &\qquad\qquad\qquad\qquad-\frac{1}{2}\sum_{a=1}^n\big(\left\{G_{aa},\bar{G}_{aa}
\right\}\big)-\sum_{i=1}^m\sum_{a=1}^n\big(\left[T_{ia},\bar{T}_{ia}\right]
\big)\Big]\ \ .
\end{align}
  Here, the dual Coxeter number is given by
  $g^{\vee}=m-2n-2$.\footnote{Since we are dealing with the fixed
    form~\eqref{eq:OSPMetric} there is a deviation from
    Table~\ref{tab:Classification} for $m\lesssim2n$. The
    discrepancy is due to different normalization conventions.} It may
  be checked that $T(z)$ reduces to the energy momentum tensor
  \eqref{eq:OSPFreeT} after using the replacements~\eqref{eq:OSPFF}.

\subsubsection{\label{sc:OSPPartitionFunction}Comments on the state
  space}

  The $\OSP(m|2n)$ WZW model can be realized as an
  orbifold of the free field theory generated by the fields $\psi^i$
  and $\beta_a,\gamma^b$. The orbifold guarantees that all free
  fields obey the same boundary conditions, i.e.\ they are either all
  periodic or all anti-periodic. This in turn ensures that the
  currents defined in \eqref{eq:OSPFF} -- which are bilinears in the
  free fields -- are all periodic. However, before we discuss the
  implementation of the orbifold, we first focus on the chiral
  representation theory of a single fermion or $\beta\gamma$-system,
  respectively.

  For a single Majorana fermion $\psi$ we need two distinct
  representations, one for periodic boundary conditions (half-integer
  moding, $\nu=0$) and one for anti-periodic boundary conditions
  (integer moding, $\nu=1/2$). The associated ground states
  $|\nu\rangle$ are defined by the highest weight conditions
\begin{align}
  \label{eq:PsiModule}
  \psi_n|\nu\rangle\ =\ 0
  \quad\text{ for }\quad n>0\ \ .
\end{align}
  It is well-known (see, e.g., \cite{FrancescoCFT}) that the two
  representations just constructed are not irreducible over the
  Virasoro algebra. The corresponding characters rather decompose as
\begin{align}
  \chi_\psi^{(0)}(q)
  \ =\ \chi_0(q)+\chi_{1/2}(q)
  \qquad\text{ and }\qquad
  \chi_\psi^{(1/2)}(q)
  \ =\ 2\chi_\sigma(q)
\end{align}
  into the characters of the Ising model. The field $1/2$ is the
  unique simple current of the latter, with fusion rules
  $1/2\otimes1/2=0$ and $1/2\otimes\sigma=\sigma$.

  The situation is slightly different for the bosonic
  $\beta\gamma$-system \cite{Lesage:2002ch}. As before, we shall
  consider sectors which differ by the choice of boundary conditions
  and we introduce a family of ground states $|\nu\rangle$ for
  $\nu\in\tfrac12\Integer$. These states are characterized by the
  conditions
\begin{equation}
 \beta_{r+\nu} |\nu\rangle\,  =\,  0 \ \ \ , \ \ \ \gamma_{r-\nu}
 |\nu\rangle \, = \, 0 \ \ \ \text{ for } \ \ \ r \, = \,
1/2,3/2,5/2, \dots
\end{equation}
From the ground states we generate the corresponding
sectors by application of raising operators. If we assign charges
$q_\beta = 1/2$ and $q_\gamma = -1/2$ to the modes of the fields
$\beta$ and $\gamma$, respectively, and $q_\nu=\nu/2$ to the
ground state $|\nu\rangle $ the generating function for the sector
$\nu$ reads,
\begin{equation} \label{chinu}
\chi^{(\nu)} (q,y) \ = \ q^{\frac{1}{24}-\frac{\nu^2}{2}}\,y^{\frac{\nu}{2}}
\prod_{n=0}^\infty \frac{1}{(1-y^{\frac12}q^{n+\frac12 -
\nu})(1-y^{-\frac12}q^{n+\frac12+\nu})} \ = \ \frac{q^{-\nu^2/2}\,y^{\frac{\nu}{2}}\,
\eta(q)}{\theta_4(q,y^{1/2}q^{-\nu})}
\end{equation}
All the constructed sectors carry an action of an affine
$\widehat{\sgl}(2)$ current algebra at level $k=-1/2$
\cite{Lesage:2002ch}. In terms of the fields $\beta$ and $\gamma$ the
three currents are constructed as follows,
\begin{equation}\label{BC1}
  E^+(z)
  \ =\ \frac12 \beta^2(z)
  \ ,\quad H(z)
  \ =\ -\frac{1}{2}\,(\beta\gamma)(z)
  \ ,\quad E^-(z)
  \ =\ -\frac12\,\gamma^2(z)\ \ .
\end{equation}
Consequently, we can decompose the generating functions
\eqref{chinu} into characters of irreducible representations of
$\widehat{\text{sl}}(2)_{-1/2}$. In case of $\chi^{(0)}$, for example,
the decomposition is given by
\begin{equation}
  \label{eq:sldeco}
  \chi^{(0)}(q,y)\, =\, \frac{\eta(q)}{\theta_4(q,y^{1/2})}\, = \,
  \chi^{k=-1/2}_0(q,y) + \chi^{k=-1/2}_{1/2}(q,y) \ .
\end{equation}
The two characters on the right hand side belong to irreducible
highest weight representations with lowest weight $h = \epsilon\in
\{0,1/2\}$,
\begin{equation}
\chi^{k=-1/2}_{\epsilon}(q,y)\ =\ \frac{\eta(q)}{2}
\left[\frac{1}{\theta_4(q,y^{1/2})}+ (-1)^{2\epsilon}
\frac{1}{\theta_3(q,y^{1/2})}\right]\ \ .
\end{equation}
  Let us note that the ground states transform in representations of
  spin $j = \epsilon$. Nevertheless, we shall continue to think of the
  subscript of $\chi$ as the conformal weight rather than the
  spin. Decomposition formulas similar to \eqref{eq:sldeco} exist for
  all the other functions \eqref{chinu}. All of them are related by
  the action of spectral flow automorphisms \cite{Lesage:2002ch}. In
  particular, we have
\begin{equation}
 \chi^{(1/2)} \ = \ \chi^{k=-1/2}_{\sigma;+} +
\chi^{k=-1/2}_{\sigma;-} \ \ \ \text{with}  \ \
\chi_{\sigma;\pm}(q,y)\ =\ \frac{y^{1/4}\eta(q)}{2}
\left[\frac{1}{i\theta_1(q,y^{-1/2})}\pm
\frac{1}{\theta_2(q,y^{-1/2})}\right] \ \ .
\end{equation}
  The two characters on the left hand side belong to the two
  irreducible lowest weight representations of the current algebra
  with spin $j = 1/4$ and $j=3/4$. Their ground states have the same
  conformal weight $h=-1/8$. In close analogy to the Ising model, the
  representation with $h=1/2$ (and $j=1/2$) is a simple current of the
  $\widehat{\sgl}(2)_{-1/2}$ theory
  \cite{Lesage:2002ch,Ridout2011jk}. Its action exchanges the two
  constituents of $\chi^{(\nu)}$.

  Restricting our attention to the bosonic subrepresentations, i.e.\
  the representations of the Ising model and of
  $\widehat{\sgl}(2)_{-1/2}$, the product of $m$ real fermions and $n$
  $\beta\gamma$-systems contains a group $\Integer_2^{m+n}$ of simple
  currents that consists of all elements $\eta$ of the form
\begin{align}
  \eta\, = \, [\epsilon_1, \dots,\epsilon_n; \epsilon_{n+1},
  \dots, \epsilon_{m+n}] \ \ \ \text{with} \ \ \epsilon_i \in
  \{0,1/2\} \ \ .
\end{align}
  The first $n$ entries of $\eta$ denote subsectors of the
  $\beta\gamma$-system while the remaining ones are representing
  sectors in the Ising models. Requiring the simple currents to have
  integer conformal dimension leads to the constraint
  $\sum_{i=1}^{m+n}\epsilon_i\,=\,0 \ \text{mod\/}\, 1$. The elements
  $\eta$ solving this equation generate the abelian group
  $\Gamma\cong\Integer_2^{m+n-1}$.

  We are now ready to discuss the bulk modular invariant for the full
  WZW model. The construction of a simple current orbifold proceeds in
  several steps. To begin with, we have to list all sectors $a$ of the
  theory which possess vanishing monodromy charge $Q_a(\eta) = h_a +
  h_\eta - h_{\eta \times a}\text{ mod }1$. These are then organized
  into orbits ${\cal O}_a$ under the action of the simple current
  group $\Gamma$. According to the standard construction (see e.g.\
  \cite{Schellekens:1990xy}), each such orbit ${\cal O}_a$ contributes
  one term $Z_a$ to the partition function of the orbifold model, with
  a coefficient $|\Gamma|/|{\cal O}_a|$ that is given by the ratio
  between the order $|\Gamma|$ of the orbifold group and the length
  $|{\cal O}_a|$ of the orbit. Unfortunately, the characters of the
  $\widehat{\sgl}(2)_{-1/2}$ theory exhibit rather peculiar properties
  such as restricted domains of convergence and relations under
  spectral flow
  \cite{Lesage:2003kn,Ridout:2008nh,Ridout:2010qx,Creutzig:2012sd}.
  For this reason, the following result on the bulk state space is
  based on a rather formal application of the orbifold
  construction. The full validity of our procedure still remains to be
  established.

  Let us first deal with the sector involving representations with
  vanishing spectral flow, $\nu=0$. Under the action of $\Gamma$, the
  sectors with vanishing monodromy charge split into two orbits of
  maximal length. Hence we are led to the following contribution to
  the state space,
\begin{equation}
  \cH^{\text{FF}}_{0}\ =\ 
\Bigl| {\bigoplus}_{\eta\in \Gamma} \cH_{\eta \times
[0,\dots,0;0, \dots,0]} \Bigr|^2  + \Bigl|{\bigoplus}_{\eta\in
\Gamma}
 \cH_{\eta \times [0,\dots,0;0,\dots,0,1/2]}\Bigr|^2\ \ .
\end{equation}
  In this expression we used $|\cH|^2=\cH\otimes\cH^\ast$ as an
  abbreviation in order to mimic the expected expression for the
  partition function. However, the total theory has to be invariant
  under the spectral flow symmetry. Hence we have to add twisted
  contributions $\cH^{\text{FF}}_{\nu}$. It was already mentioned
  above that all the bosonic ghosts and all the fermions have to have
  identical periodicity conditions in order to not to spoil the
  $\osp(m|2n)$\ symmetry. Consequently the spectral flow must act
  diagonally, i.e.\ simultaneously on all sectors, by half-integer
  shifts.\footnote{It is worth mentioning that these diagonal spectral
    flow transformations are also the only ones which commute with the
    action of the orbifold group. Note also that half-integer spectral
    flow on ghosts and fermions implies integer spectral flow on the
    currents such as those defined in eq.\ \eqref{BC1} and below.} In
  the fermionic factors, spectral flow by $\nu=1/2$ brings us from the
  $\psi$-module $(0)$ to $(1/2)$. Both subrepresentations $0$ and
  $1/2$ of $(0)$ are thus mapped to the representation $\sigma$ which,
  in turn, are fixed under a subgroup $\cS\subset\Gamma$ of the group
  of simple currents. Integer units of the spectral flow, however, do
  not give anything new since the moding is preserved. In the ghost
  sectors things works differently because the application of a
  diagonal spectral flow leads to an infinite number of new
  representations constructed from the ground states $|\nu\rangle$ for
  $\nu\in\frac{1}{2}\Integer$. Since the orbits of the half-integer
  spectral flow representations possess a stabilizer subgroup $\cS$ of
  order $|\cS| = 2^{m-1}$ with respect to the action of $\Gamma$ we
  finally end up with the state space
\begin{equation}
  \begin{split}
    \cH^{\text{WZW}}
     \ =\ \bigoplus_{\nu\in\frac{1}{2}\Integer}\cH^{\text{FF}}_{\nu}
    &\ =\ \bigoplus_{\nu\in\Integer}\Biggl[
          \biggl| {\bigoplus}_{\eta\in \Gamma}\cH^{(\nu)}_{\eta\times[0,\dots,0;0, \dots,0]}
           \biggr|^2
          + \biggl|{\bigoplus}_{\eta\in\Gamma}\cH^{(\nu)}_{\eta \times
          [0,\dots,0;0,\dots,0,1/2]}\biggr|^2 \Biggr]\\[2mm]
    &\qquad\qquad
    \oplus2^{m-1}\bigoplus_{\nu\in\Integer+\frac{1}{2}}\Bigl|\underbrace{\cH^{(\nu)}\otimes\cdots\otimes\cH^{(\nu)}}_{n\text{ factors }\beta\gamma}\otimes\underbrace{\cH_{\sigma}\otimes\cdots\otimes\cH_\sigma}_{m\text{ factors Ising}}
 \Bigr|^2\ \ . \label{PFFFn}
  \end{split}
\end{equation}
  Our proposal is admittedly somewhat sketchy. In view of the
  various subtleties related to the $\beta\gamma$-system, the
  fractional level WZW model $\widehat{\sgl}(2)_{-1/2}$ and their
  orbifolds
  \cite{Lesage:2003kn,Lesage:2003kn,Ridout:2008nh,Ridout:2010qx,Ridout2011jk,Creutzig:2012sd},
  it may well be worth revisiting it with more care. Since we will
  mainly be interested in the boundary theory of the $\OSP$ WZW models
  in this review, we postpone such an analysis to future work.

  We note that the example just discussed is one of the rare cases
  where a supergroup WZW model leads to a theory which may be regarded
  as compact to some extent since the (negative) level of the affine
  $\widehat{\spl}(2n)$ subalgebra is so close to zero that it is
  pushed into the positive regime by effects of quantum
  renormalization. Moreover, even though the model as discussed above
  is presumably not logarithmic due to the restriction to a small
  level, the theory admits a logarithmic lift by the inclusion of
  additional zero modes. This is in close analogy to the case of
  $\widehat{\su}(2)_{-1/2}$ \cite{Lesage:2003kn,Ridout:2010qx}.

\subsection{\label{sc:Deformations}Deformations}

  Supergroup WZW models may, under certain circumstances, allow for
  exactly marginal deformations that would usually spoil conformal
  invariance. These are deformations which either preserve the
  original global symmetry $\groupG\times\groupG$ or some (twisted)
  diagonal symmetry $\groupG$. We shall briefly describe the
  implementation of these deformations in terms of perturbing fields
  and discuss necessary conditions for their marginality.

\subsubsection{\texorpdfstring{$\groupG\times\groupG$}{GxG} preserving deformations}

  WZW models possess an obvious deformation which amounts to allowing
  for a different normalization of the two contributions to the action
  \eqref{eq:WZWLagrangian}. Since the coefficient of the topological term is
  quantized (at least for supergroups having a compact simple part)
  the only freedom is in fact to change the coefficient of the kinetic
  term. In most cases, especially for all simple bosonic groups, such
  a deformation would spoil conformal invariance. The only exception
  occurs if $\groupG$ is a supergroup with vanishing Killing form as
  we will now review.

  The implementation of the $\groupG\times\groupG$-preserving
  deformation in the operator language is somewhat cumbersome. We have
  to find a field $\Phi(z,\bz)$ that is invariant under
  $\groupG\times\groupG$ and possesses conformal dimension
  $(h,\bh)=(1,1)$. The deformed model would then formally be described
  by the deformed action
  $\cS_{\text{def}}=\cS^{\text{WZW}}+\xi/2\pi\int\!d^2z\,\Phi(z,\bz)$. The
  two currents $J$ and $\bJ$ transform non-trivially under either the
  left or the right action of $\groupG$ on itself. For this reason, an
  invariant can only be built by conjugating one of the two
  currents. In algebraic terms we have to consider the normal ordered
  operator
\begin{align}
  \label{eq:DefoOne}
  \Phi_{\groupG\times\groupG}(z,\bz)
  \ =\ :\bigl\langle J(z),\Ad_g\bigl(\bJ(\bz)\bigr)\bigr\rangle:
  \ =\ :J^\mu\phi_{\mu\nu}\bJ^\nu:(z,\bz)
\end{align}
  involving the non-chiral affine primary field $\phi_{\mu\nu}(z,\bz)$
  which transforms in the representation $\ad\otimes\ad$ with respect
  to the symmetry $\groupG\times\groupG$. At lowest order in
  perturbation theory, the field $\Phi_{\groupG\times\groupG}(z,\bz)$ is marginal if and
  only if the field $\phi_{\mu\nu}$ has conformal dimensions
  $(h,\bh)=(0,0)$. Since the conformal dimensions of the affine
  primary field $\phi_{\mu\nu}$ are proportional to the quadratic
  Casimir in the adjoint representation, this condition is equivalent
  to the vanishing of the Killing form of $\groupG$.

  It may be shown that the addition of the perturbing field
  \eqref{eq:DefoOne} to a WZW model leads to a non-chiral deformation
  of the original current algebra. The structure and physical
  significance of such current algebras has been discussed in a
  sequence of papers
  \cite{Ashok:2009xx,Ashok:2009jw,Benichou:2010rk}.

\subsubsection{\label{sc:WZWDef2}\texorpdfstring{$\groupG$}{G} preserving deformations}

  Supergroups with vanishing Killing form admit a second type of
  deformation. This deformation is easier to describe since it avoids
  the use of the non-chiral field $\phi_{\mu\nu}(z,\bz)$. On the other
  hand it is more difficult to interpret geometrically since the group
  structure is broken. In this case the deformation is implemented by
  the perturbing field
\begin{align}
  \label{eq:DefoTwo}
  \Phi_{\groupG}(z,\bz)
  \ =\ \bigl\langle J(z),\bJ(\bz)\bigr\rangle
  \ =\ J_\mu\bJ^\mu(z,\bz)\ \ .
\end{align}
  This field transforms non-trivially under the full
  $\groupG\times\groupG$-symmetry but trivially under the diagonal
  subgroup $\groupG$. As a consequence, the original global
  $\groupG\times\groupG$-symmetry is broken to the diagonal
  $\groupG$-symmetry as soon as the deformation is switched on. In the
  present case it is less obvious than for $\Phi_{\groupG\times\groupG}(z,\bz)$ but again
  we need to impose the vanishing of the Killing form since otherwise
  the field $\Phi_{\groupG}(z,\bz)$ would break conformal invariance at higher
  orders in perturbation theory. Similarly to the previous case,
  perturbations of a supergroup WZW model by the field
  \eqref{eq:DefoTwo} give rise to a non-chiral current algebra which
  was the subject of \cite{Konechny:2010nq}.

  In the case of the $\OSP(m|2n)$ model at level $k=1$ that was 
  discussed in Section~\ref{sc:OSP}, a slight modification of 
  $\Phi_{\groupG}$ will play an important role below. In order to define
  it, we introduce the involutive automorphism $\Omega$ such that
  the fixed point set $\{X\in\osp(m|2n)| \Omega(X)=X\}$ is
  isomorphic to $\osp(m{-}1|2n)$. On the basis we introduced in
  \eqref{eq:OSPCurrents}, $\Omega$ acts non-trivially only on $E_{ij},
  T_{ia},\bar{T}_{ia}$. In fact, it multiplies all operators with $i=1$
  by $-1$ and leaves the others invariant. If we denote the
  anti-holomorphic fields corresponding to $\psi_i,
  \beta_a,\gamma_a$ by $\bar{\psi}_i,
  \bar{\beta}_a,\bar{\gamma}_a$, the deformation operator
  $\Phi'_{\groupG} = J^{\mu}\Omega(\bar{J}_{\mu})$ can then be written 
  explicitly as
\begin{align}
  \label{deformation operator}
  J^{\mu}\Omega(\bar{J}_{\mu})
  &\ =\ \frac{1}{2}\left[\sum_{i=1}^m\varpi_{i}\psi_i
\bar{\psi}_i+\sum_{a=1}^n\big(\gamma_a\bar{\beta}_a
-\beta_a\bar{\gamma}_a\big)\right]^2
\end{align}
  where $\varpi=(-1, 1,\ldots , 1)$. In order for the expression on the
  right hand side line to make sense, we need to first expand
  the square and normal order the products. Without the contributions
  from the bosonic ghosts, the interaction introduced by $\Phi'_{\groupG}$ 
  is familiar from Gross-Neveu models, i.e.\ the deformed $\OSP$ WZW 
  model can be regarded as a Gross-Neveu model. As we have argued, 
  this Gross-Neveu model is conformally invariant if $m=2n+2$, i.e.\ 
  if the Killing form of $\osp(m|2n)$ vanishes. Our construction thus 
  generalizes the massless Thirring model (appearing, e.g., in Luttinger 
  liquid theory) to a non-abelian setting.

\subsubsection{\label{sc:Marginality}Exact marginality for
  quasi-abelian supergroups}

\begin{figure}
\begin{center}
\begin{tikzpicture}[line width=1.2pt]
  \draw (.2,.6) node {a)};
  \draw (4.2,.6) node {b)};
\begin{scope}[scale=.6]
  \draw[color=blue] (0,0) -- (1,0) to[bend left,out=80,in=100] (3,0)
  -- (4,0);
  \draw[color=blue] (1,0) to[bend right,out=-80,in=-100] (3,0);
  \draw[fill] (1,0) circle (2.3pt) node[right] {$f$};
  \draw[fill] (3,0) circle (2.3pt) node[left] {$f$};
\end{scope}
\begin{scope}[xshift=4cm,scale=.6]
  \draw[color=blue] (1.5,0) -- (4.5,0);
  \draw[color=blue] (1.5,0) .. controls (1.5,1) and (2.5,1) .. (3,1);
  \draw[color=blue] (1.5,0) ..controls (1.5,-1) and (2.5,-1) .. (3,-1);
  \draw[color=blue] (4.5,0) .. controls (4.5,1) and (3.5,1) .. (3,1);
  \draw[color=blue] (4.5,0) .. controls (4.5,-1) and (3.5,-1) .. (3,-1);
  \draw[fill] (1.5,0) circle (3pt);
  \draw[fill] (4.5,0) circle (3pt);
  \draw (1.5,0) node[left] {$f$};
  \draw (4.5,0) node[right] {$f$};
\end{scope}
\begin{scope}[xshift=8cm,scale=.6]
  \draw[color=blue] (3,0) -- (4.5,0);
  \draw[color=blue] (4.5,0) .. controls (4.5,1) and (3.5,1) .. (2.63,1);
  \draw[color=blue] (4.5,0) .. controls (4.5,-1) and (3.5,-1) .. (2.63,-1);
  \draw[fill] (4.5,0) circle (3pt);
  \draw (4.5,0) node[right] {$f$};
  \color{black}
  \draw (1.5,0) circle (1.5cm);
\begin{scope}[color=blue,line width=.6pt]
  \draw[densely dotted] (2.63,-1) .. controls (2,-1) and (2,-1) .. (1,-.8);
  \draw[densely dotted] (2.63,1) .. controls (2.2,1) and (2.2,1) .. (1.8,.8);
  \draw[densely dotted] (2.2,0) -- (1.8,.8);
  \draw[densely dotted] (2.2,0) -- (3,0);
  \draw[densely dotted] (0.7,.6) -- (1.8,.8);
  \draw[densely dotted] (0.7,.6) -- (1.5,-.2);
  \draw[densely dotted] (0.7,.6) -- (1,-.8);
  \draw[densely dotted] (1.5,-.2) -- (1,-.8);
  \draw[densely dotted] (1.5,-.2) -- (2.2,0);
  \draw[fill] (1.5,-.2) circle (2pt);
  \draw[fill] (.7,.6) circle (2pt);
  \draw[fill] (1,-.8) circle (2pt);
  \draw[fill] (1.8,.8) circle (2pt);
  \draw[fill] (2.2,0) circle (2pt);
\end{scope}
 \draw (3,.3) node[right] {$\sim f$};
\end{scope}
\end{tikzpicture}
  \caption{\label{fig:KillingForm}a) The Killing form.
    b) Vanishing contributions to the $\beta$-function.}
\end{center}
\end{figure}
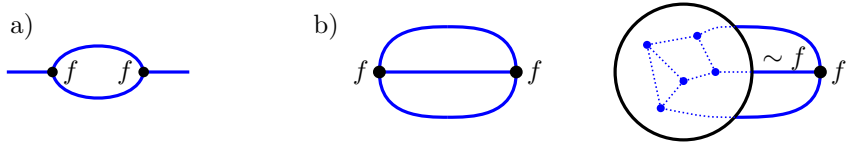

  The previous two sections showed that a vanishing Killing form of
  $\groupG$ is a necessary prerequisite for the existence of marginal
  deformation preserving either the symmetry $\groupG$ or even the
  $\groupG\times\groupG$. We will refer to such supergroups $\groupG$
  as being quasi-abelian. This name is motivated by the equation
\begin{align}
  \label{eq:VanishingKilling}
  0\ =\ K(T^\mu,T^\nu)
   \ =\ -(-1)^{d_\lambda}\,{f^{\mu\rho}}_\lambda{f^{\nu\lambda}}_\rho
\end{align}
  for the Killing form. We see that -- even though the structure
  constants themselves are not vanishing -- every contraction over two
  of its indices leads to a trivial result. This is also depicted in
  Figure~\ref{fig:KillingForm}. One can easily convince oneself that
  the Killing form coincides with the Ricci curvature of $\groupG$ up
  to a factor. Hence quasi-abelian supergroups are Ricci flat (and
  vice versa).

  Assuming the existence of a unique invariant rank three tensor (the
  structure constants), the $\beta$-function equations for
  $\groupG$-preserving deformations of supergroup WZW models on
  $\groupG$ or for the corresponding principal chiral models follow
  the combinatorics for free bosons since all structure constants can
  be dropped \cite{Bershadsky:1999hk}. This can be understood as
  follows. Since the $\groupG$-symmetry is preserved, the
  $\beta$-function is a scalar quantity without free Lie algebra
  indices. On the other hand, the perturbation series for the
  $\beta$-function will be based on the OPE \eqref{eq:WZWOPE}. The
  only ingredients are the metric (corresponding to a line) and the
  structure constants (corresponding to a 3-vertex). Let us pick out
  an arbitrary diagram which includes a structure constant
  ${f^{\mu\nu}}_\lambda$. It needs to be connected to the rest of the
  diagram by three legs. According to our assumption, the remaining
  diagram thus also needs to be proportional to the structure
  constants and hence the concatenation vanishes due to
  eq.~\eqref{eq:VanishingKilling}. As a result we are left with the
  purely abelian contributions which are known to vanish. This
  reasoning is illustrated in Figure~\ref{fig:KillingForm}.

  An alternative and less restrictive argument for the conformal
  invariance of deformed supergroup WZW models was provided in
  \cite{Quella:2007sg}. It solely rests on the general non-chiral
  structure of the bulk state space and on the vanishing of the dual
  Coxeter number. The key idea is that the perturbing field
  $\Phi(z,\bz)=\Phi_{\groupG\times\groupG}(z,\bz)$ transforms as a
  singlet $\cL_0$ under
  the right action of the supergroup $\groupG$. From the general
  analysis in Section~\ref{sc:WZWReps} we know that such a singlet can
  only arise as the socle of an atypical projective cover $\cP_0$. For
  this reason there exists a SUSY partner $Q(z,\bz)$ and a fermionic
  symmetry operation $R$ such that $\Phi(z,\bz)=RQ(z,\bz)$. This
  statement may be used to show that all $N$-point functions of the
  perturbing field $\Phi(z,\bz)$ vanish identically. Indeed, in the
  correlation function
\begin{align}
  \bigl\langle\Phi(z,\bz)\Phi(w_1,\bar{w}_1)\cdots\Phi(w_n,\bar{w}_n)\bigr\rangle
\end{align}
  one can simply replace the first instance of $\Phi(z,\bz)$ by
  $RQ(z,\bz)$ and then use the Ward identities for the
  $\groupG$-symmetry to redistribute $R$ to the other fields where it
  acts trivially. We conclude that there is no need to renormalize the
  field $\Phi(z,\bz)$ and this proves its exact marginality to all
  orders in perturbation theory. Another approach uses the
  representation of $\groupG$ as a symmetric space
  $\groupG\times\groupG/\groupG$ and cohomological reduction, see also
  Section~\ref{sc:SSConf}.

\subsubsection{\label{sc:AnDim}Anomalous dimensions}

  The perturbations discussed in the previous sections break the
  affine symmetry of the WZW model. For this reason, the rigid
  structure of affine modules is broken up, leading to a drastic
  deformation of the spectrum. The latter is most conveniently
  expressed in terms of
  anomalous dimensions, i.e.\ the difference of the conformal weight
  compared to the unperturbed theory (in this case the WZW model, not
  a free field theory) as a function of the coupling $\xi$. While
  generally valid formulas for the anomalous dimensions are currently
  still lacking, quasi-abelian perturbation theory has enabled
  progress in certain special situations which shall now briefly be
  reviewed.

  First of all, perturbations of the form \eqref{eq:DefoOne} which
  preserve the global $\groupG\times\groupG$-symmetry of the WZW model
  are very hard to deal with due to the involvement of the field
  $\phi_{\mu\nu}(z,\bz)$. A simplification only occurs if one is
  considering the boundary spectrum of a D0-brane. In that case, the
  field $\phi_{\mu\nu}(z,\bz)$ effectively becomes localized and may
  be replaced by a constant \cite{Quella:2007sg}. Under these
  circumstances, the perturbation series for the boundary two-point
  functions may be summed up
  completely to all orders for a certain subset of fields, and it
  leads to a ``Casimir evolution'' where the anomalous dimensions are
  proportional to the Casimir $C_\lambda$ of the representation under
  consideration. The Casimir is multiplied by a geometric series in
  the coupling constant $\xi$ which resembles the result for the
  radius deformation of a free field theory. Surprisingly, the
  anomalous dimension only depends on the transformation properties of
  the field under the global symmetry $\groupG$ but not on its
  original location in the affine representation of the WZW model.

  The situation is more favorable in case of $\groupG$-preserving
  deformations. In that case, formulas for bulk and boundary anomalous
  dimensions have been proposed in
  \cite{Mitev:2008yt,Obuse:2008nc,Candu:2012xc}, based on similar
  ideas. In both cases, one recovers a Casimir evolution which, for
  the boundary case, leads to the formula
\begin{align}
  \label{eq:AnomalousDimension2}
  \delta h_\lambda(\xi)
  \ =\ -\frac{k\xi}{1+k\xi}\,\frac{C_\lambda}{k}\ \ .
\end{align}
  Here, $C_\lambda$ refers to the (generalized) eigenvalue of the
  quadratic Casimir acting on the representation $\lambda$. The 
  recent analysis in \cite{Candu:2012xc} proves formula 
  \eqref{eq:AnomalousDimension2} for a specific subset of fields. 
  Even though no general proof for generic fields exists, formula
  \eqref{eq:AnomalousDimension2} has been backed up in several 
  models by numerical checks in lattice
  discretizations~\cite{Candu:2008yw,Candu:2009ep}. Assuming its
  general validity leads to remarkable statements about spectral
  equivalences between deformed WZW models and supersphere
  $\sigma$-models which are likely too non-trivial to be just
  accidental. A more detailed discussion of these conjectural
  dualities can be found in Section~\ref{sc:SupersphereDuality}.
  We finally wish to mention that the combination of Casimir evolution
  with cohomological reduction has been used in \cite{Candu:2009ep} in
  order to propose a formula for boundary spectra of space-filling
  branes with monopole charge in $\CPone$.

\section{\label{sc:String}Superspace sigma models and strings}

  Up to now our discussion of logarithmic conformal field theories
  with internal supersymmetry was focused on WZW models and their
  deformations. On the other hand, applications e.g. to string theory
  in Anti-de Sitter backgrounds, often involve superspace
  $\sigma$-models. These will be the focus of the final chapter. After
  reviewing the definition of $\sigma$-models we shall discuss
  existing results on conformal invariance of these model. We continue
  with some comments on embeddings into string theory using either the
  hybrid or pure spinor approach. In the third subsection we study the
  $\sigma$-model on the supersphere $\Sphere^{3|2}$ and establish an
  intriguing link with an $\OSP(4|2)$ WZW model at level $k=1$.

\subsection{Sigma models on coset superspaces}

\subsubsection{Basic formulation}

We want to consider non-linear $\sigma$-models on homogeneous
superspaces $\groupG/\groupGp$, where the quotient is defined as the
set of right cosets of $\groupGp$ in $\groupG$ through the
identification
\begin{align} g\ \sim \ gh \ \ \text{ for all } \ \  h\ \in
\ \groupGp\ \subset\  \groupG\ \ .
\end{align}
Let $\g$ be the Lie superalgebra associated to $\groupG$. We
assume that $\g$ comes equipped with a non-degenerate
invariant bilinear form $\langle\cdot,\cdot\rangle$. Examples
include $\g = \gl(m|n)$, $\sgl(m|n)$, $\psl(n|n)$  or
$\osp(m|2n)$.\footnote{We
exclude $\sgl(n|n)$ and $\pgl(n|n)$, since they do not have a
non-degenerate metric.} Similarly,
let $\g'$ be the Lie superalgebra associated to
$\groupGp$. We assume that the restriction of
$\langle\cdot,\cdot\rangle$ to $\g'$ is non-degenerate.
In this case, the orthogonal complement $\mathfrak{m}$ of
$\g'$ in $\g$ is a $\g'$-module and
one can write the following $\g'$-module decomposition
$\g=\g'\oplus \mathfrak{m}$. In
particular, this means that there are projectors $P'$ onto
$\g'$ and $P=1-P'$ onto
$\mathfrak{m}$ which commute with the action of $\g'$.

With the above requirements, the quotient $\groupG/\groupGp$ can be endowed
with a $\groupG$-invariant metric~$\mathsf{g}$. This metric is by no
means unique and generally depends on a number of continuous
parameters. The square root of the superdeterminant of $\mathsf{g}$
provides in the standard way a $\groupG$-invariant measure $\mu$ on
$\groupG/\groupGp$.
With these two structures one can already
write down a purely  kinetic Lagrangian for the $\sigma$-model on
$\groupG/\groupGp$ and quantize it in the path integral formalism. Inclusion
of $\theta$-terms,  WZW terms or $B$-fields requires a better
understanding of the geometry of the $\groupG/\groupGp$ superspace. In fact,
the $\theta$ and WZW terms are associated to $\groupG$-invariant closed
but not exact 2- and 3-forms, respectively. $B$-fields, on the
other hand, are written in terms of $\groupG$-invariant exact 2-forms.
Every such linearly independent form comes with its own coupling
constant. We shall only consider Lagrangians with a kinetic term
and a $B$-field. Let $\mathsf{b}$ be some general $\groupG$-invariant
exact 2-form. Then the most general Lagrangian we consider can be
written in the form
\begin{equation}\label{eq:lag_origin}
 \mathcal{L} = \eta^{ab}\mathsf{g}_{\mu\nu}(\phi)
 \partial_a \phi^\mu\partial_b\phi^\nu+ \epsilon^{ab}
 \mathsf{b}_{\mu\nu}(\phi) \partial_a \phi^\mu\partial_b
 \phi^\nu\ ,
\end{equation}
where $\eta^{ab}$ is the constant world sheet metric,
$\epsilon^{ab}$ the antisymmetric tensor with $\epsilon^{01}=1$. The
Lagrangian is obviously evaluated on maps $\phi$ from the worldsheet
$\Sigma$ to the superspace $\groupG/\groupGp$.

There is a different way to formulate the $\sigma$-model on $\groupG/\groupGp$,
which makes its coset nature manifest and allows to explicitly
construct the metric $\mathsf{g}$ and the $B$-field $\mathsf{b}$
in eq.~\eqref{eq:lag_origin}. For that purpose, instead of maps $\phi$
from the worldsheet to the target space $\groupG/\groupGp$, we consider more
general maps $g:\Sigma\rightarrow\groupG$ from the world sheet to the
Lie supergroup $\groupG$. A basis set of  1-forms on $\groupG$ which are
invariant under the global left $\groupG$-action is provided by the
so-called Maurer-Cartan forms
\begin{align} \label{eq:currdef} J(x)\ =\
g^{-1}(x) d g(x) \ =\
\bigl[ \omega^i_\mu(\phi) H_i + E^m_\mu X_m\bigr] d\phi^\mu
\ . \end{align}
Here $H_i, i = 1, \dots, \dim \g',$ denotes a basis in
the Lie superalgebra $\g'$ and $X_m$ is one for the
complement of $\g'$ in $\g$. The coefficient
functions $\omega^i_\mu$ and $E^m_\mu$ are the spin connection
and the vielbein, respectively. In terms of these currents, we
can write
\begin{align} \label{generalLagrangian} \cL\ =\
\eta^{ab}\tG\bigl\langle P(J_{a}),P(J_{b})\bigr\rangle+\epsilon^{
ab} \tB\bigl\langle P(J_{a}),P(J_{b})\bigr\rangle\ ,
\end{align}
where $P: \g \rightarrow \mathfrak{m} = \g/\g'$
is the projection and
\begin{equation}\label{eq:GBdef}
\tG\ \in\
\text{Hom}_{\g'}\left(\mathfrak{m}\circ\mathfrak{m},
\mathbb{C}\right) \ \ \ \text{ and } \ \ \  \tB\ \in\
\text{Hom}_{\g'}\left(\mathfrak{m}\wedge\mathfrak{m},
\mathbb{C}\right)
\end{equation}
are taken from the symmetric, respectively antisymmetric tensor
product of $\mathfrak{m}$ with itself to the trivial representation.
The relation with eq.\ \eqref{eq:lag_origin} is provided by
\begin{align}
\mathsf{g}_{\mu\nu}(\phi) = \tG_{mn} E^m_\mu(\phi) E^n_\nu (\phi)
\quad , \quad
\mathsf{b}_{\mu\nu}(\phi) = \tB_{mn} E^m_\mu(\phi) E^n_\nu (\phi)\ .
\end{align}
Any model of the form \eqref{generalLagrangian} may be considered
as a consistent $\sigma$-model on the coset space $\groupG/\groupGp$. Under right
$\groupGp$-gauge transformations $g':\Sigma\mapsto\groupGp$
the Maurer-Cartan forms $J_\mu$ transform as
\begin{align} \label{rightaction} g(x)\ \mapsto\  g(x)g'(x)\ ,\qquad
J_{\mu}(x)\ \mapsto\
g'(x)^{-1}J_{\mu}(x)g'(x)+g'(x)^{-1}\partial_{\mu
} g'(x)\ . \end{align}
Since the projection $P$ on $\mathfrak{m}$ commutes with the
action of $\g'$, the projected forms $P(J_{\mu})$
transform by conjugation with $g'$. Hence, the Lagrangian
\eqref{generalLagrangian} is independent of how we choose
representatives in the coset space $\groupG/\groupGp$. Global left
$\groupG$-invariance
of the Lagrangian~\eqref{generalLagrangian} is automatic since
Maurer-Cartan forms $J_{\mu}(x)$ are left $\groupG$-invariant by
construction.

\subsubsection{\label{sc:SSConf}Symmetric superspaces and conformal invariance}

As we discussed in the previous paragraph, $\sigma$-models on coset spaces
may be defined in terms of the basic objects \eqref{eq:GBdef}. There are
many examples. If we choose $\groupG = \SO(4)$ and $\groupGp=\SO(3)$, the space
$\mathfrak{m}$ is 3-dimensional and it transforms in the adjoint
representation of $\SO(3)$. In this case, there exists a single invariant
in the symmetric tensor product of $\mathfrak{m}$ with itself and no
invariant in the anti-symmetric case. Hence, there is a one-parameter
family of metrics $\tG$ and no $\tB$ field. The corresponding family
of $\sigma$-models has a target space $\Sphere^3 = \SO(4)/\SO(3)$ with radius
$R$. Because of the non-vanishing curvature, the model is not
conformal, unless we add a WZW term.

A very similar analysis applies to the pair $\groupG = \OSP(4|2)$ and
$\groupGp=\OSP(3|2)$, only that in this case the $\sigma$-model on the
quotient $\Sphere^{3|2} = \OSP(4|2)/\OSP(3|2)$ turns out to be conformal.
In the case of symmetric superspaces a complete list of conformal
invariant $\sigma$-models is actually known. For simplicity, let us assume
that $\groupG$ is simple. Then the $\sigma$-model on the symmetric superspace
$\groupG/\groupGp$, where $\groupGp = \groupG^{\Integer_2}$ is a subgroup 
$\groupGp \subset \groupG$ of elements that are left invariant by the 
action of an automorphism of order two on $\groupG$, and $\tB =0$ is
conformal if and only if
\begin{align}
& C^{(2)}_{\groupG} (\g) = 0 \label{eq:CI1loop} \\[2mm]
& C^{(2)}_{\groupGp_\sigma} (\mathfrak{m}) = 0  \label{eq:CIallloop}
\end{align}
Here $C^{(2)}_\groupG(L)$ denotes the value of the quadratic Casimir 
element of $\groupG$ on the module $L$ and we have decomposed 
$\groupGp$ into simple factors $\groupGp = \prod_\sigma\groupGp_{\sigma}$. 
The first condition on the dual Coxeter number $g^\vee = C^{(2)}_{\groupG}(\g)/2$ 
of the numerator supergroup $\groupG$ guarantees that the perturbative beta
function vanishes at one loop. In fact, the Ricci tensor of a coset
space $\groupG/\groupGp$ is given by
\begin{align}
R_{mn}(\groupG/\groupGp) = 2 R_{mn}(\groupG) + \frac14 (-1)^{|k|} {f^k}_{ml}
    {f^l}_{nk} \ .
\end{align}
We shall use in addition that
\begin{align}
R_{mn}(\groupG) = - \frac14 K^{\groupG}_{mn} = - \frac14 (-1)^{|D|}
 {f^{D}}_{mC} {f^C}_{nD}\ .
\end{align}
Here, capital letters $C,D$ run over a basis in $\g$
while lower case letters $m,n$ are restricted to basis elements
in $\mathfrak{m} \subset \g$. For symmetric superspaces,
the structure constants of $\groupG$ must be consistent with the
$\Integer_2$ automorphism and hence ${f^k}_{ml}=0$. Consequently,
the Ricci tensor for symmetric spaces is determined by the Killing
form of $\groupG$. If the latter vanishes, so does the beta function at
one loop. The condition \eqref{eq:CIallloop} arises from the two
loop beta function. It is possible to show that all models that
pass this condition possess a vanishing beta functions to all
loop order, see \cite{Candu:2008PhD}.

Let us note that the principal chiral model on a group or
supergroup $\groupU$ could also be formulated as a symmetric space
$\sigma$-model. Without any further thought one might be tempted
to describe this model through $\groupG = \groupU$ and $\groupGp = \{ e\}$.
But as our introductory comments suggest, we prefer to rewrite the
group manifold $\groupU$ as a coset superspace $\groupU = \groupU
\times \groupU/\groupU$ and hence to set
\begin{align}
 \groupG\ =\ \bigl\{\left(x,y\right)\bigl|x,y \in\groupU\bigr\}\qquad ,
\qquad\groupGp\ =\ \bigl\{\left(x,x\right)\bigl|x\in\groupU\bigr\}\ .
\end{align}
The left and right action of $\groupG$ on itself is given by
componentwise multiplication. The right coset superspace
$\groupG/\groupGp\cong\groupU$ is considered as the space of equivalence
classes under the equivalence relation $\left(x,y\right)\sim
\left(xz,yz\right)$, for all $z\in\groupU$. In particular,
$\left(xy^{-1},1\right)$ is the canonical representative of the
equivalence class of $\left(x,y\right)$. Hence, the currents
$J_{\mu}$ and the projection map $P:\g\rightarrow \mathfrak{m}$
are given by
\begin{align} J_{\mu}\ =\
\left(x^{-1}\partial_{\mu}x,y^{-1}\partial_{\mu}y\right)\ ,
\qquad  P:\left(v,w\right)\ \mapsto\
\left(\frac{v-w}{2},-\frac{v-w}{2}\right)\ . \end{align}
If $\langle\cdot,\cdot\rangle$ is the invariant form on the Lie
superalgebra of $\groupU$ and we take $\tG$ to be given by
\begin{align} \tG\bigl((v_1,w_1)\otimes_s
(v_2,w_2)\bigr)\ = \ \left\langle v_1,v_2\right\rangle+
\left\langle w_1,w_2\right\rangle\
\end{align}
we obtain the usual principal chiral model for $\groupU$. In fact, one
may easily show that
$$\tG\bigl(P(J_{\mu}),P(J_{\nu})\bigr)\eta^{\mu\nu}
\ =\ \frac{1}{2}\left(u^{-1}\partial_{\mu} u, u^{-1}\partial_{\nu}
u\right)\eta^{\mu\nu}\ \ , $$where $u=xy^{-1}\in\groupU$. Thereby we
have established the standard geometric result that allows us to
treat the principal chiral model on $\groupU$ as a $\groupG/\groupGp$
coset superspace model.

\subsubsection{Remarks on
  \texorpdfstring{$\groupG/\groupG^{\Integer_N}$}{G/Inv(G)} coset
  superspaces}

While the results for conformal $\sigma$-models on symmetric superspaces
are complete, much less is known for some of the extensions that appear
e.g. in the context of AdS compactifications, see below. In fact, for many
cases of interest, the Lie sub-superalgebra $\g'$ in
$\g$ consists of elements that are invariant under some
automorphism $\Omega:\g \to \g$ of order $N >2$.
An automorphism of order $N$ defines a decomposition
\begin{align} \label{eq:decOmeig} \g\ =\ \g'\oplus
\bigoplus_{i=1}^{N-1}\mathfrak{m}_i\ , \quad
\Omega|_{\g'}\ =\ 1 \ , \quad
\Omega(\mathfrak{m}_k)\ =\ e^{\frac{2\pi i}{N}k}\,\mathfrak{m}_k
\end{align}
of the superalgebra $\g$ into eigenspaces of $\Omega$.
Extending our previous notation, we denote by $P_i$ the projection
maps onto $\mathfrak{m}_i$. Thanks to the properties of
$\Omega$, we find
\begin{align} \label{properties1} \left[\mathfrak{m}_i,\mathfrak{m}_j\right]\
\subset\  \mathfrak{m}_{i+j\text{ mod } N}\ ,\qquad
\left\langle\mathfrak{m}_i,\mathfrak{m}_j\right\rangle\ =\ 0\ \ \text{ if } \ \ i+j\
\neq \ 0 \text{ mod } N\ , \end{align}
where we have set $\mathfrak{m}_0\equiv \g'$.
Consequently, the subalgebra $\g'$ acts on the
$\Omega$-eigenspaces $\mathfrak{m}_i$. Note that the spaces
$\mathfrak{m}_i$ need
not be indecomposable under $\g'$ in which case the
decomposition into $\g'$-modules is finer than the
decomposition \eqref{eq:decOmeig} into eigenspaces of $\Omega$.

Whenever a coset superspace $\groupG/\groupGp$ is defined by an automorphism
$\Omega$ of order $N$ we shall use the alternative notation
$\groupG/\groupG^{\Integer_N}$. The cases when the grading induced by
$\Omega$ is compatible with the $\Integer_2$ superalgebra
grading, that is $\mathfrak{m}_{2i}\in \g_{\bar{0}}$ and
$\mathfrak{m}_{2i-1}\in \g_{\bar{1}}$, were considered
by Kagan and Young in \cite{Kagan:2005wt}. They restricted to a
family of Lagrangians for which $\tG$ and $\tB$ take the following
special form
\begin{align} \label{eq:KYGB} \tG(X,Y)\ =\
\sum_{i=1}^{N-1}p_i\,\bigl\langle P_i(X),P_{N-i}(Y)\bigr\rangle\quad , \quad
\tB(X,Y)\ =\ \sum_{i=1}^{N-1}q_i\,\bigl\langle P_i(X),P_{N-i}(Y)\bigr\rangle\ ,
\end{align}
where the $p_i$ and $q_i$ are constants obeying the additional
constraints
\begin{align} p_i=p_{N-i}\quad\text{ and }\quad q_i=-q_{N-i}\ . \end{align}
The forms of $\tG$ and $\tB$ in eq.~\eqref{eq:KYGB} do not give
rise to the most general Lagrangian for coset superspaces
$\groupG/\groupGp$.
As an example consider the famous $\Integer_4$ quotient
$\text{PSU}{(2,2|4)}/\SO(1,4)\times \SO(5)$.
Its metric has two radii because its bosonic base is
 $\AdS_5\times \Sphere^5$.
On the other hand, the special form of $\tG$ in
eq.~\eqref{eq:KYGB} allows for only two parameters $p_1=p_3$ and
$p_2$, among which $p_1$ is redundant because of the purely
fermionic nature of $\mathfrak{m}_1$ and $\mathfrak{m}_3$. In this
example, the form that $\tG$ takes in eq.~\eqref{eq:KYGB}
restricts the radii of $\AdS_5$ and $\Sphere^5$ to be equal.

The properties of the theory defined by eqs.\ \eqref{eq:KYGB}
certainly depend on the precise choice of the parameters $p_i$
and $q_i$. In particular, it was shown in \cite{Young:2005jv} and
\cite{Kagan:2005wt} that one loop conformal invariance requires
\begin{align} p_i\ =\ 1\qquad q_i\ =\ 1-\frac{2i}{N}\qquad \text{ for } i\
\neq\  0\ , \end{align}
for all even $N$. We believe, however, that in most cases these
conditions are not sufficient to guarantee the vanishing of the
full beta function.

\subsection{Embedding into string theory}

All superspace $\sigma$-models that appear in string theory are constructed
in terms of generalized symmetric spaces for automorphisms of order $N=4$.
Here we shall describe one of these models in the so-called hybrid approach
before we give an overview over related models and their application to
strings in AdS backgrounds.

\subsubsection{The coset space \texorpdfstring{$\PSU(1,1|2)/\U(1) \times \U(1)$}{PSU(1,1|2)/U(1)xU(1)}}

The coset space we want to look at in this subsection takes the form
\begin{align} \label{eq:PSUUU}
  \groupG/\groupGp
  = \PSU(1,1|2)/\U(1) \times \U(1) \ .
\end{align}
We will show below that the denominator subgroup $\groupGp=\U(1) \times
\U(1)$ is kept  fixed by an order $N=4$ automorphisms of the numerator
supergroup $\groupG = \PSU(1,1|2)$. The bosonic submanifold of the quotient
is a product of two factors
\begin{align}
\PSU(1,1|2)^{(0)}/\U(1) \times \U(1) = \SU(1,1)/\U(1) \times \SU(2)/\U(1)\ .
\end{align}
These two factors possess a simple geometric interpretation
\begin{align}
\SU(2)/\U(1) & = \SO(3)/\SO(2) = \Sphere^2 = \{ x_1^2 + x_2^2 + x_3^2 = 1\} \\[2mm]
\SU(1,1)/\U(1) & = \SO(1,2)/\SO(1,1) = \AdS_2 = \{ y_{-1}^2 + y_0^2 - y_1^2 = 1\}\ .
\end{align}
The Lie superalgebra $\psl(2|2)$ was first introduced in
Section~\ref{sc:LieDef}. It is generated by six bosonic elements
$K_{ab}= -K_{ba},\, a,b =0,1,2,3$ along with eight fermionic
ones. These are denoted by $S_{a\alpha}, \alpha = 1,2$ and they obey
the relations
\begin{align}
  \label{eq:PSLRelations}
[K_{ab}, K_{cd}] & = \eta_{ac}K_{bd} - \eta_{ad}K_{bc} -
\eta_{bc} K_{ad} + \eta_{bd} K_{ac}  \\[3mm]
[K_{ab}, S_{c\alpha} ] & = \eta_{ac} S_{b\alpha} -
\eta_{bc} S_{a\alpha} \\[2mm]
[S_{a\alpha},S_{b\beta}] & = \frac12 \epsilon_{abcd}
\epsilon_{\alpha\beta} K_{cd} \ .
\end{align}
As we know from our discussion in Section~\ref{sc:Classification},
this Lie superalgebra possesses vanishing dual Coxeter number. Let us
now define the action of $\Integer_4$ on $\g$ by
\begin{align}
\gamma(\mathfrak{m}_r) = e^{\frac{\pi i r}{2}} \mathfrak{m}_r
\end{align}
where
\begin{align}
\mathfrak{m}_0 & = \langle K_{01},K_{23}\rangle \subset \g_\den \\[2mm]
    \mathfrak{m}_1 & = \langle S_{0\alpha}+i S_{1\alpha}, S_{2\alpha} + i S_{3\alpha}
    \rangle \subset \g_\deo \\[2mm]
    \mathfrak{m}_2 & = \langle K_{03},K_{12},K_{13},K_{02}\rangle \subset \g_\den \\[2mm]
     \mathfrak{m}_3 & = \langle S_{0\alpha}-i S_{1\alpha}, S_{2\alpha} - i S_{3\alpha}\rangle
     \subset \g_\deo\ \ .
\end{align}
  It is easy to check that this definition respects the graded
  commutation relations \eqref{eq:PSLRelations} that define
  $\psl(2|2)$. The invariant subalgebra is generated by the two
  commuting $\U(1)$ charges $K_{01}$ and $K_{23}$. Hence, the coset
  \eqref{eq:PSUUU} is a generalized symmetric superspace with an
  automorphism group of order $N=4$. Therefore, we can apply the
  results discussed in the previous subsection to build a conformally
  invariant $\sigma$-model, at least to one loop order.

\subsubsection{String theory on \texorpdfstring{$\mathbf{\AdS_2 \times
      \Sphere^2 \times \CY_6}$}{AdS(2)xS(2)xCY(6)}}

In this subsection we want to illustrate in one example how to embed
one of our $\sigma$-models into a full string background. For the $\AdS_2
\times \Sphere^2$ $\sigma$-model this was explained by Berkovits et al.\ in
\cite{Berkovits:1999zq}. The proposal is to add any $\sigma$-model on a
Calabi-Yau superspace $\CY_6$ along with an additional free boson $\rho$.
Let us discuss these ingredients separately. The factors $\AdS_2 \times
\Sphere^2$ bring in a $\sigma$-model on $\PSU(1,1|2)/\U(1)\times \U(1)$. According
to our general discussion in the previous subsection, this model has a
single free parameter once we require conformal invariance,
\begin{align} \label{eq:PSUcoset}
S_{\groupG/\groupGp} = \frac{R^2}{4\pi} \int d^2z\,\left(
\bigl\langle J^{(2)},\bar J^{(2)}\bigr\rangle
+\frac{3}{2}\bigl\langle J^{(1)},\bar J^{(3)}\bigr\rangle +
\frac12 \bigl\langle J^{(3)},\bar J^{(1)}\bigr\rangle\right) \ .
\end{align}
As we discussed above, the beta function of this theory vanishes at
one loop. We believe that in this particular case, it vanishes to
all loops. It would be interesting to check this claim directly.

Next, let us turn to the free boson $\rho$. This boson is assumed to
be compactified to the self-dual radius and to possess time-like signature
$$ \rho(z, \bar z) \rho(w,\bar w) \sim - \log |z-w|^2 \ . $$
As usual, we can decompose $\rho(z,\bar z)$ into its chiral components
$\rho(z,\bar z) = \rho(z) + \bar \rho(z)$ and build vertex operators of
the form
$$ V(z) = e^{i \rho(z)} \quad , \quad
    \bar V(\bar z) = e^{i \bar \rho(\bar z)}\ .
$$
These are fermionic local (anti-)holomorphic operators of conformal weight
$(h,\bar h) = (-1/2,0)$ and $(h,\bar h) = (0,-1/2)$.

Finally, $\CY_6$ can stand for any conformal field theory with
an $N=2$ superconformal symmetry of central charge $c=9$. Sigma
models on Calabi-Yau manifolds of complex dimension $d=3$ provide
many examples. Exact conformal field theories with the desired
properties have been obtained through Gepner's famous construction
\cite{Gepner:1987qi}.

Assuming that the $\sigma$-model on the coset superspace \eqref{eq:PSUUU}
is conformal the total central charge of the resulting theory is
\begin{align} \label{eq:cctot}
c^{\text{tot}} = (-2 -1 - 1) + 1 + 9 = 6
\end{align}
Berkovits et al.\ were able to construct an $N=2$ superconformal algebra out
of these models. For the $\sigma$-model on $\CY_6$ this comes with the Calabi-Yau
condition. The challenge was therefore to build one out of the $\sigma$-model
on the superspace \eqref{eq:PSUUU} and the boson $\rho$. The bosonic elements
take the form
\begin{align}
T = T_{\groupG/\groupGp} + \frac12 \partial \rho \partial \rho + T_{\CY}\quad ,
\quad   J = i \partial \rho + J_{\CY}
   \end{align}
and there exist two fermionic currents $G^\pm$ in which the boson $\rho$
is coupled to a particular holomorphic combination of the currents
$S_{a\alpha}$, see \cite{Berkovits:1999zq} for details.

  The string spectrum is obtained by gauging the $N=2$ superconformal
  algebra. To this end one introduces fermionic/bosonic ghosts for the
  bosonic/fermionic fields $T,J/G^\pm$ that form the $N=2$
  algebra. The $b/\beta$ ghosts of these four ghost systems must
  possess conformal weights $h_{b^T} = 2, h_{b^J} = 1$ and $h_{\beta^\pm} 
  = 3/2$. Given the usual rules for computing the central charge,
\begin{align}
  c_{b} = -3(2h_{b}-1)^2+1
  \qquad\text{ and }\qquad
  c_{\beta} = 3(2h_{\beta}-1)^2-1\ ,
\end{align}
  for $b=b^T,b^J$ and $\beta = \beta^\pm$, the total $c^{\text{gh}}$ of 
  the four ghost systems adds up to
\begin{align} \label{eq:ccgh}
c^{\text{gh}} = - 26 - 2 + 2 \times 11 = -6 \ .
\end{align}
  Hence, the sum of the central charges \eqref{eq:cctot} from the
  matter sector and $c^{\text{gh}}$ from the ghost sector add up to
  $c=0$. The BRST operator for the gauging of an $N=2$ superconformal
  algebra is given in \cite{Giveon:1993ew}. When the supercoset
  $\groupG/\groupGp$ is replaced by a 4-dimensional Minkowski space,
  the cohomology of the corresponding BRST operator may be shown to
  coincide with the spectrum of a Calabi-Yau compactification of type
  II superstring theory \cite{Berkovits:1994wr}. In the case of the
  $\sigma$-model on $\PSU(1,1|2)/\U(1)^2$, the cohomology of $Q_{\text{BRST}}$
  provides the spectrum of type II superstring theory on $\AdS_2\times
  \Sphere^2 \times \CY_6$. Note that the underlying model is just a
  product of the superspace $\sigma$-model with the chiral boson and the
  $c=9$ CFT describing strings on $\CY_6$. These models are only
  coupled through the BRST operator.

\subsubsection{Other string backgrounds - an overview}

A discussion similar to the one we have outlined here for strings
in  $\AdS_2\times \Sphere^2$ can be performed for $\AdS_3 \times \Sphere^3$. The
associated $\Integer_4$ coset model is given by $\groupG/\groupGp =
\PSU(1,1|2)^2/\SO(1,2)\times \SO(3)$, see \cite{Berkovits:1999zq}.
The full model requires two chiral bosons and a number of ghost
fields for harmonic constraints. The latter may be used to reduce
the model to an alternative formulation involving a $\sigma$-model on
$\PSU(1,1|2)$ that was described first in \cite{Berkovits:1999du}.
Conformal invariance of the $\sigma$-model on $\PSU(1,1|2)$ has been
established to all orders in \cite{Berkovits:1999im,Bershadsky:1999hk}.
In case of $\AdS_3 \times \Sphere^3$ it is possible to switch on an NSNS 3-form
flux $H$ in addition to the RR-flux that is usually considered to obtain a
consistent string background. On the world-sheet, the NSNS-flux corresponds
to a (bosonic) WZW term in the $\sigma$-model on $\PSU(1,1|2)$. This model is
known to be classically integrable \cite{Cagnazzo:2012se} and quantum
conformally invariant. At the so-called WZ point, the theory possesses
additional holomorphic currents that render it solvable by standard
conformal field theory techniques. The solution has been worked out
in \cite{Gotz:2006qp}, using methods similar to the ones presented in
Section~\ref{sc:FFR}. With the WZW model being under good control,
it has been possible to analyse the spectrum of physical states, i.e.\
the cohomology of the BRST operator of the hybrid formalism, see
\cite{Gaberdiel:2011vf} for the massless spectrum and \cite{Gerigk:2012cq}
for an extension to massive states. As has been argued in a toy model by
Troost \cite{Troost:2011fd} the resulting spectrum is a direct sum of
irreducibles.

For higher dimensional $\AdS$ backgrounds with maximal supersymmetry, i.e.\
$\AdS_5 \times \Sphere^5$ and $\AdS_4 \times\CP^{3}$, covariant string theoretic
models may be constructed in the so-called pure spinor formalism. In the case
of $\AdS_5$, the model was first proposed by Berkovits \cite{Berkovits:2000fe}.
Once again, it is based on a $\sigma$-model with a generalized symmetric
target space $\groupG/\groupGp = \PSU(2,2|4)/\SO(1,4)\times \SO(5)$ where the
denominator is kept fixed by an automorphism of order $N=4$. In this
case, one needs to add $16$ bosonic ghost fields subject to five
independent pure spinor constraints. The pure spinor sector alone
would hence contribute a total central charge $c^{\text{gh}} = 32-10=22$.
In the action of the pure spinor string theory, the ghost sector and
the $\sigma$-model are coupled and once the interaction terms are included,
the combined theory is believed to be conformal to all loop orders
\cite{Berkovits:2004xu}. Its central charge $c = -22+22=0$ is obtained
by adding the superdimension $\sdim\groupG/\groupGp = -2-10-10 = -22$
to the central charge of the ghost sector.   Classical integrability of the
model was established in \cite{Vallilo:2003nx}. Properties of the monodromy
of the Lax connection were studied more recently, see \cite{Benichou:2011ch}
and references therein. Similar developments exist in the case of $\AdS_4
\times\CP^3$. In the latter case, the pure spinor model is based on
the generalized symmetric superspace $\groupG/\groupGp =
\OSP(6|2,2)/\U(3) \times \SO(1,3)$ \cite{Fre:2008qc}.

Before we conclude this short overview on superspace $\sigma$-models in exact
string backgrounds, we would like to add a few comments on the relation
with the Green-Schwarz formalism. For all the backgrounds we discussed
in the previous two paragraphs, a corresponding Green-Schwarz description
is known. In the case of $\AdS_5 \times \Sphere^5$ this was proposed by Metsaev
and Tseytlin, see \cite{Metsaev:1998it}. Similar models exist for $\AdS_2
\times \Sphere^2 \times T^6$, see  \cite{Sorokin:2011rr} and references therein,
$\AdS_3 \times \Sphere^3 $ \cite{Rahmfeld:1998zn} and $\AdS_4 \times\CP^6$
\cite{Arutyunov:2008if}. All these models employ $\sigma$-models on the
very same generalized symmetric superspaces we have described before.
On the other hand, the fermionic WZ terms need to be re-adjusted compared to
the values that appeared in our discussion of one-loop conformal invariance.
For consistency of the Green-Schwarz superstring, $\kappa$ symmetry is
crucial. This feature requires a degenerate fermionic metric with the
parameter $p_1 =0$ rather than $p_1 = 1$ as in the case of the pure
spinor formalism. After $\kappa$ symmetry has been fixed, one obtains
conformal field theories with $c=26$ and for physical state selection
one needs to impose the usual Virasoro constraints. All consistent AdS
backgrounds  that can be obtained in this way have been classified in
\cite{Zarembo:2010sg}. In all known examples the Green-Schwarz and pure
spinor/hybrid formulations are believed to be equivalent, see
\cite{Cagnazzo:2012uq} for a recent analysis at one loop.

\subsection{\label{sc:SupersphereDuality}A duality between Gross-Neveu and supersphere sigma models}

Most of the results we have reported about in this review deal with WZW
models and their deformations. On the other hand, $\sigma$-models certainly
play an important role for applications in physics. Hence it is of interest
 to understand the possible relations between these two types of models. In
 string theory, there exists a beautiful relation between $\sigma$-models on
 Calabi-Yau spaces and so-called Gepner models \cite{Gepner:1987qi}. The
 latter are obtained from products of WZW coset models and both types of
 models are related by exactly marginal deformations \cite{Witten:1993yc}.
 Therefore, one may think of the non-geometric Gepner models as describing
 string-size Calabi-Yau compactifications. Witten's explanation of the
 duality between Gepner-models and Calabi-Yau $\sigma$-models heavily relies
 on unitarity and hence no such techniques are at our disposal in the
 context of conformal field theories with internal supersymmetry.

On the other hand,   a very similar duality between $\OSP(2S+2|2S)$
Gross-Neveu models and $\sigma$-models on the superspheres $\Sphere^{2S+1|2S}$
has been proposed in \cite{Candu:2008yw} and then supported by a smooth
interpolation of certain boundary partition functions \cite{Mitev:2008yt}.
In the following we shall recall the formulation of the Gross-Neveu model
as a deformed WZW model
and explain how boundary spectra can be calculated along the lines of
Section~\ref{sc:Deformations}. For simplicity we restrict ourselves to
the case of $S=1$.
Then we deform the boundary spectrum away from the WZW point until we reach the
large volume limit in which the partition function describes the counting of
operators for a $\sigma$-model on $\Sphere^{3|2}$.  We think of the relation
between the Gross-Neveu and the $\sigma$-model as a prototype of a large class
of similar dualities for non-unitary models with internal supersymmetry. It is
quite feasible that similar relations also exist for $\sigma$-models on the coset
spaces we listed in the previous subsection. If this were the case, it could
play a central role in a possible world-sheet derivation of Maldacena's AdS/CFT
correspondence.

\subsubsection{\label{sc:GrossNeveu}The
  \texorpdfstring{$\OSP(2S+2|2S)$}{OSP(2S+2|2S)} Gross-Neveu model}

  We are now presenting arguments
  in favor of a duality between non-linear $\sigma$-models on
  superspheres $\Sphere^{2S+1|2S}$ and $\OSP(2S+2|2S)$ Gross-Neveu models
  \cite{Mitev:2008yt}. In the case $S=0$, this duality reduces to the
  well-known correspondence between the massless Thirring model (also
  known as Luttinger liquid in the condensed matter community) and the
  free compactified boson. All cases $S\geq1$ can be thought of as non-abelian
  generalizations of this equivalence.

  The $\OSP(4|2)$ Gross-Neveu model is a non-geometric theory defined
  by the following Lagrangian
\begin{align}
  \cS^{\text{GN}}[\Psi]
  &\ =\ \frac{1}{2\pi}\int
  d^2z\Bigl[\langle\Psi,\bartial\Psi\rangle+\langle\bar{\Psi},\partial\bar{\Psi}\rangle+g^2\langle\Psi,\bar{\Psi}\rangle^2\Bigr]\ \ .
\end{align}
  Here, $\Psi=(\psi_1,\ldots,\psi_4,\beta,\gamma)$ is a fundamental
  $\OSP(4|2)$ multiplet with four fermions and two bosons, all having
  conformal dimension $h=1/2$, as discussed in
  Section~\ref{sc:OSP}. The theory has a single coupling constant $g$
  which determines the strength of the quartic potential. According to
  our discussion in Section~\ref{sc:OSP} we can think of the
  $\OSP(4|2)$ Gross-Neveu model as a deformed $\OSP(4|2)$ WZW model at
  level $k=1$ and construct the interaction in terms of the currents
  as
\begin{align}
  \cS^{\text{GN}}
  &\ =\ \cS^{\text{WZW}}+g^2\cS_{\text{def}}&
  &\text{with}&
  \cS_{\text{def}}
  &\ =\ \frac{1}{2\pi}\int\!d^2z\,\bigl\langle J,\Omega(\bJ)\bigr\rangle\ \ ,
\end{align}
where $\Omega$ is induced from the exchange automorphism of
$\widehat{\text{SU}}(2)_1\times\widehat{\text{SU}}(2)_1$. This
kind of deformation is covered by our discussion in
Section~\ref{sc:WZWDef2}.

In Section~\ref{sc:OSP} we discussed the spectrum of the $\OSP(4|2)$ 
WZW model in the bulk. Alternatively, this theory can be formulated 
as an orbifold WZW model
\begin{align}
  \widehat{\OSP}(4|2)_1
  \ =\ \Bigl(\widehat{\text{SU}}(2)_{-\frac{1}{2}}\times\widehat{\text{SU}}(2)_1\times
  \widehat{\text{SU}}(2)_1\Bigr)\Bigr/\Integer_2
\end{align}
of purely bosonic WZW models. The two copies of $\text{SU}(2)_1$
arise from the two pairs of fermions, while the
$\widehat{\text{SU}}(2)_{-\frac{1}{2}}$ arises from the bosonic
$\beta\gamma$ system. Now we want to place the theory on the upper
half-plane and impose $\OSP(4|2)$ symmetry preserving boundary
conditions. These are associated to trivial gluing conditions in
the $\widehat{\text{SU}}(2)_{-\frac{1}{2}}$ part and permutation
gluing conditions in the $\widehat{\text{SU}}(2)_1\times
\widehat{\text{SU}}(2)_1$ factors. It turns out that there is
a unique boundary condition whose spectrum takes the form
\cite{Mitev:2008yt}
\begin{align}
  \label{eq:WZWSpectrum}
  Z_{\text{GN}}(q,y|0)
  &\ =\ \frac{\eta(q)}{\theta_4(y_1)}\Biggl[\frac{\theta_2(q^2,y_2^2)\theta_2(q^2,y_3^2)}{\eta(q)^2}
        +\frac{\theta_3(q^2,y_2^2)\theta_3(q^2,y_3^2)}{\eta(q)^2}\Biggr]\
        \ .
\end{align}
  It is just the sum of the affine $\widehat{\OSP}(4|2)_1$
  characters based on the trivial and the fundamental representation
  of $\OSP(4|2)$.

\subsubsection{Deformed boundary spectrum}

\begin{figure}
\begin{center}
\begin{tikzpicture}{0cm}{0cm}{10cm}{5cm}
  \draw[-latex] (.7,.8) -- (4,.8);
  \draw[-latex] (5.7,.8) -- (9,.8);
  \draw[-latex] (1,0.5) -- (1,4.5);
  \draw[-latex] (6,0.5) -- (6,4.5);
  \draw (2.5,0) node {WZW model ($g=0$)};
  \draw (7.5,0) node {Strong
        deformation ($g=\infty$)};
  \draw (1,4.3) node[left] {$E$};
  \draw (1,1.8) node[left] {$\frac{1}{2}$};
  \draw (1,2.8) node[left] {$1$};
  \draw (6,4.3) node[left] {$E$};
  \draw (6,1.8) node[left] {$\frac{1}{2}$};
  \draw (6,2.8) node[left] {$1$};
  \draw[dashed] (1,1.8) -- (3.7,1.8);
  \draw[dashed] (6,1.8) -- (8.7,1.8);
  \draw[dashed] (1,2.8) -- (3.7,2.8);
  \draw[dashed] (6,2.8) -- (8.7,2.8);
  \draw[dashed] (1,3.8) -- (3.7,3.8);
  \draw[dashed] (6,3.8) -- (8.7,3.8);
  \draw[color=red,fill] (2.5,.8) circle (2pt);
  \draw[color=blue,fill] (2.5,1.8) ellipse (.3 and .1);
  \draw[color=green,fill] (2.5,2.8) ellipse (.5 and .1);
  \draw[color=brown,fill] (3,3.8) ellipse (.4 and .1);
  \draw[color=pink,fill] (2,3.8) ellipse (.6 and .1);
  \draw (2.5,3.1) node {\tiny Adjoint};
  \draw (2.5,2.1) node {\tiny Fundamental};
  \draw (2.5,1.1) node {\tiny Trivial};
  \draw (2.5,4.1) node {\tiny Fundamental$\otimes$Adjoint};
  \draw[color=yellow,fill] (7.5,.8) ellipse (1.3 and .1);
  \draw[color=yellow,fill] (7.5,2.8) ellipse (1.3 and .1);
  \draw[color=violet,fill] (8,2.8) ellipse (.4 and .1);
  \draw[color=green,fill] (7.0,2.8) ellipse (.5 and .1);
  \draw[fill] (7.3,.8) ellipse (.5 and .1);
  \draw[color=blue,fill] (6.9,.8) ellipse (.3 and .1);
  \draw[color=red,fill] (8,.8) circle (2pt);
  \draw (7.5,1.1) node {\tiny Algebra of functions on $S^{3|2}$};
  \draw (7.5,3.1) node {\tiny $\infty$
      many representations};
  \draw[color=red,densely dotted,line width=.8pt] (2.5,.8) -- (8,.8);
  \draw[color=green,densely dotted,line width=.8pt] (2.5,2.8) -- (7.0,2.8);
  \draw[color=blue,densely dotted,line width=.8pt] (2.5,1.8) -- (6.9,.8);
\end{tikzpicture}
\end{center}
  \caption{\label{fig:Spectrum}(Color online) A distinguished boundary
    spectrum of the $\OSP(4|2)$ Gross-Neveu model at zero and
    infinite coupling. The interpolation between these two spectra for
    other values of the coupling $g$ is described by formula
    \eqref{eq:Interpolation}.}
\end{figure}
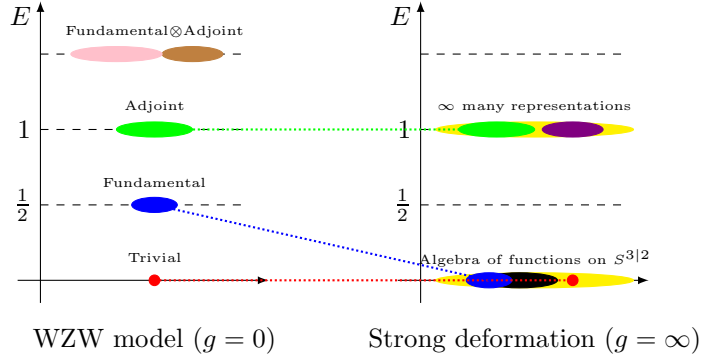

  Once the WZW model has been solved, it is straightforward to
  determine the deformed boundary spectrum using the results of
  Section~\ref{sc:WZWDef2}. According to formula
  \eqref{eq:AnomalousDimension2} the anomalous dimensions of a
  boundary field only depend on the transformation properties with
  respect to the global $\OSP(4|2)$-symmetry. In order to use this
  formula, we must first decompose the spectrum \eqref{eq:WZWSpectrum}
  at the WZW point into $\osp(4|2)$ multiplets,
\begin{align}
  \label{eq:Decomposition}
  Z_{\text{GN}}(q,y|g^2=0)
  \ =\ \sum_{\Lambda}
       \,\psi_\Lambda^{\text{WZW}}(q)\,\chi_\Lambda(y)\ \ .
\end{align}
Here, $\Lambda$ runs over weights of finite dimensional representations
of $\osp(4|2)$, $\chi_\Lambda(y) = \chi_\Lambda(y_1,y_2,y_3)$ denote the
associated characters  and $\psi_\Lambda^{\text{WZW}}$ are the branching functions
that are defined by the decomposition \eqref{eq:Decomposition}. For the
case at hand, the branching functions can be computed explicitly
\cite{Mitev:2008yt},
\begin{equation}
  \begin{split}
  \psi_{[j_1,j_2,j_3]}^{\text{WZW}}(q) &\  = \
\frac{1}{\eta(q)^4}\sum_{n,m=0}^{\infty}(-1)^{n+m}
q^{\frac{m}{2}(m+4j_1+2n+1)+j_1+\frac{n}{2}-\frac{1}{8}}
\\[2mm]& \hspace*{1cm}\times\
\Bigl[q^{(j_2-\frac{n}{2})^2}-q^{(j_2+\frac{n}{2}+1)^2}\Bigr]
\Bigl[q^{(j_3-\frac{n}{2})^2}-q^{(j_3+\frac{n}{2}+1)^2}\Bigr]\ \ .
  \end{split}
\end{equation}
Using eq.\ \eqref{eq:AnomalousDimension2}, we are now prepared to
construct the partition function at finite coupling $g^2$:
\begin{align}
  \label{eq:Interpolation}
  Z_{\text{GN}}(q,y|g^2)
  \ =\ \sum_{\Lambda}q^{-\frac{1}{2}\frac{g^2}{1+g^2}C_\Lambda}
       \,\psi_\Lambda^{\text{WZW}}(q)\,\chi_\Lambda(y)\ \ .
\end{align}
  The validity of this formula has been checked against numerical
  results from lattice discretizations with impressive agreement
  \cite{Candu:2008yw}.

Let us discuss the consequences of formula \eqref{eq:Interpolation}
in more detail, see also Figure~\ref{fig:Spectrum}. At
  zero coupling, the spectrum is characterized by the following
  features: All states have either integer or half-integer energy and
  at each energy level there is only a finite number of states. As
  mentioned above, these states are accounted for by the two affine
  $\widehat{\OSP}(4|2)_1$ representations built on top of the
  vacuum (with $h=0$) and the fundamental representation (with
  $h=1/2$), respectively. Once the deformation is switched on, the
  affine symmetry is broken and the states will receive an anomalous
  dimension depending on their transformation behavior under global
  $\OSP(4|2)$ transformations (the zero-modes of the current
  algebra). In particular, multiplets belonging to a representation
  with vanishing Casimir do not receive any correction. These are all
  protected BPS representations.\footnote{It
    should be noted, however, that there are short/BPS representations for
    $\OSP(4|2)$ which are not protected in this sense.} This applies
  in particular to the adjoint representation and ensures that the
  currents stay at conformal dimension $h=1$.

  At intermediate coupling the spectrum is very complicated,
  exhibiting almost no sign of an underlying organizing
  principle, except for the preserved global $\OSP(4|2)$ and the
  Virasoro symmetry. However, at infinite coupling we again recover a
  special situation. The energy of a multiplet $\Lambda$ is shifted by
  $-C_\Lambda/2$ in this case. It can be shown that despite this shift
  all conformal dimensions remain non-negative. Even more surprising,
  the spectrum is very regular again, exhibiting an integer level
  spacing (as opposed to the half-integer spacing at
  $g=0$). Nevertheless the spectrum now has entirely different
  characteristics than at zero coupling. Indeed, at infinite coupling
  we find an infinite number of states on each energy level, see again
  Figure~\ref{fig:Spectrum}.

\subsubsection{\label{sc:SphereSpectrum}Identification with
supersphere sigma model}

\definecolor{c334070}{RGB}{51,64,112}
\newcommand{\wiresphere}[1]{
\begin{scope}[y=0.80pt, x=0.8pt,yscale=-1, inner sep=0pt]
\begin{scope}[xscale=#1,yscale=-#1,shift={(0,-.5)}]
  \begin{scope}[rotate=180.0]
      \begin{scope}[fill=c334070!25!white]
        \path[fill] (-0.0309,0.9995){[rotate=-88.4775] arc(180.000:0.000:0.999990 and
          0.035)}arc(271.272:271.773:1.000){[rotate=-88.4775] arc(0.000:180.000:0.999990
          and 0.035)}arc(91.272:91.773:1.000) -- cycle;
        \path[fill] (-0.1381,0.9904){[rotate=-82.3139] arc(180.028:-0.028:0.999990 and
          0.172)}arc(277.434:277.941:1.000){[rotate=-82.3139] arc(0.028:179.972:0.999990
          and 0.172)}arc(97.434:97.941:1.000) -- cycle;
        \path[fill] (-0.2489,0.9685){[rotate=-75.8475] arc(180.062:-0.062:0.999990 and
          0.304)}arc(283.894:284.418:1.000){[rotate=-75.8475] arc(0.062:179.938:0.999990
          and 0.304)}arc(103.894:104.418:1.000) -- cycle;
        \path[fill] (-0.3661,0.9306){[rotate=-68.8025] arc(180.103:-0.104:0.999990 and
          0.426)}arc(290.921:291.473:1.000){[rotate=-68.8025] arc(0.103:179.896:0.999990
          and 0.426)}arc(110.921:111.473:1.000) -- cycle;
        \path[fill] (-0.4917,0.8708){[rotate=-60.846] arc(180.152:-0.152:0.999990 and
          0.536)}arc(298.860:299.452:1.000){[rotate=-60.846] arc(0.152:179.848:0.999990
          and 0.536)}arc(118.860:119.452:1.000) -- cycle;
        \path[fill] (-0.6259,0.7799){[rotate=-51.5732] arc(180.196:-0.195:0.999990 and
          0.629)}arc(308.105:308.749:1.000){[rotate=-51.5732] arc(0.196:179.805:0.999990
          and 0.629)}arc(128.105:128.749:1.000) -- cycle;
        \path[fill] (-0.7640,0.6452){[rotate=-40.5346] arc(180.242:-0.242:0.999990 and
          0.704)}arc(319.112:319.816:1.000){[rotate=-40.5346] arc(0.242:179.758:0.999990
          and 0.704)}arc(139.112:139.816:1.000) -- cycle;
        \path[fill] (-0.8910,0.4540){[rotate=-27.3812] arc(180.283:-0.283:0.999990 and
          0.757)}arc(332.233:332.998:1.000){[rotate=-27.3812] arc(0.283:179.717:0.999990
          and 0.757)}arc(152.233:152.998:1.000) -- cycle;
        \path[fill] (-0.9789,0.2045){[rotate=-12.203] arc(180.312:-0.312:0.999990 and
          0.787)}arc(347.392:348.202:1.000){[rotate=-12.203] arc(0.312:179.688:0.999990
          and 0.787)}arc(167.392:168.202:1.000) -- cycle;
        \path[fill] (-0.9969,-0.0787){[rotate=4.10228] arc(180.324:-0.324:0.999990 and
          0.793)}arc(3.693:4.513:1.000){[rotate=4.10228] arc(0.324:179.676:0.999990 and
          0.793)}arc(183.693:184.513:1.000) -- cycle;
        \path[fill] (-0.9372,-0.3487){[rotate=20.0103] arc(180.305:-0.305:0.999990 and
          0.775)}arc(19.616:20.406:1.000){[rotate=20.0103] arc(0.305:179.695:0.999990
          and 0.775)}arc(199.616:200.406:1.000) -- cycle;
        \path[fill] (-0.8231,-0.5678){[rotate=34.2322] arc(180.262:-0.263:0.999990 and
          0.733)}arc(33.863:34.598:1.000){[rotate=34.2322] arc(0.262:179.737:0.999990
          and 0.733)}arc(213.863:214.598:1.000) -- cycle;
        \path[fill] (-0.6866,-0.7270){[rotate=46.3017] arc(180.220:-0.220:0.999990 and
          0.669)}arc(45.968:46.641:1.000){[rotate=46.3017] arc(0.220:179.780:0.999990
          and 0.669)}arc(225.968:226.641:1.000) -- cycle;
        \path[fill] (-0.5489,-0.8359){[rotate=56.4021] arc(180.174:-0.174:0.999990 and
          0.585)}arc(56.090:56.707:1.000){[rotate=56.4021] arc(0.174:179.826:0.999990
          and 0.585)}arc(236.090:236.707:1.000) -- cycle;
        \path[fill] (-0.4187,-0.9081){[rotate=64.9618] arc(180.123:-0.123:0.999990 and
          0.483)}arc(64.674:65.245:1.000){[rotate=64.9618] arc(0.123:179.877:0.999990
          and 0.483)}arc(244.674:245.245:1.000) -- cycle;
        \path[fill] (-0.2976,-0.9547){[rotate=72.4165] arc(180.077:-0.077:0.999990 and
          0.366)}arc(72.150:72.688:1.000){[rotate=72.4165] arc(0.077:179.923:0.999990
          and 0.366)}arc(252.150:252.688:1.000) -- cycle;
        \path[fill] (-0.1841,-0.9829){[rotate=79.1347] arc(180.044:-0.044:0.999990 and
          0.239)}arc(78.877:79.391:1.000){[rotate=79.1347] arc(0.044:179.956:0.999990
          and 0.239)}arc(258.877:259.391:1.000) -- cycle;
        \path[fill] (-0.0755,-0.9971){[rotate=85.4178] arc(180.000:0.000:0.999991 and
          0.104)}arc(85.166:85.669:1.000){[rotate=85.4178] arc(0.000:180.000:0.999991
          and 0.104)}arc(265.166:265.669:1.000) -- cycle;
      \end{scope}
      \begin{scope}[fill=c334070!25!white]
        \path[fill] (-0.2729,-0.9620)arc(245.013:294.987:0.646124 and
          0.393)arc(285.840:284.281:1.000)arc(292.689:247.311:0.639439 and
          0.389)arc(255.719:254.160:1.000) -- cycle;
        \path[fill] (-0.5919,-0.8060)arc(219.656:320.344:0.768842 and
          0.468)arc(306.293:305.470:1.000)arc(319.489:220.511:0.763232 and
          0.465)arc(234.530:233.707:1.000) -- cycle;
        \path[fill] (-0.7803,-0.6255)arc(206.012:333.988:0.868199 and
          0.529)arc(321.286:320.583:1.000)arc(333.417:206.583:0.863836 and
          0.526)arc(219.417:218.714:1.000) -- cycle;
        \path[fill] (-0.9048,-0.4259)arc(195.992:344.008:0.941176 and
          0.573)arc(334.788:334.131:1.000)arc(343.556:196.444:0.938191 and
          0.571)arc(205.869:205.212:1.000) -- cycle;
        \path[fill] (-0.9770,-0.2135)arc(187.576:352.424:0.985556 and
          0.600)arc(347.675:347.039:1.000)arc(352.024:187.976:0.984041 and
          0.599)arc(192.961:192.325:1.000) -- cycle;
        \path[fill] (-1.0000,0.0055)arc(-180.181:0.181:0.999990 and
          0.609)arc(0.315:-0.315:1.000)arc(359.819:180.181:0.999990 and
          0.609)arc(180.315:179.685:1.000) -- cycle;
        \path[fill] (-0.9745,0.2243)arc(-187.976:7.976:0.984041 and
          0.599)arc(12.961:12.325:1.000)arc(7.576:-187.576:0.985556 and
          0.600)arc(167.675:167.039:1.000) -- cycle;
        \path[fill] (-0.8998,0.4363)arc(-196.444:16.444:0.938191 and
          0.571)arc(25.869:25.212:1.000)arc(15.992:-195.992:0.941176 and
          0.573)arc(154.788:154.131:1.000) -- cycle;
        \path[fill] (-0.7725,0.6350)arc(-206.583:26.583:0.863836 and
          0.526)arc(39.417:38.714:1.000)arc(26.012:-206.012:0.868199 and
          0.529)arc(141.286:140.583:1.000) -- cycle;
        \path[fill] (-0.5803,0.8144)arc(-220.511:40.511:0.763232 and
          0.465)arc(54.530:53.707:1.000)arc(39.656:-219.656:0.768842 and
          0.468)arc(126.293:125.470:1.000) -- cycle;
        \path[fill] (-0.2467,0.9691)arc(-247.311:67.311:0.639439 and
          0.389)arc(75.719:74.160:1.000)arc(65.013:-245.013:0.646124 and
          0.393)arc(105.840:104.281:1.000) -- cycle;
        \path[fill] (-0.4962,0.6888)arc(180.000:0.000:0.496217 and
          0.302)arc(0.000:-180.000:0.496217 and 0.302) --
          cycle(-0.5038,0.6853)arc(-180.000:0.000:0.503774 and
          0.307)arc(0.000:180.000:0.503774 and 0.307) -- cycle;
        \path[fill] (-0.3379,0.7467)arc(180.000:0.000:0.337917 and
          0.206)arc(0.000:-180.000:0.337917 and 0.206) --
          cycle(-0.3461,0.7443)arc(-180.000:0.000:0.346117 and
          0.211)arc(0.000:180.000:0.346117 and 0.211) -- cycle;
        \path[fill] (-0.1694,0.7819)arc(180.000:0.000:0.169350 and
          0.103)arc(0.000:-180.000:0.169350 and 0.103) --
          cycle(-0.1779,0.7807)arc(-180.000:0.000:0.177944 and
          0.108)arc(0.000:180.000:0.177944 and 0.108) -- cycle;
      \end{scope}
  \end{scope}
      \begin{scope}[fill=c334070]
        \path[fill] (-0.0309,0.9995){[rotate=-88.4775] arc(180.000:0.000:0.999990 and
          0.035)}arc(271.272:271.773:1.000){[rotate=-88.4775] arc(0.000:180.000:0.999990
          and 0.035)}arc(91.272:91.773:1.000) -- cycle;
        \path[fill] (-0.1381,0.9904){[rotate=-82.3139] arc(180.028:-0.028:0.999990 and
          0.172)}arc(277.434:277.941:1.000){[rotate=-82.3139] arc(0.028:179.972:0.999990
          and 0.172)}arc(97.434:97.941:1.000) -- cycle;
        \path[fill] (-0.2489,0.9685){[rotate=-75.8475] arc(180.062:-0.062:0.999990 and
          0.304)}arc(283.894:284.418:1.000){[rotate=-75.8475] arc(0.062:179.938:0.999990
          and 0.304)}arc(103.894:104.418:1.000) -- cycle;
        \path[fill] (-0.3661,0.9306){[rotate=-68.8025] arc(180.103:-0.104:0.999990 and
          0.426)}arc(290.921:291.473:1.000){[rotate=-68.8025] arc(0.103:179.896:0.999990
          and 0.426)}arc(110.921:111.473:1.000) -- cycle;
        \path[fill] (-0.4917,0.8708){[rotate=-60.846] arc(180.152:-0.152:0.999990 and
          0.536)}arc(298.860:299.452:1.000){[rotate=-60.846] arc(0.152:179.848:0.999990
          and 0.536)}arc(118.860:119.452:1.000) -- cycle;
        \path[fill] (-0.6259,0.7799){[rotate=-51.5732] arc(180.196:-0.195:0.999990 and
          0.629)}arc(308.105:308.749:1.000){[rotate=-51.5732] arc(0.196:179.805:0.999990
          and 0.629)}arc(128.105:128.749:1.000) -- cycle;
        \path[fill] (-0.7640,0.6452){[rotate=-40.5346] arc(180.242:-0.242:0.999990 and
          0.704)}arc(319.112:319.816:1.000){[rotate=-40.5346] arc(0.242:179.758:0.999990
          and 0.704)}arc(139.112:139.816:1.000) -- cycle;
        \path[fill] (-0.8910,0.4540){[rotate=-27.3812] arc(180.283:-0.283:0.999990 and
          0.757)}arc(332.233:332.998:1.000){[rotate=-27.3812] arc(0.283:179.717:0.999990
          and 0.757)}arc(152.233:152.998:1.000) -- cycle;
        \path[fill] (-0.9789,0.2045){[rotate=-12.203] arc(180.312:-0.312:0.999990 and
          0.787)}arc(347.392:348.202:1.000){[rotate=-12.203] arc(0.312:179.688:0.999990
          and 0.787)}arc(167.392:168.202:1.000) -- cycle;
        \path[fill] (-0.9969,-0.0787){[rotate=4.10228] arc(180.324:-0.324:0.999990 and
          0.793)}arc(3.693:4.513:1.000){[rotate=4.10228] arc(0.324:179.676:0.999990 and
          0.793)}arc(183.693:184.513:1.000) -- cycle;
        \path[fill] (-0.9372,-0.3487){[rotate=20.0103] arc(180.305:-0.305:0.999990 and
          0.775)}arc(19.616:20.406:1.000){[rotate=20.0103] arc(0.305:179.695:0.999990
          and 0.775)}arc(199.616:200.406:1.000) -- cycle;
        \path[fill] (-0.8231,-0.5678){[rotate=34.2322] arc(180.262:-0.263:0.999990 and
          0.733)}arc(33.863:34.598:1.000){[rotate=34.2322] arc(0.262:179.737:0.999990
          and 0.733)}arc(213.863:214.598:1.000) -- cycle;
        \path[fill] (-0.6866,-0.7270){[rotate=46.3017] arc(180.220:-0.220:0.999990 and
          0.669)}arc(45.968:46.641:1.000){[rotate=46.3017] arc(0.220:179.780:0.999990
          and 0.669)}arc(225.968:226.641:1.000) -- cycle;
        \path[fill] (-0.5489,-0.8359){[rotate=56.4021] arc(180.174:-0.174:0.999990 and
          0.585)}arc(56.090:56.707:1.000){[rotate=56.4021] arc(0.174:179.826:0.999990
          and 0.585)}arc(236.090:236.707:1.000) -- cycle;
        \path[fill] (-0.4187,-0.9081){[rotate=64.9618] arc(180.123:-0.123:0.999990 and
          0.483)}arc(64.674:65.245:1.000){[rotate=64.9618] arc(0.123:179.877:0.999990
          and 0.483)}arc(244.674:245.245:1.000) -- cycle;
        \path[fill] (-0.2976,-0.9547){[rotate=72.4165] arc(180.077:-0.077:0.999990 and
          0.366)}arc(72.150:72.688:1.000){[rotate=72.4165] arc(0.077:179.923:0.999990
          and 0.366)}arc(252.150:252.688:1.000) -- cycle;
        \path[fill] (-0.1841,-0.9829){[rotate=79.1347] arc(180.044:-0.044:0.999990 and
          0.239)}arc(78.877:79.391:1.000){[rotate=79.1347] arc(0.044:179.956:0.999990
          and 0.239)}arc(258.877:259.391:1.000) -- cycle;
        \path[fill] (-0.0755,-0.9971){[rotate=85.4178] arc(180.000:0.000:0.999991 and
          0.104)}arc(85.166:85.669:1.000){[rotate=85.4178] arc(0.000:180.000:0.999991
          and 0.104)}arc(265.166:265.669:1.000) -- cycle;
      \end{scope}
      \begin{scope}[fill=c334070]
        \path[fill] (-0.2729,-0.9620)arc(245.013:294.987:0.646124 and
          0.393)arc(285.840:284.281:1.000)arc(292.689:247.311:0.639439 and
          0.389)arc(255.719:254.160:1.000) -- cycle;
        \path[fill] (-0.5919,-0.8060)arc(219.656:320.344:0.768842 and
          0.468)arc(306.293:305.470:1.000)arc(319.489:220.511:0.763232 and
          0.465)arc(234.530:233.707:1.000) -- cycle;
        \path[fill] (-0.7803,-0.6255)arc(206.012:333.988:0.868199 and
          0.529)arc(321.286:320.583:1.000)arc(333.417:206.583:0.863836 and
          0.526)arc(219.417:218.714:1.000) -- cycle;
        \path[fill] (-0.9048,-0.4259)arc(195.992:344.008:0.941176 and
          0.573)arc(334.788:334.131:1.000)arc(343.556:196.444:0.938191 and
          0.571)arc(205.869:205.212:1.000) -- cycle;
        \path[fill] (-0.9770,-0.2135)arc(187.576:352.424:0.985556 and
          0.600)arc(347.675:347.039:1.000)arc(352.024:187.976:0.984041 and
          0.599)arc(192.961:192.325:1.000) -- cycle;
        \path[fill] (-1.0000,0.0055)arc(-180.181:0.181:0.999990 and
          0.609)arc(0.315:-0.315:1.000)arc(359.819:180.181:0.999990 and
          0.609)arc(180.315:179.685:1.000) -- cycle;
        \path[fill] (-0.9745,0.2243)arc(-187.976:7.976:0.984041 and
          0.599)arc(12.961:12.325:1.000)arc(7.576:-187.576:0.985556 and
          0.600)arc(167.675:167.039:1.000) -- cycle;
        \path[fill] (-0.8998,0.4363)arc(-196.444:16.444:0.938191 and
          0.571)arc(25.869:25.212:1.000)arc(15.992:-195.992:0.941176 and
          0.573)arc(154.788:154.131:1.000) -- cycle;
        \path[fill] (-0.7725,0.6350)arc(-206.583:26.583:0.863836 and
          0.526)arc(39.417:38.714:1.000)arc(26.012:-206.012:0.868199 and
          0.529)arc(141.286:140.583:1.000) -- cycle;
        \path[fill] (-0.5803,0.8144)arc(-220.511:40.511:0.763232 and
          0.465)arc(54.530:53.707:1.000)arc(39.656:-219.656:0.768842 and
          0.468)arc(126.293:125.470:1.000) -- cycle;
        \path[fill] (-0.2467,0.9691)arc(-247.311:67.311:0.639439 and
          0.389)arc(75.719:74.160:1.000)arc(65.013:-245.013:0.646124 and
          0.393)arc(105.840:104.281:1.000) -- cycle;
        \path[fill] (-0.4962,0.6888)arc(180.000:0.000:0.496217 and
          0.302)arc(0.000:-180.000:0.496217 and 0.302) --
          cycle(-0.5038,0.6853)arc(-180.000:0.000:0.503774 and
          0.307)arc(0.000:180.000:0.503774 and 0.307) -- cycle;
        \path[fill] (-0.3379,0.7467)arc(180.000:0.000:0.337917 and
          0.206)arc(0.000:-180.000:0.337917 and 0.206) --
          cycle(-0.3461,0.7443)arc(-180.000:0.000:0.346117 and
          0.211)arc(0.000:180.000:0.346117 and 0.211) -- cycle;
        \path[fill] (-0.1694,0.7819)arc(180.000:0.000:0.169350 and
          0.103)arc(0.000:-180.000:0.169350 and 0.103) --
          cycle(-0.1779,0.7807)arc(-180.000:0.000:0.177944 and
          0.108)arc(0.000:180.000:0.177944 and 0.108) -- cycle;
      \end{scope}
\end{scope}
\end{scope}
}

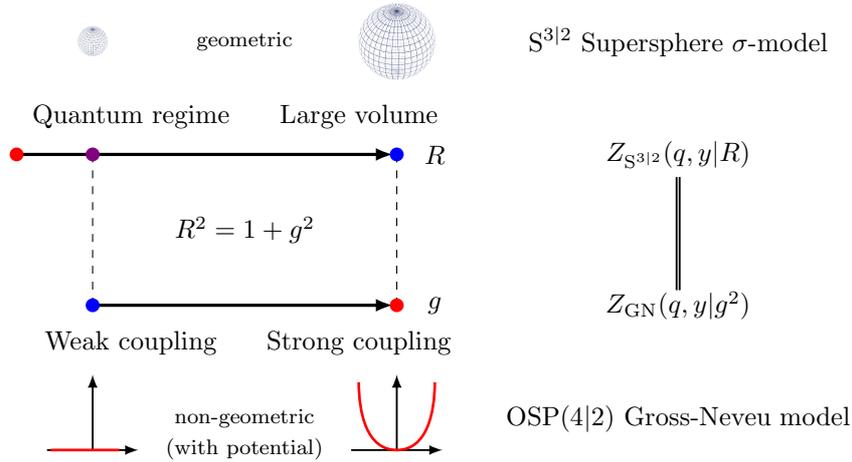
\begin{figure}
\begin{center}
\begin{tikzpicture}[line width=1.2pt]
\begin{scope}[xshift=0cm,yshift=5.1cm]
  \wiresphere{7}
\end{scope}
\begin{scope}[xshift=4cm,yshift=5.25cm]
  \wiresphere{18}
\end{scope}
  \draw[-latex] (-1,3.5) -- (3.95,3.5);
  \draw (7.7,5) node {$\Sphere^{3|2}$ Supersphere $\sigma$-model};
  \draw (4.5,3.5) node {$R$};
  \draw (3.5,4) node {Large volume};
  \draw (.5,4) node {Quantum regime};
  \draw (2,5) node {\footnotesize geometric};
  \draw (7.7,3.5) node {$Z_{\Sphere^{3|2}}(q,y|R)$};
  \draw[-latex] (0,1.5) -- (3.95,1.5);
  \draw (4.5,1.5) node {$g$};
  \draw (7.7,0) node {$\OSP(4|2)$ Gross-Neveu model};
  \draw (7.7,1.5) node {$Z_{\text{GN}}(q,y|g^2)$};
  \draw (3.5,1) node {Strong coupling};
  \draw (.5,1) node {Weak coupling};
  \draw (2,0) node {\footnotesize non-geometric};
  \draw (2,-.4) node {\footnotesize(with potential)};
  \draw[line width=.4pt,dashed] (4,3.5) -- (4,1.5);
  \draw[line width=.4pt,dashed] (0,3.5) -- (0,1.5);
  \draw (2,2.5) node {$R^2=1+g^2$};
  \draw[color=blue,fill] (0,1.5) circle (2pt);
  \draw[color=red,fill] (4,1.5) circle (2pt);
  \draw[color=blue,fill] (4,3.5) circle (2pt);
  \draw[color=red,fill] (-1,3.5) circle (2pt);
  \draw[violet,fill] (0,3.5) circle (2pt);
  \draw[line width=.7pt] (7.68,1.7) -- (7.68,3.2) (7.72,1.7) --
  (7.72,3.2);
\begin{scope}[xshift=-.6cm,yshift=-.42cm]
  \draw[-latex,line width=.7pt] (0,0) -- (1.2,0);
  \draw[-latex,line width=.7pt] (.6,0) -- (.6,1);
  \draw[color=red,line width=1pt] (.05,0) -- (.95,0);
\end{scope}
\begin{scope}[xshift=3.4cm,yshift=-.42cm]
  \draw[-latex,line width=.7pt] (0,0) -- (1.2,0);
  \draw[-latex,line width=.7pt] (.6,0) -- (.6,1);
  \draw[color=red,line width=1pt] (.1,.9) .. controls (.1,.5) and (.2,0)
  .. (.6,0) .. controls (1,0) and (1.1,.5) .. (1.1,.9);
\end{scope}
\end{tikzpicture}
\end{center}
  \caption{\label{fig:Duality}(Color online) The duality between the
    $\OSP(4|2)$ Gross-Neveu model and the $\Sphere^{3|2}$
    supersphere $\sigma$-model in pictures.}
\end{figure}

  We now wish to argue that the spectrum of the Gross-Neveu model
  discussed in the previous section coincides with the large volume
  partition function of the $\sigma$-model on the supersphere
  $\Sphere^{3|2}$ when we send the coupling $g$ to infinity. At infinite
  volume the partition function is easy to write down since the fields
  $\vec{X}$ become free. The most general field is obtained by
  considering the normal ordered products
  $\prod\partial^{n_i}\bartial^{m_i}X^{a_i}$ of the fields $X^a$ and
  their derivatives and the energy (scaling dimension) of such a field
  is just given by the number of derivatives.

  We assume Neumann boundary conditions, i.e.\ a freely moving open
  string. In this case we are only left with one type of
  derivative. In close analogy to the harmonic analysis on
  $\Sphere^{3|2}$ (see Section~\ref{sc:HASupercoset}) we can write
  down the open string partition function
\begin{multline}
  \label{eq:ZSSS}
  Z_{\Sphere^{3|2}}(q,y|R=\infty)\\[2mm]
  \ =\ \lim_{t\to1}\prod_{n=1}^\infty\frac{(1-t^2q^n)(1+tq^ny_3)(1+tq^n/y_3)}{(1-tq^ny_1y_2)(1-tq^ny_1/y_2)(1-tq^ny_2/y_1)(1-tq^n/y_1y_2)}\
  \ .
\end{multline}
  The only difference compared to eq.~\eqref{eq:ZSS1} are the
  additional terms involving powers $q^n$. These correspond to counting
  derivatives $\partial^n\vec{X}$ instead of plain coordinates
  $\vec{X}$. Since the constraint $\vec{X}^2=R^2$ also leads to
  constraints on derivatives of $\vec{X}$, also the first term in the
  numerator had to be extended to an infinite product.

  Even though it is by no means obvious, the decomposition of the
  partition function \eqref{eq:ZSSS} into irreducible characters of
  $\OSP(4|2)$ precisely agrees with the limit $g\to\infty$ of the
  expression \eqref{eq:Interpolation}
  \cite{Mitev:2008yt}. This suggests that the moduli spaces of the two
  theories indeed overlap -- and employing their common symmetry --
  actually coincide, see Figure~\ref{fig:Duality} for an
  illustration. Complementary calculations based on either lattice
  models \cite{Candu:2008yw}, background field methods or a
  cohomological reduction \cite{Candu:2010yg} confirm this picture and
  predict that the couplings should in fact be related as
  $R^2=1+g^2$.

\section{Conclusions and Outlook}

  Our goal in this review was to present conformal field theories with
  internal supersymmetries as a vast and intriguing branch of
  logarithmic conformal field theory. We have seen a large family of
  models, including WZW models on supergroups and $\sigma$-models on
  coset superspaces, along with their deformations and some
  dualities. Many of these models possess profound applications to
  problems in different areas of theoretical physics.

  To make our presentation reasonably self-contained we included some
  basic material, in particular on superalgebras and their
  representation theory. This also meant that we had to skip over
  quite a few further developments. These include a systematic
  discussion of boundary conditions. For WZW models on type I
  supergroups, these are quite will understood. At the example of the
  $\GL(1|1)$ WZW model it has been shown that Cardy's construction
  basically carries over to the logarithmic setting
  \cite{Creutzig:2007jy}. Besides there also exist indications that
  the geometric characterization of D-branes in group manifolds
  extends from the purely bosonic case
  \cite{Quella:2007sg,Creutzig:2007jy,Creutzig:2008ag}.

  Another topic of considerable interest are GKO-type coset models for
  supergroups with and without world-sheet supersymmetry. Some first
  studies can be found in \cite{Creutzig:2009fh} and
  \cite{Candu:2011hu}. It turns out that supergroup GKO coset models
  include several families of models which possess exactly marginal
  integrable deformations, generalizing the current-current
  deformations of WZW models we sketched in
  Section~\ref{sc:Deformations}. These models along with their
  deformations and potential dualities certainly deserve further
  study.

  On the more mathematical side, the relation of the representation
  theory of affine Lie superalgebras with Mock modular forms warrants
  more exploration, see \cite{Semikhatov:2003uc} and
  \cite{Folsom:MR2719689,Bringmann:MR2534107,Alfes:2012pa} for some
  existing results. In our opinion, however, the most pressing
  questions concern the study of dualities between WZW and
  $\sigma$-models. The only example that is reasonably well understood
  right now is the duality between the $\OSP(4|2)$ WZW model and the
  $\sigma$-model on the supersphere $S^{3|2}$, see
  Section~\ref{sc:SupersphereDuality}. It is likely that further
  progress requires a better understanding of the interplay between
  conformal invariance and integrability.


\subsubsection*{Acknowledgments}
  Much of the material presented in this review is based on joint work
  with a number of different collaborators. We would like to thank
  Constantin Candu, Thomas Creutzig, Gerhard G\"otz, Vladimir Mitev,
  Peter R{\o}nne and Hubert Saleur for many discussions and pleasant
  collaboration. Besides we are indebted to the anonymous referees for
  providing extremely detailed feedback which helped improve our
  presentation. Thomas Quella is funded by the German Research Council
  (DFG) through M.\ Zirnbauer's Leibniz Prize, DFG grant no.\ ZI
  513/2-1. Additional support is received from the DFG through the
  SFB$|$TR\,12 ``Symmetries and Universality in Mesoscopic Systems''
  and the Center of Excellence ``Quantum Matter and Materials''. The
  research of Volker Schomerus is supported in part by the SFB 676
  ``Particles, Strings, and the Early Universe'' and the Marie Curie
  network GATIS (gatis.desy.eu) of the European Union's Seventh
  Framework Programme FP7/2007-2013/ under REA Grant Agreement No
  317089.


\bibliographystyle{utphysTQ}
\bibliography{bibliographyreview}

\def\cprime{$'$}
\providecommand{\href}[2]{#2}\begingroup\raggedright\begin{thebibliography}{10%
0}

\bibitem{Aharony:1999ti}
O.~Aharony, S.~S. Gubser, J.~M. Maldacena, H.~Ooguri, and Y.~Oz, ``Large {N}
  field theories, string theory and gravity,'' {\em Phys. Rept.} {{\em 323}}
  (2000) 183--386,
\href{http://arxiv.org/abs/hep-th/9905111}{{\ttfamily hep-th/9905111}}.

\bibitem{Parisi:1979ka}
G.~Parisi and N.~Sourlas, ``Random magnetic fields, supersymmetry and negative
  dimensions,''
{\em Phys. Rev. Lett.} {{\em 43}} (1979) 744.

\bibitem{Efetov1983:MR708812}
K.~B. Efetov, ``Supersymmetry and theory of disordered metals,'' {\em Adv.
  Phys.} {{\em 32}} (1983) 53--127.

\bibitem{Bernard:1995as}
D.~Bernard, ``{(Perturbed)} conformal field theory applied to {2D} disordered
  systems: {An} introduction,'' in {\em Low-dimensional applications of quantum
  field theory}, pp.~19--61.
\newblock 1995.
\newblock \href{http://arxiv.org/abs/hep-th/9509137}{{\ttfamily
  hep-th/9509137}}.
\newblock
Cargese Lectures in Theoretical Physics.

\bibitem{Parisi:1982ud}
G.~Parisi and N.~Sourlas, ``Supersymmetric field theories and stochastic
  differential equations,''
{\em Nucl. Phys.} {{\em B206}} (1982) 321.

\bibitem{Read:2001pz}
N.~Read and H.~Saleur, ``Exact spectra of conformal supersymmetric nonlinear
  sigma models in two dimensions,'' {\em Nucl. Phys.} {{\em B613}} (2001) 409,
\href{http://arxiv.org/abs/hep-th/0106124}{{\ttfamily hep-th/0106124}}.

\bibitem{Gruzberg:1999dk}
I.~A. Gruzberg, A.~W.~W. Ludwig, and N.~Read, ``Exact exponents for the spin
  quantum {Hall} transition,'' {\em Phys. Rev. Lett.} {{\em 82}} (1999) 4524,
\href{http://arxiv.org/abs/cond-mat/9902063}{{\ttfamily cond-mat/9902063}}.

\bibitem{Essler:2005ag}
F.~H.~L. Essler, H.~Frahm, and H.~Saleur, ``Continuum limit of the integrable
  {$sl(2|1)$} {$3-\bar{3}$} superspin chain,'' {\em Nucl. Phys.} {{\em B712}}
  (2005) 513--572,
\href{http://arxiv.org/abs/cond-mat/0501197}{{\ttfamily cond-mat/0501197}}.

\bibitem{Weidenmuller:1987gi}
H.~Weidenm{\"u}ller, ``Single electron in a random potential and a strong
  magnetic field,''
\href{http://dx.doi.org/10.1016/0550-3213(87)90179-9}{{\em Nucl. Phys.} {{\em
  B290}} (1987) 87--110}.

\bibitem{Zirnbauer:1999ua}
M.~R. Zirnbauer, ``Conformal field theory of the integer quantum {Hall} plateau
  transition,''
\href{http://arxiv.org/abs/hep-th/9905054}{{\ttfamily hep-th/9905054}}.

\bibitem{Bhaseen:1999nm}
M.~J. Bhaseen, I.~I. Kogan, O.~A. Solovev, N.~Tanigichi, and A.~M. Tsvelik,
  ``Towards a field theory of the plateau transitions in the integer quantum
  {Hall} effect,'' {\em Nucl. Phys.} {{\em B580}} (2000) 688--720,
\href{http://arxiv.org/abs/cond-mat/9912060}{{\ttfamily cond-mat/9912060}}.

\bibitem{Tsvelik:2007dm}
A.~M. Tsvelik, ``Evidence for the {$PSL(2|2)$} {Wess-Zumino-Novikov-Witten}
  model as a model for the plateau transition in quantum {Hall} effect:
  {Evaluation} of numerical simulations,'' {\em Phys. Rev.} {{\em B75}} (2007)
  184201,
\href{http://arxiv.org/abs/cond-mat/0702611}{{\ttfamily cond-mat/0702611}}.

\bibitem{Chalker:1988}
J.~Chalker and P.~Coddington, ``Percolation, quantum tunnelling and the integer
  {Hall} effect,'' {\em J. Phys.} {{\em C21}} (1988) 2665--2679.

\bibitem{Gaberdiel:1998ps}
M.~R. Gaberdiel and H.~G. Kausch, ``A local logarithmic conformal field
  theory,'' {\em Nucl. Phys.} {{\em B538}} (1999) 631--658,
\href{http://arxiv.org/abs/hep-th/9807091}{{\ttfamily hep-th/9807091}}.

\bibitem{Semikhatov:2003uc}
A.~M. Semikhatov, A.~Taormina, and I.~Y. Tipunin, ``Higher level {Appell}
  functions, modular transformations, and characters,'' {\em Comm. Math. Phys.}
  {{\em 255}} (2005) 469--512,
\href{http://arxiv.org/abs/math.qa/0311314}{{\ttfamily math.qa/0311314}}.

\bibitem{Fuchs:2003yu}
J.~Fuchs, S.~Hwang, A.~M. Semikhatov, and I.~Y. Tipunin, ``Nonsemisimple fusion
  algebras and the {Verlinde} formula,'' {\em Commun. Math. Phys.} {{\em 247}}
  (2004) 713--742,
\href{http://arxiv.org/abs/hep-th/0306274}{{\ttfamily hep-th/0306274}}.

\bibitem{Gaberdiel:2006pp}
M.~R. Gaberdiel and I.~Runkel, ``The logarithmic triplet theory with
  boundary,'' {\em J. Phys.} {{\em A39}} (2006) 14745--14780,
\href{http://arxiv.org/abs/hep-th/0608184}{{\ttfamily hep-th/0608184}}.

\bibitem{Metsaev:1998it}
R.~R. Metsaev and A.~A. Tseytlin, ``Type {IIB} superstring action in
  {$AdS_5\times S^5$} background,'' {\em Nucl. Phys.} {{\em B533}} (1998)
  109--126,
\href{http://arxiv.org/abs/hep-th/9805028}{{\ttfamily hep-th/9805028}}.

\bibitem{Berkovits:1999zq}
N.~Berkovits, M.~Bershadsky, T.~Hauer, S.~Zhukov, and B.~Zwiebach,
  ``Superstring theory on {$AdS_2\times S^2$} as a coset supermanifold,'' {\em
  Nucl. Phys.} {{\em B567}} (2000) 61--86,
\href{http://arxiv.org/abs/hep-th/9907200}{{\ttfamily hep-th/9907200}}.

\bibitem{Kagan:2005wt}
D.~Kagan and C.~A.~S. Young, ``Conformal sigma-models on supercoset targets,''
  {\em Nucl. Phys.} {{\em B745}} (2006) 109--122,
\href{http://arxiv.org/abs/hep-th/0512250}{{\ttfamily hep-th/0512250}}.

\bibitem{Babichenko:2006uc}
A.~Babichenko, ``Conformal invariance and quantum integrability of sigma models
  on symmetric superspaces,'' {\em Phys. Lett.} {{\em B648}} (2007) 254--261,
\href{http://arxiv.org/abs/hep-th/0611214}{{\ttfamily hep-th/0611214}}.

\bibitem{Berkovits:1999im}
N.~Berkovits, C.~Vafa, and E.~Witten, ``Conformal field theory of {AdS}
  background with {Ramond-Ramond} flux,'' {\em JHEP} {{\em 03}} (1999) 018,
\href{http://arxiv.org/abs/hep-th/9902098}{{\ttfamily hep-th/9902098}}.

\bibitem{Bershadsky:1999hk}
M.~Bershadsky, S.~Zhukov, and A.~Vaintrob, ``{$PSL(n|n)$} sigma model as a
  conformal field theory,'' {\em Nucl. Phys.} {{\em B559}} (1999) 205--234,
\href{http://arxiv.org/abs/hep-th/9902180}{{\ttfamily hep-th/9902180}}.

\bibitem{Gotz:2006qp}
G.~{G{\"o}tz}, T.~Quella, and V.~Schomerus, ``The {WZNW} model on
  {$PSU(1,1|2)$},'' {\em JHEP} {{\em 03}} (2007) 003,
\href{http://arxiv.org/abs/hep-th/0610070}{{\ttfamily hep-th/0610070}}.

\bibitem{Bena:2003wd}
I.~Bena, J.~Polchinski, and R.~Roiban, ``Hidden symmetries of the {$AdS_5\times
  S^5$} superstring,'' {\em Phys. Rev.} {{\em D69}} (2004) 046002,
\href{http://arxiv.org/abs/hep-th/0305116}{{\ttfamily hep-th/0305116}}.

\bibitem{Young:2005jv}
C.~A.~S. Young, ``Non-local charges, {$Z_m$} gradings and coset space
  actions,'' {\em Phys. Lett.} {{\em B632}} (2006) 559--565,
\href{http://arxiv.org/abs/hep-th/0503008}{{\ttfamily hep-th/0503008}}.

\bibitem{Gurarie:1993xq}
V.~Gurarie, ``Logarithmic operators in conformal field theory,'' {\em Nucl.
  Phys.} {{\em B410}} (1993) 535--549,
\href{http://arxiv.org/abs/hep-th/9303160}{{\ttfamily hep-th/9303160}}.

\bibitem{Kausch:2000fu}
H.~G. Kausch, ``Symplectic fermions,'' {\em Nucl. Phys.} {{\em B583}} (2000)
  513--541,
\href{http://arxiv.org/abs/hep-th/0003029}{{\ttfamily hep-th/0003029}}.

\bibitem{Gaberdiel:2001tr}
M.~R. Gaberdiel, ``An algebraic approach to logarithmic conformal field
  theory,'' {\em Int. J. Mod. Phys.} {{\em A18}} (2003) 4593--4638,
\href{http://arxiv.org/abs/hep-th/0111260}{{\ttfamily hep-th/0111260}}.

\bibitem{Flohr:2001zs}
M.~Flohr, ``Bits and pieces in logarithmic conformal field theory,'' {\em Int.
  J. Mod. Phys.} {{\em A18}} (2003) 4497--4592,
\href{http://arxiv.org/abs/hep-th/0111228}{{\ttfamily hep-th/0111228}}.

\bibitem{Rozansky:1992rx}
L.~Rozansky and H.~Saleur, ``Quantum field theory for the multivariable
  {Alexander-Conway} polynomial,''
{\em Nucl. Phys.} {{\em B376}} (1992) 461--509.

\bibitem{Rozansky:1992td}
L.~Rozansky and H.~Saleur, ``{S} and {T} matrices for the {$U(1|1)$} {WZW}
  model: {Application} to surgery and three manifolds invariants based on the
  {Alexander-Conway} polynomial,'' {\em Nucl. Phys.} {{\em B389}} (1993)
  365--423,
\href{http://arxiv.org/abs/hep-th/9203069}{{\ttfamily hep-th/9203069}}.

\bibitem{Maassarani:1996jn}
Z.~Maassarani and D.~Serban, ``Non-unitary conformal field theory and
  logarithmic operators for disordered systems,'' {\em Nucl. Phys.} {{\em
  B489}} (1997) 603--625,
\href{http://arxiv.org/abs/hep-th/9605062}{{\ttfamily hep-th/9605062}}.

\bibitem{Guruswamy:1999hi}
S.~Guruswamy, A.~Le{\,}Clair, and A.~W.~W. Ludwig, ``{$gl(N|N)$} super-current
  algebras for disordered {Dirac} fermions in two dimensions,'' {\em Nucl.
  Phys.} {{\em B583}} (2000) 475--512,
\href{http://arxiv.org/abs/cond-mat/9909143}{{\ttfamily cond-mat/9909143}}.

\bibitem{Ludwig:2000em}
A.~W.~W. Ludwig, ``A free field representation of the {$osp(2|2)$} current
  algebra at level {$k=-2$}, and {Dirac} fermions in a random {$SU(2)$} gauge
  potential,''
\href{http://arxiv.org/abs/cond-mat/0012189}{{\ttfamily cond-mat/0012189}}.

\bibitem{Schomerus:2005bf}
V.~Schomerus and H.~Saleur, ``The {$GL(1|1)$} {WZW} model: {From} supergeometry
  to logarithmic {CFT},'' {\em Nucl. Phys.} {{\em B734}} (2006) 221--245,
\href{http://arxiv.org/abs/hep-th/0510032}{{\ttfamily hep-th/0510032}}.

\bibitem{Saleur:2006tf}
H.~Saleur and V.~Schomerus, ``On the {$SU(2|1)$} {WZW} model and its
  statistical mechanics applications,'' {\em Nucl. Phys.} {{\em B775}} (2007)
  312--340,
\href{http://arxiv.org/abs/hep-th/0611147}{{\ttfamily hep-th/0611147}}.

\bibitem{Quella:2007hr}
T.~Quella and V.~Schomerus, ``Free fermion resolution of supergroup {WZNW}
  models,'' \href{http://dx.doi.org/10.1088/1126-6708/2007/09/085}{{\em JHEP}
  {{\em 09}} (2007) 085},
\href{http://arxiv.org/abs/0706.0744}{{\ttfamily arXiv:0706.0744}}.

\bibitem{Mitev:2008yt}
V.~Mitev, T.~Quella, and V.~Schomerus, ``Principal chiral model on
  superspheres,'' \href{http://dx.doi.org/10.1088/1126-6708/2008/11/086}{{\em
  JHEP} {{\em 11}} (2008) 086},
\href{http://arxiv.org/abs/0809.1046}{{\ttfamily arXiv:0809.1046}}.

\bibitem{Kac:1977em}
V.~G. Kac, ``Lie superalgebras,''
{\em Adv. Math.} {{\em 26}} (1977) 8--96.

\bibitem{Scheunert:1976uf}
M.~Scheunert, W.~Nahm, and V.~Rittenberg, ``Classification of all simple graded
  {Lie} algebras whose {Lie} algebra is reductive. 1,''
{\em J. Math. Phys.} {{\em 17}} (1976) 1626.

\bibitem{FrancescoCFT}
P.~{Di Francesco}, P.~Mathieu, and D.~Senechal, {\em {Conformal Field Theory}}.
\newblock Graduate Texts in Contemporary Physics. Springer, New York, 1999.

\bibitem{Fuchs:1995}
J.~Fuchs, {\em Affine Lie algebras and quantum groups}.
\newblock Cambridge Monographs on Mathematical Physics. Cambridge University
  Press, Cambridge, 1995.

\bibitem{FuchsSchweigert}
J.~Fuchs and C.~Schweigert, {\em Symmetries, {Lie} algebras and
  representations}.
\newblock Cambridge Monographs on Mathematical Physics. Cambridge University
  Press, Cambridge, 1997.

\bibitem{Frappat:1996pb}
L.~Frappat, P.~Sorba, and A.~Sciarrino, {\em Dictionary on {Lie} algebras and
  superalgebras}.
\newblock Academic Press Inc., San Diego, CA, 2000.
\newblock \href{http://arxiv.org/abs/hep-th/9607161}{{\ttfamily
  hep-th/9607161}}.
\newblock
Extended and corrected version of the E-print [hep-th/9607161].

\bibitem{Kac:1994kn}
V.~G. Kac and M.~Wakimoto, ``Integrable highest weight modules over affine
  superalgebras and number theory,'' in {\em Lie theory and geometry}, vol.~123
  of {\em Progr. Math.}, pp.~415--456.
\newblock Birkh{\"a}user Boston, Boston, MA, 1994.
\newblock
\href{http://arxiv.org/abs/hep-th/9407057}{{\ttfamily hep-th/9407057}}.
\newblock

\bibitem{Nahm:1977tg}
W.~Nahm, ``Supersymmetries and their representations,''
\href{http://dx.doi.org/10.1016/0550-3213(78)90218-3}{{\em Nucl. Phys.} {{\em
  B135}} (1978) 149}.

\bibitem{Berkovits:2008ga}
N.~Berkovits, ``Simplifying and extending the {$AdS_5\times S^5$} pure spinor
  formalism,'' \href{http://dx.doi.org/10.1088/1126-6708/2009/09/051}{{\em
  JHEP} {{\em 09}} (2009) 051},
\href{http://arxiv.org/abs/0812.5074}{{\ttfamily arXiv:0812.5074}}.

\bibitem{Kac1977:MR0444725}
V.~G. Kac, ``Characters of typical representations of classical {L}ie
  superalgebras,'' {\em Comm. Algebra} {{\em 5}} (1977) 889--897.

\bibitem{Kac:1977MR519631}
V.~Kac, ``Representations of classical {L}ie superalgebras,'' in {\em
  Differential geometrical methods in mathematical physics, {II} ({P}roc.
  {C}onf., {U}niv. {B}onn, {B}onn, 1977)}, vol.~676 of {\em Lecture Notes in
  Math.}, pp.~597--626.
\newblock Springer, Berlin, 1978.

\bibitem{Cheng:2012MR3012224}
S.-J. Cheng and W.~Wang, {\em Dualities and representations of {L}ie
  superalgebras}, vol.~144 of {\em Graduate Studies in Mathematics}.
\newblock American Mathematical Society, Providence, RI, 2012.

\bibitem{Germoni1998:MR1659915}
J.~Germoni, ``Indecomposable representations of special linear {L}ie
  superalgebras,'' {\em J. Algebra} {{\em 209}} (1998) 367--401.

\bibitem{Serganova1998:MR1648107}
V.~Serganova, ``Characters of irreducible representations of simple {L}ie
  superalgebras,'' in {\em Proceedings of the International Congress of
  Mathematicians, Vol. II (Berlin, 1998)}, no.~Extra Vol. II, pp.~583--593
  (electronic).
\newblock 1998.

\bibitem{Brundan2001:MR1937204}
J.~Brundan, ``Kazhdan-{L}usztig polynomials and character formulae for the
  {L}ie superalgebra {$gl(m|n)$},'' {\em J. Amer. Math. Soc.} {{\em 16}} (2003)
  185--231 (electronic), \href{http://arxiv.org/abs/math.RT/0203011}{{\ttfamily
  math.RT/0203011}}.

\bibitem{Brundan:MR2881300}
J.~Brundan and C.~Stroppel, ``Highest weight categories arising from
  {K}hovanov's diagram algebra {IV}: the general linear supergroup,''
  \href{http://dx.doi.org/10.4171/JEMS/306}{{\em J. Eur. Math. Soc. (JEMS)}
  {{\em 14}} (2012) 373--419}, \href{http://arxiv.org/abs/0907.2543}{{\ttfamily
  arXiv:0907.2543}}.

\bibitem{Gruson:2010MR2734963}
C.~Gruson and V.~Serganova, ``Cohomology of generalized supergrassmannians and
  character formulae for basic classical {Lie} superalgebras,''
  \href{http://dx.doi.org/10.1112/plms/pdq014}{{\em Proc. Lond. Math. Soc.}
  {{\em 101}} (2010) 852--892},
  \href{http://arxiv.org/abs/0906.0918}{{\ttfamily arXiv:0906.0918}}.

\bibitem{Musson:2011MR2806501}
I.~M. Musson and V.~Serganova, ``Combinatorics of character formulas for the
  {L}ie superalgebra {$gl(m,n)$},''
  \href{http://dx.doi.org/10.1007/s00031-011-9147-4}{{\em Transform. Groups}
  {{\em 16}} (2011) 555--578}, \href{http://arxiv.org/abs/1104.1668}{{\ttfamily
  arXiv:1104.1668}}.

\bibitem{Cheng:2011MR2755062}
S.-J. Cheng, N.~Lam, and W.~Wang, ``Super duality and irreducible characters of
  ortho-symplectic {L}ie superalgebras,''
  \href{http://dx.doi.org/10.1007/s00222-010-0277-4}{{\em Invent. Math.} {{\em
  183}} (2011) 189--224}, \href{http://arxiv.org/abs/0911.0129}{{\ttfamily
  arXiv:0911.0129}}.

\bibitem{Germoni2000:MR1840448}
J.~Germoni, ``Indecomposable representations of {$osp(3|2)$}, {$D(2,1;\alpha)$}
  and {$G(3)$},'' {\em Bol. Acad. Nac. Cienc. (C{\'o}rdoba)} {{\em 65}} (2000)
  147--163. Colloquium on Homology and Representation Theory (Spanish)
  (Vaquer{\'i} as, 1998).

\bibitem{Gotz:2005jz}
G.~{G{\"o}tz}, T.~Quella, and V.~Schomerus, ``Representation theory of
  {$sl(2|1)$},'' \href{http://dx.doi.org/10.1016/j.jalgebra.2007.03.012}{{\em
  J. Algebra} {{\em 312}} (2007) 829--848},
\href{http://arxiv.org/abs/hep-th/0504234}{{\ttfamily hep-th/0504234}}.

\bibitem{Gotz:2005ka}
G.~{G{\"o}tz}, T.~Quella, and V.~Schomerus, ``Tensor products of {$psl(2|2)$}
  representations,''
\href{http://arxiv.org/abs/hep-th/0506072}{{\ttfamily hep-th/0506072}}.

\bibitem{Dobrev:2004tk}
V.~Dobrev, ``{Characters of the positive energy UIRs of D=4 conformal
  supersymmetry},'' \href{http://dx.doi.org/10.1134/S1063779607050024}{{\em
  Phys.Part.Nucl.} {{\em 38}} (2007) 564--609},
\href{http://arxiv.org/abs/hep-th/0406154}{{\ttfamily arXiv:hep-th/0406154}}.

\bibitem{Bianchi:2006ti}
M.~Bianchi, F.~Dolan, P.~Heslop, and H.~Osborn, ``{N=4 superconformal
  characters and partition functions},''
  \href{http://dx.doi.org/10.1016/j.nuclphysb.2006.12.005}{{\em Nucl.Phys.}
  {{\em B767}} (2007) 163--226},
\href{http://arxiv.org/abs/hep-th/0609179}{{\ttfamily arXiv:hep-th/0609179}}.

\bibitem{Dolan:2008vc}
F.~Dolan, ``{On superconformal characters and partition functions in three
  dimensions},'' \href{http://dx.doi.org/10.1063/1.3211091}{{\em J.Math.Phys.}
  {{\em 51}} (2010) 022301},
\href{http://arxiv.org/abs/0811.2740}{{\ttfamily arXiv:0811.2740}}.

\bibitem{Dobrev:2012me}
V.~Dobrev, ``{Explicit character formulae for positive energy unitary
  irreducible representations of D = 4 conformal supersymmetry},''
  \href{http://dx.doi.org/10.1088/1751-8113/46/40/405202}{{\em J.Phys.} {{\em
  A46}} (2013) 405202},
\href{http://arxiv.org/abs/1208.6250}{{\ttfamily arXiv:1208.6250}}.

\bibitem{AndersonFuller}
F.~W. Anderson and K.~R. Fuller, {\em Rings and categories of modules}.
\newblock Springer, 1992.

\bibitem{Brundan2004:MR2100468}
J.~Brundan, ``Tilting modules for {L}ie superalgebras,'' {\em Comm. Algebra}
  {{\em 32}} (2004) 2251--2268,
  \href{http://arxiv.org/abs/math.RT/0209235}{{\ttfamily math.RT/0209235}}.

\bibitem{Humphreys:2008MR2428237}
J.~E. Humphreys, {\em Representations of semisimple {L}ie algebras in the {BGG}
  category {$O$}}, vol.~94 of {\em Graduate Studies in Mathematics}.
\newblock American Mathematical Society, Providence, RI, 2008.

\bibitem{Musson:2012MR2906817}
I.~M. Musson, {\em Lie superalgebras and enveloping algebras}, vol.~131 of {\em
  Graduate Studies in Mathematics}.
\newblock American Mathematical Society, Providence, RI, 2012.

\bibitem{Zou1996:MR1378540}
Y.~M. Zou, ``Categories of finite-dimensional weight modules over type {I}
  classical {L}ie superalgebras,'' {\em J. Algebra} {{\em 180}} (1996)
  459--482.

\bibitem{Serganova:Blocks}
V.~Serganova, ``Blocks in the category of finite-dimensional representations of
  {$gl(m|n)$}.'' Preprint available on V.\ Serganova's homepage, 2006.

\bibitem{Mitev:2011zza}
V.~Mitev, T.~Quella, and V.~Schomerus, ``{Conformal superspace sigma-models},''
  \href{http://dx.doi.org/10.1016/j.geomphys.2010.11.004}{{\em J. Geom. Phys.}
  {{\em 61}} (2011) 1703--1716},
  \href{http://arxiv.org/abs/1210.8159}{{\ttfamily arXiv:1210.8159}}.

\bibitem{Witten:2012bg}
E.~Witten, ``Notes on supermanifolds and integration,''
\href{http://arxiv.org/abs/1209.2199}{{\ttfamily arXiv:1209.2199}}.

\bibitem{Leites:1980MR565567}
D.~A. Leites, ``Introduction to the theory of supermanifolds,''
  \href{http://dx.doi.org/10.1070/RM1980v035n01ABEH001545}{{\em Russian
  Mathematical Surveys} {{\em 35}} (1980) 1--64}.

\bibitem{Varadarajan:2004MR2069561}
V.~S. Varadarajan, {\em Supersymmetry for mathematicians: an introduction},
  vol.~11 of {\em Courant Lecture Notes in Mathematics}.
\newblock New York University Courant Institute of Mathematical Sciences, New
  York, 2004.

\bibitem{Carmeli:2011MR2840967}
C.~Carmeli, L.~Caston, and R.~Fioresi,
  \href{http://dx.doi.org/10.4171/097}{{\em Mathematical foundations of
  supersymmetry}}.
\newblock EMS Series of Lectures in Mathematics. European Mathematical Society
  (EMS), Z\"urich, 2011.

\bibitem{Alldridge:Book}
A.~Alldridge, J.~Hilgert, and T.~Wurzbacher, {\em Calculus on supermanifolds}.
\newblock Book in preparation.

\bibitem{Manin:1998}
Y.~I. Manin, {\em Gauge Field Theory and Complex Geometry}.
\newblock Grundlehren der mathematischen Wissenschaften. Springer, 1998.

\bibitem{Deligne:1999MR1701597}
P.~Deligne and J.~W. Morgan, ``Notes on supersymmetry (following {J}oseph
  {B}ernstein),'' in {\em Quantum fields and strings: a course for
  mathematicians, {V}ol. 1, 2 ({P}rinceton, {NJ}, 1996/1997)}, pp.~41--97.
\newblock Amer. Math. Soc., Providence, RI, 1999.

\bibitem{Bartocci:1991MR1175751}
C.~Bartocci, U.~Bruzzo, and D.~Hern{\'a}ndez~Ruip{\'e}rez,
  \href{http://dx.doi.org/10.1007/978-94-011-3504-7}{{\em The geometry of
  supermanifolds}}, vol.~71 of {\em Mathematics and its Applications}.
\newblock Kluwer Academic Publishers Group, Dordrecht, 1991.

\bibitem{DeWitt:1992MR1172996}
B.~DeWitt, \href{http://dx.doi.org/10.1017/CBO9780511564000}{{\em
  Supermanifolds}}.
\newblock Cambridge Monographs on Mathematical Physics. Cambridge University
  Press, Cambridge, 1992.

\bibitem{Tuynman:2004MR2102797}
G.~M. Tuynman, {\em Supermanifolds and supergroups -- Basic theory}, vol.~570
  of {\em Mathematics and its Applications}.
\newblock Kluwer Academic Publishers, Dordrecht, 2004.

\bibitem{Rogers:2007MR2320438}
A.~Rogers, \href{http://dx.doi.org/10.1142/9789812708854}{{\em Supermanifolds
  -- Theory and applications}}.
\newblock World Scientific Publishing, 2007.

\bibitem{Coulembier:2012arXiv1202.0668C}
K.~{Coulembier}, ``{The orthosymplectic supergroup in harmonic analysis},''
  {\em J. Lie Theory} {{\em 23}} (2013) 55--83,
  \href{http://arxiv.org/abs/1202.0668}{{\ttfamily arXiv:1202.0668}}.

\bibitem{Candu:2010yg}
C.~Candu, T.~Creutzig, V.~Mitev, and V.~Schomerus, ``Cohomological reduction of
  sigma models,'' \href{http://dx.doi.org/10.1007/JHEP05(2010)047}{{\em JHEP}
  {{\em 05}} (2010) 047},
\href{http://arxiv.org/abs/1001.1344}{{\ttfamily arXiv:1001.1344}}.

\bibitem{Candu:2009ep}
C.~Candu, V.~Mitev, T.~Quella, H.~Saleur, and V.~Schomerus, ``The sigma model
  on complex projective superspaces,''
  \href{http://dx.doi.org/10.1007/JHEP02(2010)015}{{\em JHEP} {{\em 02}} (2010)
  015},
\href{http://arxiv.org/abs/0908.0878}{{\ttfamily arXiv:0908.0878}}.

\bibitem{Witten:1983ar}
E.~Witten, ``Nonabelian bosonization in two dimensions,''
\href{http://dx.doi.org/10.1007/BF01215276}{{\em Commun. Math. Phys.} {{\em
  92}} (1984) 455--472}.

\bibitem{Gates:1983nr}
S.~J. Gates, M.~T. Grisaru, M.~Rocek, and W.~Siegel, ``Superspace, or one
  thousand and one lessons in supersymmetry,'' {\em Front. Phys.} {{\em 58}}
  (1983) 1--548,
\href{http://arxiv.org/abs/hep-th/0108200}{{\ttfamily hep-th/0108200}}.

\bibitem{Polyakov:1984et}
A.~M. Polyakov and P.~B. Wiegmann, ``Goldstone fields in two-dimensions with
  multivalued actions,''
{\em Phys. Lett.} {{\em B141}} (1984) 223--228.

\bibitem{Kac:2001}
V.~G. Kac and M.~Wakimoto, ``Integrable highest weights modules over affine
  superalgebras and {Appell's} function,'' {\em Comm. Math. Phys.} {{\em 215}}
  (2001) 631--682, \href{http://arxiv.org/abs/math-ph/0006007}{{\ttfamily
  math-ph/0006007}}.

\bibitem{Serganova:2011MR2743764}
V.~Serganova,
  \href{http://dx.doi.org/10.1007/978-0-8176-4741-4_6}{``Kac-{M}oody
  superalgebras and integrability,''} in {\em Developments and trends in
  infinite-dimensional {L}ie theory}, vol.~288 of {\em Progr. Math.},
  pp.~169--218.
\newblock Birkh\"auser Boston Inc., Boston, MA, 2011.

\bibitem{Gorelik:2011MR2796063}
M.~Gorelik, ``Weyl denominator identity for affine {L}ie superalgebras with
  non-zero dual {C}oxeter number,''
  \href{http://dx.doi.org/10.1016/j.jalgebra.2011.04.011}{{\em J. Algebra}
  {{\em 337}} (2011) 50--62}, \href{http://arxiv.org/abs/0911.5594}{{\ttfamily
  arXiv:0911.5594}}.

\bibitem{Gorelik:2012MR2968633}
M.~Gorelik and S.~Reif, ``A denominator identity for affine {L}ie superalgebras
  with zero dual {C}oxeter number,''
  \href{http://dx.doi.org/10.2140/ant.2012.6.1043}{{\em Algebra Number Theory}
  {{\em 6}} (2012) 1043--1059},
  \href{http://arxiv.org/abs/1012.5879}{{\ttfamily arXiv:1012.5879}}.

\bibitem{Serganova:2012MR2866847}
V.~Serganova, \href{http://dx.doi.org/10.1007/978-0-8176-8274-3_3}{``Structure
  and representation theory of {K}ac-{M}oody superalgebras,''} in {\em
  Highlights in {L}ie algebraic methods}, vol.~295 of {\em Progr. Math.},
  pp.~65--102.
\newblock Birkh\"auser/Springer, New York, 2012.

\bibitem{Zwegers:2008}
S.~{Zwegers}, {\em Mock theta functions}.
\newblock PhD thesis, Utrecht University, 2002.
\newblock \href{http://arxiv.org/abs/0807.4834}{{\ttfamily arXiv:0807.4834}}.

\bibitem{Folsom:MR2719689}
A.~Folsom, ``Kac-{W}akimoto characters and universal mock theta functions,''
  \href{http://dx.doi.org/10.1090/S0002-9947-2010-05181-5}{{\em Trans. Amer.
  Math. Soc.} {{\em 363}} (2011) 439--455}.

\bibitem{Bringmann:MR2534107}
K.~Bringmann and K.~Ono, ``Some characters of {K}ac and {W}akimoto and
  nonholomorphic modular functions,''
  \href{http://dx.doi.org/10.1007/s00208-009-0364-2}{{\em Math. Ann.} {{\em
  345}} (2009) 547--558}.

\bibitem{Creutzig:2007jy}
T.~Creutzig, T.~Quella, and V.~Schomerus, ``Branes in the {$GL(1|1)$} {WZNW}
  model,'' \href{http://dx.doi.org/10.1016/j.nuclphysb.2007.09.014}{{\em Nucl.
  Phys.} {{\em B792}} (2008) 257--283},
\href{http://arxiv.org/abs/0708.0583}{{\ttfamily arXiv:0708.0583}}.

\bibitem{Maldacena:2000hw}
J.~M. Maldacena and H.~Ooguri, ``Strings in {$AdS_3$} and the {$SL(2,R)$} {WZW}
  model. {I},'' {\em J. Math. Phys.} {{\em 42}} (2001) 2929--2960,
\href{http://arxiv.org/abs/hep-th/0001053}{{\ttfamily hep-th/0001053}}.

\bibitem{Gepner:1986wi}
D.~Gepner and E.~Witten, ``String theory on group manifolds,''
{\em Nucl. Phys.} {{\em B278}} (1986) 493.

\bibitem{Feigin:1990qn}
B.~L. Feigin and E.~V. Frenkel, ``Representations of affine {Kac-Moody}
  algebras, bosonization and resolutions,''
{\em Lett. Math. Phys.} {{\em 19}} (1990) 307--317.

\bibitem{Rasmussen:1998cc}
J.~Rasmussen, ``Free field realizations of affine current superalgebras,
  screening currents and primary fields,'' {\em Nucl. Phys.} {{\em B510}}
  (1998) 688--720,
\href{http://arxiv.org/abs/hep-th/9706091}{{\ttfamily hep-th/9706091}}.

\bibitem{Candu:2008vw}
C.~Candu and H.~Saleur, ``A lattice approach to the conformal {$OSp(2S+2|2S)$}
  supercoset sigma model. {Part I:} {Algebraic} structures in the spin chain.
  {The} {Brauer} algebra,''
  \href{http://dx.doi.org/10.1016/j.nuclphysb.2008.09.034}{{\em Nucl. Phys.}
  {{\em B808}} (2009) 441--486},
\href{http://arxiv.org/abs/0801.0430}{{\ttfamily arXiv:0801.0430}}.

\bibitem{Candu:2008yw}
C.~Candu and H.~Saleur, ``A lattice approach to the conformal {$OSp(2S+2|2S)$}
  supercoset sigma model. {Part II:} {The} boundary spectrum,''
  \href{http://dx.doi.org/10.1016/j.nuclphysb.2008.08.015}{{\em Nucl. Phys.}
  {{\em B808}} (2009) 487--524},
\href{http://arxiv.org/abs/0801.0444}{{\ttfamily arXiv:0801.0444}}.

\bibitem{Candu:2008PhD}
C.~Candu, {\em Discr{\'e}tisation des mod{\`e}les sigma invariants conformes
  sur des supersph{\`e}res et superespaces projectifs}.
\newblock PhD thesis, University Pierre and Marie Curie -- Paris 6, 2008.

\bibitem{Lesage:2002ch}
F.~Lesage, P.~Mathieu, J.~Rasmussen, and H.~Saleur, ``The
  {$\widehat{su}(2)_{-1/2}$} {WZW} model and the {$\beta\gamma$} system,'' {\em
  Nucl. Phys.} {{\em B647}} (2002) 363--403,
\href{http://arxiv.org/abs/hep-th/0207201}{{\ttfamily hep-th/0207201}}.

\bibitem{Ridout2011jk}
D.~Ridout, ``Fusion in fractional level {$\widehat{sl}(2)$}-theories with
  {$k=-1/2$},'' \href{http://dx.doi.org/10.1016/j.nuclphysb.2011.02.015}{{\em
  Nucl. Phys.} {{\em B848}} (2011) 216--250},
\href{http://arxiv.org/abs/1012.2905}{{\ttfamily arXiv:1012.2905}}.

\bibitem{Schellekens:1990xy}
A.~N. Schellekens and S.~Yankielowicz, ``Simple currents, modular invariants
  and fixed points,''
{\em Int. J. Mod. Phys.} {{\em A5}} (1990) 2903--2952.

\bibitem{Lesage:2003kn}
F.~Lesage, P.~Mathieu, J.~Rasmussen, and H.~Saleur, ``Logarithmic lift of the
  {$\widehat{su}(2)_{-1/2}$} model,'' {\em Nucl. Phys.} {{\em B686}} (2004)
  313,
\href{http://arxiv.org/abs/hep-th/0311039}{{\ttfamily hep-th/0311039}}.

\bibitem{Ridout:2008nh}
D.~Ridout, ``{$sl(2)_{-1/2}$}: {A} case study,''
  \href{http://dx.doi.org/10.1016/j.nuclphysb.2009.01.008}{{\em Nucl. Phys.}
  {{\em B814}} (2009) 485--521},
\href{http://arxiv.org/abs/0810.3532}{{\ttfamily arXiv:0810.3532}}.

\bibitem{Ridout:2010qx}
D.~Ridout, ``{$sl(2)_{-1/2}$} and the triplet model,''
  \href{http://dx.doi.org/10.1016/j.nuclphysb.2010.03.018}{{\em Nucl. Phys.}
  {{\em B835}} (2010) 314--342},
\href{http://arxiv.org/abs/1001.3960}{{\ttfamily arXiv:1001.3960}}.

\bibitem{Creutzig:2012sd}
T.~Creutzig and D.~Ridout, ``Modular data and {Verlinde} formulae for
  fractional level {WZW} models {I},''
  \href{http://dx.doi.org/10.1016/j.nuclphysb.2012.07.018}{{\em Nucl. Phys.}
  {{\em B865}} (2012) 83--114},
\href{http://arxiv.org/abs/1205.6513}{{\ttfamily arXiv:1205.6513}}.

\bibitem{Ashok:2009xx}
S.~K. Ashok, R.~Benichou, and J.~Troost, ``Conformal current algebra in two
  dimensions,'' \href{http://dx.doi.org/10.1088/1126-6708/2009/06/017}{{\em
  JHEP} {{\em 06}} (2009) 017},
\href{http://arxiv.org/abs/0903.4277}{{\ttfamily arXiv:0903.4277}}.

\bibitem{Ashok:2009jw}
S.~K. Ashok, R.~Benichou, and J.~Troost, ``Asymptotic symmetries of string
  theory on {$AdS_3\times S^3$} with {Ramond-Ramond} fluxes,''
  \href{http://dx.doi.org/10.1088/1126-6708/2009/10/051}{{\em JHEP} {{\em 10}}
  (2009) 051},
\href{http://arxiv.org/abs/0907.1242}{{\ttfamily arXiv:0907.1242}}.

\bibitem{Benichou:2010rk}
R.~Benichou and J.~Troost, ``The conformal current algebra on supergroups with
  applications to the spectrum and integrability,''
  \href{http://dx.doi.org/10.1007/JHEP04(2010)121}{{\em JHEP} {{\em 04}} (2010)
  121},
\href{http://arxiv.org/abs/1002.3712}{{\ttfamily arXiv:1002.3712}}.

\bibitem{Konechny:2010nq}
A.~Konechny and T.~Quella, ``{Non-chiral current algebras for deformed
  supergroup WZW models},''
  \href{http://dx.doi.org/10.1007/JHEP03(2011)124}{{\em JHEP} {{\em 1103}}
  (2011) 124},
\href{http://arxiv.org/abs/1011.4813}{{\ttfamily arXiv:1011.4813}}.

\bibitem{Quella:2007sg}
T.~Quella, V.~Schomerus, and T.~Creutzig, ``Boundary spectra in superspace
  sigma models,'' \href{http://dx.doi.org/10.1088/1126-6708/2008/10/024}{{\em
  JHEP} {{\em 10}} (2008) 024},
\href{http://arxiv.org/abs/0712.3549}{{\ttfamily arXiv:0712.3549}}.

\bibitem{Obuse:2008nc}
H.~Obuse, A.~R. Subramaniam, A.~Furusaki, I.~A. Gruzberg, and A.~W.~W. Ludwig,
  ``Boundary multifractality at the integer quantum {Hall} plateau transition:
  Implications for the critical theory,'' {\em Phys. Rev. Lett.} {{\em 101}}
  (2008) 116802,
\href{http://arxiv.org/abs/0804.2409}{{\ttfamily arXiv:0804.2409}}.

\bibitem{Candu:2012xc}
C.~Candu, V.~Mitev, and V.~Schomerus, ``Anomalous dimensions in deformed {WZW}
  models on supergroups,''
  \href{http://dx.doi.org/10.1007/JHEP03(2013)003}{{\em JHEP} {{\em 1303}}
  (2013) 003},
\href{http://arxiv.org/abs/1211.2238}{{\ttfamily arXiv:1211.2238}}.

\bibitem{Gepner:1987qi}
D.~Gepner, ``Space-time supersymmetry in compactified string theory and
  superconformal models,''
\href{http://dx.doi.org/10.1016/0550-3213(88)90397-5}{{\em Nucl. Phys.} {{\em
  B296}} (1988) 757}.

\bibitem{Giveon:1993ew}
A.~Giveon and M.~Rocek, ``On the {BRST} operator structure of the {N=2}
  string,'' \href{http://dx.doi.org/10.1016/0550-3213(93)90401-A}{{\em Nucl.
  Phys.} {{\em B400}} (1993) 145--160},
\href{http://arxiv.org/abs/hep-th/9302049}{{\ttfamily arXiv:hep-th/9302049}}.

\bibitem{Berkovits:1994wr}
N.~Berkovits, ``Covariant quantization of the {Green-Schwarz} superstring in a
  {Calabi-Yau} background,''
  \href{http://dx.doi.org/10.1016/0550-3213(94)90106-6}{{\em Nucl. Phys.} {{\em
  B431}} (1994) 258--272},
\href{http://arxiv.org/abs/hep-th/9404162}{{\ttfamily arXiv:hep-th/9404162}}.

\bibitem{Berkovits:1999du}
N.~Berkovits, ``Quantization of the type {II} superstring in a curved six-
  dimensional background,'' {\em Nucl. Phys.} {{\em B565}} (2000) 333--344,
\href{http://arxiv.org/abs/hep-th/9908041}{{\ttfamily hep-th/9908041}}.

\bibitem{Cagnazzo:2012se}
A.~Cagnazzo and K.~Zarembo, ``B-field in {$AdS_3/CFT_2$} correspondence and
  integrability,'' \href{http://dx.doi.org/10.1007/JHEP11(2012)133}{{\em JHEP}
  {{\em 1211}} (2012) 133},
\href{http://arxiv.org/abs/1209.4049}{{\ttfamily arXiv:1209.4049}}.

\bibitem{Gaberdiel:2011vf}
M.~R. Gaberdiel and S.~Gerigk, ``The massless string spectrum on {$AdS_3\times
  S^3$} from the supergroup,''
  \href{http://dx.doi.org/10.1007/JHEP10(2011)045}{{\em JHEP} {{\em 1110}}
  (2011) 045},
\href{http://arxiv.org/abs/1107.2660}{{\ttfamily arXiv:1107.2660}}.

\bibitem{Gerigk:2012cq}
S.~Gerigk, ``String states on {$AdS_3\times S^3$} from the supergroup,''
  \href{http://dx.doi.org/10.1007/JHEP10(2012)084}{{\em JHEP} {{\em 1210}}
  (2012) 084},
\href{http://arxiv.org/abs/1208.0345}{{\ttfamily arXiv:1208.0345}}.

\bibitem{Troost:2011fd}
J.~Troost, ``Massless particles on supergroups and {$AdS_3\times S^3$}
  supergravity,'' \href{http://dx.doi.org/10.1007/JHEP07(2011)042}{{\em JHEP}
  {{\em 07}} (2011) 042},
\href{http://arxiv.org/abs/1102.0153}{{\ttfamily arXiv:1102.0153}}.

\bibitem{Berkovits:2000fe}
N.~Berkovits, ``Super-{Poincar{\'e}} covariant quantization of the
  superstring,'' {\em JHEP} {{\em 04}} (2000) 018,
\href{http://arxiv.org/abs/hep-th/0001035}{{\ttfamily hep-th/0001035}}.

\bibitem{Berkovits:2004xu}
N.~Berkovits, ``Quantum consistency of the superstring in {$AdS_5\times S^5$}
  background,'' \href{http://dx.doi.org/10.1088/1126-6708/2005/03/041}{{\em
  JHEP} {{\em 0503}} (2005) 041},
\href{http://arxiv.org/abs/hep-th/0411170}{{\ttfamily arXiv:hep-th/0411170}}.

\bibitem{Vallilo:2003nx}
B.~C. Vallilo, ``Flat currents in the classical {$AdS_5\times S^5$} pure spinor
  superstring,'' \href{http://dx.doi.org/10.1088/1126-6708/2004/03/037}{{\em
  JHEP} {{\em 0403}} (2004) 037},
\href{http://arxiv.org/abs/hep-th/0307018}{{\ttfamily arXiv:hep-th/0307018}}.

\bibitem{Benichou:2011ch}
R.~Benichou, ``First-principles derivation of the {AdS/CFT} {Y-systems},''
  \href{http://dx.doi.org/10.1007/JHEP10(2011)112}{{\em JHEP} {{\em 1110}}
  (2011) 112},
\href{http://arxiv.org/abs/1108.4927}{{\ttfamily arXiv:1108.4927}}.

\bibitem{Fre:2008qc}
P.~Fre and P.~A. Grassi, ``Pure spinor formalism for {$Osp(N|4)$}
  backgrounds,'' \href{http://dx.doi.org/10.1142/S0217751X12501850}{{\em Int.
  J. Mod. Phys.} {{\em A27}} (2012) 1250185},
\href{http://arxiv.org/abs/0807.0044}{{\ttfamily arXiv:0807.0044}}.

\bibitem{Sorokin:2011rr}
D.~Sorokin, A.~Tseytlin, L.~Wulff, and K.~Zarembo, ``Superstrings in
  {$AdS_2\times S^2\times T^6$},''
  \href{http://dx.doi.org/10.1088/1751-8113/44/27/275401}{{\em J. Phys.} {{\em
  A44}} (2011) 275401},
\href{http://arxiv.org/abs/1104.1793}{{\ttfamily arXiv:1104.1793}}.

\bibitem{Rahmfeld:1998zn}
J.~Rahmfeld and A.~Rajaraman, ``The {GS} string action on {$AdS_3\times S^3$}
  with {Ramond-Ramond} charge,'' {\em Phys. Rev.} {{\em D60}} (1999) 064014,
\href{http://arxiv.org/abs/hep-th/9809164}{{\ttfamily hep-th/9809164}}.

\bibitem{Arutyunov:2008if}
G.~Arutyunov and S.~Frolov, ``Superstrings on {$AdS_4 \times CP^3$} as a coset
  sigma-model,'' \href{http://dx.doi.org/10.1088/1126-6708/2008/09/129}{{\em
  JHEP} {{\em 09}} (2008) 129},
\href{http://arxiv.org/abs/0806.4940}{{\ttfamily arXiv:0806.4940}}.

\bibitem{Zarembo:2010sg}
K.~Zarembo, ``Strings on semisymmetric superspaces,''
  \href{http://dx.doi.org/10.1007/JHEP05(2010)002}{{\em JHEP} {{\em 05}} (2010)
  002},
\href{http://arxiv.org/abs/1003.0465}{{\ttfamily arXiv:1003.0465}}.

\bibitem{Cagnazzo:2012uq}
A.~Cagnazzo, D.~Sorokin, A.~A. Tseytlin, and L.~Wulff, ``Semiclassical
  equivalence of {Green-Schwarz} and pure-spinor/hybrid formulations of
  superstrings in {$AdS_5\times S^5$} and {$AdS_2\times S^2\times T^6$},''
  \href{http://dx.doi.org/10.1088/1751-8113/46/6/065401}{{\em J. Phys.} {{\em
  A46}} (2013) 065401},
\href{http://arxiv.org/abs/1211.1554}{{\ttfamily arXiv:1211.1554}}.

\bibitem{Witten:1993yc}
E.~Witten, ``Phases of {$N=2$} theories in two-dimensions,''
  \href{http://dx.doi.org/10.1016/0550-3213(93)90033-L}{{\em Nucl. Phys.} {{\em
  B403}} (1993) 159--222},
\href{http://arxiv.org/abs/hep-th/9301042}{{\ttfamily arXiv:hep-th/9301042}}.

\bibitem{Creutzig:2008ag}
T.~Creutzig, ``Geometry of branes on supergroups,''
  \href{http://dx.doi.org/10.1016/j.nuclphysb.2008.10.006}{{\em Nucl. Phys.}
  {{\em B812}} (2009) 301--321},
\href{http://arxiv.org/abs/0809.0468}{{\ttfamily arXiv:0809.0468}}.

\bibitem{Creutzig:2009fh}
T.~Creutzig, P.~B. R{\o}nne, and V.~Schomerus, ``{N=2} superconformal symmetry
  in super coset models,''
  \href{http://dx.doi.org/10.1103/PhysRevD.80.066010}{{\em Phys. Rev.} {{\em
  D80}} (2009) 066010},
\href{http://arxiv.org/abs/0907.3902}{{\ttfamily arXiv:0907.3902}}.

\bibitem{Candu:2011hu}
C.~Candu and V.~Schomerus, ``Exactly marginal parafermions,''
  \href{http://dx.doi.org/10.1103/PhysRevD.84.051704}{{\em Phys. Rev.} {{\em
  D84}} (2011) 051704},
\href{http://arxiv.org/abs/1104.5028}{{\ttfamily arXiv:1104.5028}}.

\bibitem{Alfes:2012pa}
C.~Alfes and T.~Creutzig, ``The {Mock} modular data of a family of
  superalgebras,''
\href{http://arxiv.org/abs/1205.1518}{{\ttfamily arXiv:1205.1518}}.

\end{thebibliography}\endgroup

\end{document}